\documentclass{jfm}
\usepackage{graphicx}
\usepackage{float}
\usepackage{epstopdf, epsfig}
\usepackage{upgreek}
\usepackage{subfigure}
\usepackage{amsmath}
\usepackage{amssymb}
\usepackage{setspace}
\usepackage[round]{natbib}
\usepackage{soul}
\usepackage{wasysym}
\usepackage{multirow}
\usepackage{placeins}
\usepackage[usenames,dvipsnames]{xcolor}
\usepackage{comment}

\def\lline{\vrule width14pt height2.5pt depth -2pt}
\def\mline{\vrule width4pt height2.5pt depth -2pt}
\def\m1line{\vrule width3pt height2.5pt depth -2pt}

\def\bdot{\raise.2em\hbox to .15em{.}}
\def\dashed{\m1line\hskip3.5pt\m1line\hskip3.5pt\m1line\thinspace}

\def\chaindot{\mline\ \bdot\ \mline\thinspace}

\def\dotted{\bdot\ \bdot\ \bdot\ \bdot\thinspace}

\allowdisplaybreaks[1]

\title{Concave-wall turbulent boundary layers without and with free-stream turbulence} 
\author{Jiho You, David A.\,Buchta
\and Tamer A.\,Zaki\thanks{Email address for correspondence: t.zaki@jhu.edu}}

\shorttitle{Turbulent boundary layers beneath free-stream turbulence on curved walls }
\shortauthor{J. You, D. A. Buchta and T. A. Zaki}

\affiliation{Department of Mechanical Engineering, Johns Hopkins University, Baltimore, MD 21218, USA}

\begin{document}
\maketitle 

\begin{abstract}
Direct numerical simulations are performed to contrast turbulent boundary layers over a concave wall without and with free-stream turbulence.
Adverse pressure gradient near the onset of curvature leads to sharp decrease in skin friction and intermittent separation. 
The presence of free-stream turbulence reduces the probability of reverse flow, accelerates the recovery of the boundary layer in the downstream zero-pressure gradient region, and leads to a sustained and appreciable increase in the skin friction.  
The forcing also promotes the amplification of coherent G\"ortler structures in the logarithmic layer of the curved-wall boundary layer. Statistically, the spanwise and wall-normal Reynolds stresses intensify and the radial distance between their peaks increases downstream as the G\"ortler structures expand. 
The Reynolds shear stress coefficient also increases in the logarithmic layer in contrast to a decrease when a flat-plate boundary layer is exposed to free-stream turbulence. 
In addition, the more coherent and energetic roll motions in the forced flow promote mixing of free-stream and boundary-layer fluids, where the former is seen more often deep within the buffer layer.
\end{abstract}

\section{Introduction}

In practical flow configurations, turbulent boundary layers (TBLs) often develop over curved surfaces.  
The focus in the present study is on concave curvature which induces centrifugal effects in addition to the external pressure gradient.
Another important practical consideration is the presence of environmental disturbances, or free-stream forcing, which can appreciably alter the dynamics within the underlying boundary layer. 
For example, when a flat-plate TBL is buffeted by free-stream turbulence (FST), its thickness and skin-friction coefficient increase appreciably, even when the free-stream fluid has not breached the logarithmic layer \citep{You_2019}.  
The present work uses direct numerical simulations (DNS) to probe the combined effects of the concave curvature and FST on the boundary layer.  Two simulations are contrasted, and correspond to the turbulent boundary layer on the concave wall developing beneath quiescent and vortical free streams (see figure \ref{fig_config}).

\subsection{Turbulent boundary layer on concave wall}

Curved-wall turbulent boundary layers are influenced by three effects~\citep{floryan1991gortler}: i)~turbulence changes due to the mean-flow turning, ii) potential formation of coherent G\"ortler vortices, and iii) the impact of these coherent vortices on the turbulence.
The second mechanism is triggered when the boundary-layer thickness (e.g.~momentum thickness $\theta$) is comparable to the radius of curvature~($R$). In laminar flows, an instability, due to the gradient of angular momentum in the radial direction, leads to the formation of longitudinal G\"ortler vortices~\citep{saric1994gortler,Schrader_2011}. In turbulent flows, however, G\"ortler structures are notoriously difficult to identify, which obfuscates precise description. 
The complexity is compounded when TBLs are exposed to free-stream turbulence: 
Whether the external forcing will decorrelate the structures or enhance them is uncertain, and how the coupling between the G\"ortler structures and the finer-scale boundary-layer turbulence will be affected is unknown.

Much of the literature on curved-wall turbulence has focused on conditions with quiescent free streams.  
\citet{Barlow_1988a} performed experiments to probe the response of the boundary layer to a sudden onset of concave curvature.  They reported that higher momentum eddies move toward the wall, while lower momentum eddies travel away from it. Despite efforts to visualize the streamwise vortices using the colored-dye and laser-induced-fluorescence methods, direct observation of G\"ortler vortices was difficult; yet enhanced shear stress in the outer BL region suggests their presence. 
In other works, \citet{Hoffmann_1985} and \citet{Barlow_1988b} used a vortex generator to induce large-scale, time-stable roll cells, which produced a coupling between the large-scale outer fluid motions and near-wall turbulence.
\citet{Barlow_1988b} showed that the downwash suppresses the bursting, which induces interactions between sublayer structures and outer-layer eddies, while the upwash enhances the process.
However, the connection between naturally occurring G\"ortler vortices in turbulence (irregularly spaced and temporally intermittent) to those produced by synthetic means is unknown \citep[e.g.][]{Patel_1997}.

\citet{Lund_1996} and \citet{Arolla_2015} numerically modelled the experiment by \citet{Barlow_1988a}.
They used large-eddy simulations (LES) to directly probe the longitudinal vortices due to the centrifugal effect. Spatially filtering the turbulence field revealed streamwise-aligned structures inside the TBL above the curved wall, which supports the existence of G\"ortler structures within the boundary-layer turbulence~\citep{Lund_1996,Arolla_2015}. 
However, the degree that the external pressure gradient influences the TBL on the curved surface was not addressed\textemdash an effect that depends on the flow setup.
At the onset of mean-streamline curvature, an adverse pressure gradient is established and the opposite effect takes place at the end of the curved section. 
Top walls can be contoured to isolate curvature effects from streamwise pressure gradient~\citep{Barlow_1988a, Barlow_1988b}, but in practical configurations the pressure gradient is present and impacts the behaviour of the flow on the curved wall.  

A related configuration is the flow in an `S'-shaped duct, where the boundary layer is subject to pressure gradients and curvature. Experiments show that the skin friction has local extrema after curvature changes, which are concomitant with streamwise pressure variations~\citep{Band_1993}. 
Corresponding LES by \citet{Lopes_2006} shows that the mean flow was separated near the convex-to-concave transition due to the strong adverse pressure gradient. 
Intermittent separation was also observed as the flow progressed from the convex to flat region, but the phenomenon was seldom observed in the flat-to-concave transition~\citep{Lopes_2006}.

Studies that have examined FST-TBL interactions over curved walls have mainly focused on the flow recovery from the curved section on the downstream flat wall.  
\citep{Kestoras_1995} compared the boundary layers at low and high inlet free-stream turbulence intensities, $Tu = 0.6\%$ and $8\%$, and reported an increase in skin friction by up to $20\%$ in the latter case.
The authors also contrasted their ability to identify the G\"ortler vortices in the first case but not in presence of free-stream forcing, which is unlike the established observation that free-stream forcing promotes the generation of outer large-scale motions in flat-plate boundary layers  \citep[e.g.][]{You_2019}.  
In the same experimental configuration, \citet{Kestoras_1998} used the temperature field over the constant heat-flux plate in order to evaluate the probability of observing boundary-layer fluid, which was their definition of intermittency.
The forced case exhibited enhanced mixing, a trend that is in agreement with flat-plate boundary layers exposed to free-stream turbulence \citep{Hancock_1989}.  
Their intermittency curves were not however presented in viscous scaling, so it is not possible to quantify the extent to which the free-stream fluid penetrated towards the wall.


\subsection{Effects of free-stream turbulence on flat-plate turbulent boundary layers}

The influence of FST on TBL has much more commonly been studied in zero-pressure-gradient, flat-plate boundary layers. 
Early experiments were performed using grid-generated FST with intensities $Tu\lesssim 7\%$ \citep{Simonich_1978, Hancock_1983, Hancock_1989, Castro_1984}.
\citet{Simonich_1978} reported that the drag and also the heat-transfer rate increase proportionally to the $Tu$.
\citet{Hancock_1983} demonstrated a marked effect of the free-stream turbulence lengthscale on TBL response.
They introduced the non-dimensional parameter $b \equiv Tu (\%)/(L_u/\delta +2)$, where $L_u$ and $\delta$ are the dissipation length scale and boundary layer thickness, and showed that skin friction increases with $b$.  
\citet{Hancock_1989} thermally `tagged' the boundary-layer fluid in order to perform conditional sampling. They reported that FST increases the standard deviation of the intermittency profiles, and that the isotropy of FST reduces the Reynolds-shear-stress correlation coefficient 
in the boundary layer.
\citet{Ames_1990} and \citet{Thole_1995, Thole_1996} investigated the influence of higher free-stream intensities $Tu\gtrsim 10\%$. The former effort reported that under these intense conditions, skin friction depends not only on $b$, but also on the Reynolds number.
In contrast, \citet{Thole_1995} affirmed that the skin friction depends mainly on $b$, up to the turbulent intensity $Tu = 28 \%$.
Based on simulations of forced temporal boundary layers, \citet{Kozul_2020} reported that another relevant parameter is the ratio of eddy-turnover timescales of the free-stream turbulence and boundary layer.  
Too small a value leads to weak interactions since the external turbulence decays quickly and cannot influence of the boundary layer.

The changes in the spectra and flow structures in the boundary layer, when exposed to free-stream forcing, are noteworthy.
Absent free-stream disturbances, a peak in the pre-multiplied energy spectra in the outer part of the boundary layer signals the formation of large-scale motions\citep{Hutchins_2007, Mathis_2009}\textemdash a behaviour that is observed at Reynolds numbers $Re_{\tau} \ge 2000$.
Under the influence of FST, that outer peak is observed at lower Reynolds numbers, in both the streamwise and spanwise energy spectra \citep{Thole_1996, You_2019} .

Free-stream turbulence forcing also leads to an increase in the near-wall streamwise velocity fluctuations.
Using a scale-decomposition analysis, \citet{Dogan_2016} attributed this increase to large scales relative to a cutoff wavelength between $1$-$2\delta$. 
\citet{Hearst_2018} divided the spectrogram of the streamwise velocity fluctuations into four regions based on the wall-normal height and wavelength.
Near the wall, the large-wavelength region is significantly affected by the free-stream turbulence, although the small-scale inner peak was relatively insensitive to external forcing. Based on these results, \citet{Hearst_2018} concluded that the FST is directly observed in the near-wall region. 

The recent direct numerical simulations by \cite{You_2019,You_2020} provided a detailed analysis of the interaction of free-stream turbulence with underlying flat-plate boundary layers at Reynolds numbers exceeding $Re_\theta \simeq 3200$.  A levelset approach was embedded in the simulations to objectively distinguish the free-stream and boundary-layer fluids, and to quantify the degree of penetration of the former into the later (and vice versa). 
The levelset approach thus provided an unambiguous description of the respective roles of the free-stream and boundary-layer turbulence, and their respective contributions to observations through conditional sampling.
The results showed that only the low-frequency component of the FST penetrates the logarithmic layer, which is consistent with the phenomenon of shear sheltering \citep{Hunt_1999, Zaki_2009}.  
The outcomes are a direct increase in the turbulence energy in this region, and the formation of large-scale motions at lower Reynolds numbers than in canonical unforced boundary layers.  
In contrast to the logarithmic region, the FST did not directly reach the buffer layer and the increase in the near-wall TKE was due to an indirect effect: The formation and amplification of the outer large-scale motions modulated the near-wall structures and led to the increase in their turbulence energy.

In contrast to the recent discoveries in context of FST interactions with flat-plate boundary layers, much less is known regarding how such interaction unfolds on curved walls. The present work will highlight that free-stream turbulence has important implications as early as the onset of curvature where the flow experiences an adverse pressure gradient, which can lead to intermittent separation. The external forcing also appreciably alters the boundary-layer statistics and the G\"ortler structures on the curved section and has important practical implications on the wall stress. We perform two simulations: (i) a reference case with a quiescent
free stream and (ii) a forced case with 10\% free-stream turbulence intensity at the inlet
plane. The setup of the simulations is described in section 2. Key statistical results are presented in section 3 and changes to boundary-layer structures are discussed in section 4. The conclusions are provided in the final section.


\section{Simulation setup}

\begin{figure}
\begin{center}
\includegraphics[width=0.90\columnwidth]{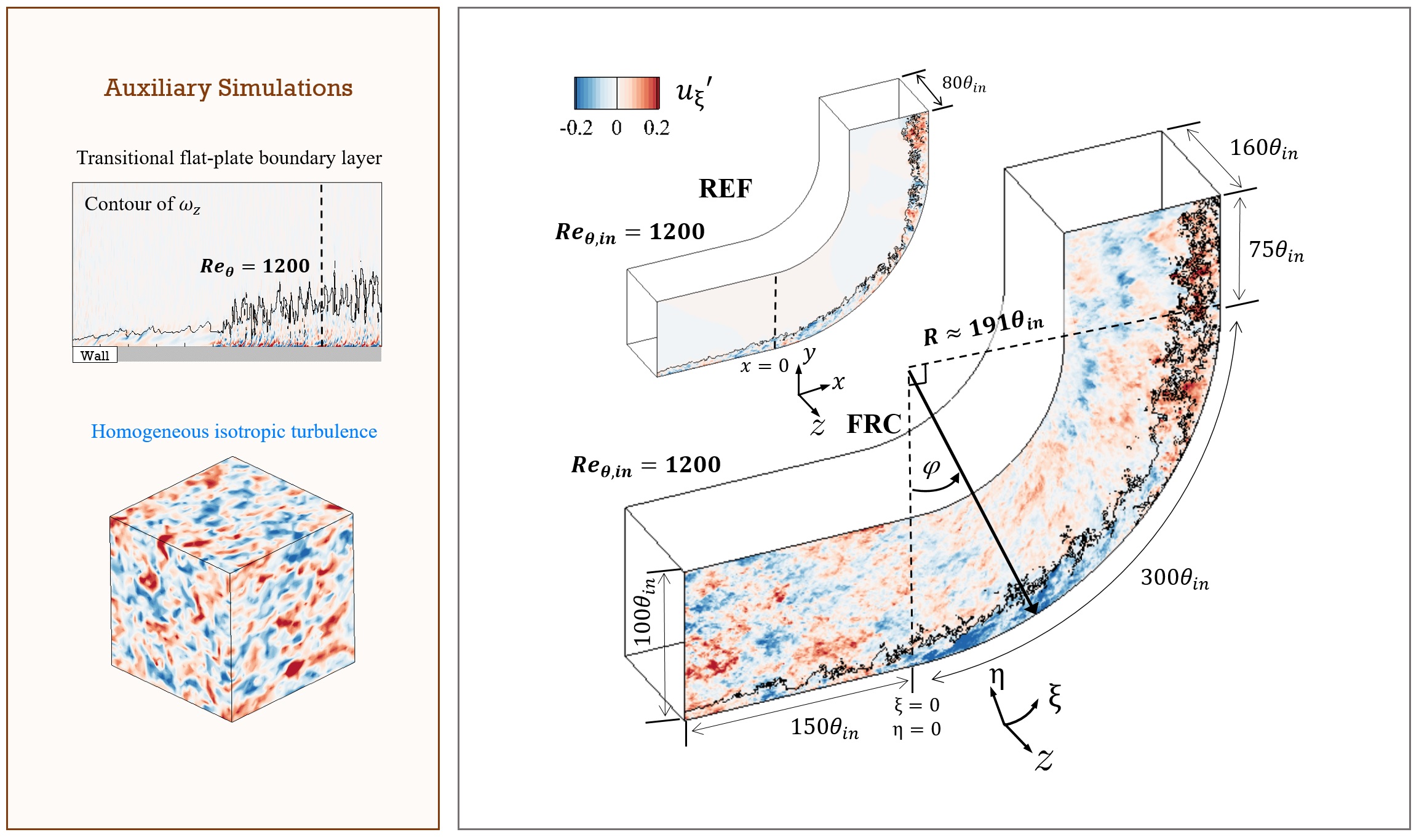}
\end{center}
\vspace{-6pt}
	\caption{Configurations of two auxiliary computations to generate the inflow conditions, and the two main computations of turbulent boundary layer over curved wall without and with free-stream turbulence. } 
\label{fig_config}
\end{figure}

The flow configuration adopted in the present study is shown in figure \ref{fig_config}.  The flow is governed by the incompressible Navier-Stokes and continuity equations which, expressed in non-dimensional form, are 
\begin{align}
\label{eqn_n01}
\frac{\partial u_i}{\partial t} + \frac{\partial u_i u_j}{\partial x_j} &= - \frac{\partial p}{\partial x_i}
+ \frac{1}{Re_{\theta_{in}}} \frac{\partial^2 u_i}{\partial x_j^2}\\\label{eqn_n02}\frac{\partial u_j}{\partial x_j} &= 0. 
\end{align}
The reference scales are the free-stream velocity $U_\infty$ and the boundary-layer momentum thickness $\theta_{in}$ at the inflow of the main simulation domain.
The momentum-thickness Reynolds number at the inflow is $Re_{\theta_{in}} \equiv \rho U_\infty \theta_{in} / \mu=1200$, where $\rho$ and $\mu$ are the density and dynamic viscosity, respectively.
The velocity components in the streamwise ($\xi$), wall-normal ($\eta$) and spanwise ($z$) directions are $u_\xi$, $u_\eta$ and $w$, respectively, and the pressure is $p$.
Note that $x$ and $y$ indicate the horizontal and vertical Cartesian coordinates.

The flow equations are solved using a fractional step algorithm on a staggered grid with a local volume-flux formulation \citep{Rosenfeld_1991}.  The algorithm was extensively validated and adopted in DNS of transitional \citep{Schrader_2011,zaki_v103_2010} and turbulent flows \citep{Jelly_2014, WangM_2019}.
The viscous terms are integrated in time using the implicit Crank--Nicolson method, and the convective terms are treated explicitly using the Adams-Bashforth scheme.
The pressure equation is solved using Fourier transform in the periodic spanwise direction and geometric multigrid for the resulting Helmholtz equation, then used to project the intermediate velocity onto a divergence-free field.  

Two main simulations are contrasted: a reference (REF) case where the curved-wall boundary layer develops beneath a quiescent free stream and a forced (FRC) case where the free stream is turbulent. 
In both cases, the flow domains include an initial flat section ($150\,\theta_{in}$), a curved section ($300\,\theta_{in}$), and a recovery flat section ($75\,\theta_{in}$). 
The quarter-circular section has radius $R=191 \,\theta_{in}$. The spanwise domain of the forced configuration is two times larger than the reference case, in order to accommodate the formation of large-scale structures which are anticipated based on earlier studies \citep{You_2019}. In addition, the spanwise two-point velocity correlations were evaluated and confirmed that the widths of the domains are sufficiently large.  
Table \ref{tab01} summarizes the domain sizes and grid resolutions. 
The grids are uniform in the span and stretched in the wall-normal direction using a hyperbolic tangent function.
In the streamwise direction, the grid spacing is uniform on the bottom wall ($\eta=0$).   
On the top surface, the grid is uniform on the curved section and is adjusted smoothly near the changes in curvature.  
An elliptic grid generation technique \citep[e.g.][]{Thompson_1985} is adopted to reduce strong variations in mesh spacing in those regions; the ratio of successive streamwise grid spacing was less than $3\%$.

\begin {table}
\begin{center}
\begin{tabular}{ c | c | c | c | c } \hline
 \multirow{2}{*}{Designation}  &  \multirow{2}{*}{Inflow}  &  Domain size ($\theta_{in}$)	    &  No. of grid points		 &  Resolution  \\
              &              &  $L_\xi \times L_\eta \times L_z$  &  $N_\xi \times N_\eta \times N_z$ &  $\Delta \xi^+,\Delta \eta^+,\Delta z^+,\Delta t^+$   \\\hline
    REF       &  TBL         &  $525 \times 100 \times 80$   &  $2688\times 896 \times 768$ &  10.5, 0.29-7.8, 5.6, 0.048  \\
    FRC       &  TBL+HIT     &  $525 \times 100 \times 160$  &  $2688\times 896 \times1536$ &  10.5, 0.29-7.8, 5.6, 0.041  \\\hline
\end{tabular}
\end{center}
\caption{Computational domain sizes, and spatial and temporal resolutions at the inflow plane expressed in viscous `$\scriptscriptstyle{+}$' units.}
\label{tab01}
\end{table}

No-slip conditions are applied at the bottom wall, while impermeability and no-stress conditions are imposed at the parallel top boundary. 
The domains are periodic in the spanwise direction, and convective outflow conditions are imposed at the exit planes.
In order to seed the inflow TBL and FST in the main computations, two auxiliary simulations are performed and their full details are provided in \citet{You_2019}. 
The first auxiliary simulation is an independent DNS of transitional flat-plate boundary layer, where a cross-flow plane was stored in the fully turbulent regime at $Re_\theta = 1200$ as a function of time; the same data were used for inflow conditions in a number of studies of TBL \citep{Lee_2017, You_2019, motoori_goto_2019}. 
The second auxiliary simulations is a pseudo-spectral DNS of homogeneous isotropic turbulence (HIT) in a periodic domain with dimensions $\{L_{\xi} , L_{\eta} , L_{z}\}_{\scriptscriptstyle{\mathrm{HIT}}} =$ $\{80, 80, 160\}$, which generates the free-stream turbulence. 
The inflow HIT has intensity $Tu = 0.1$
and length scale $L_k \equiv {k^{\scriptscriptstyle{3/2}}}/{\epsilon} \approx 10.8$, where $k$ is the turbulent kinetic energy and $\epsilon$ is the dissipation rate. 

At the inflow plane of the main computations, either the time-dependent turbulent boundary layer is applied alone (REF case) or it is superposed with free-stream turbulence (FRC case).  In the latter case, the HIT box is introduced above the edge of the boundary layer which was identified using a normalized vorticity threshold, $\frac{|\mathbf{\omega}|}{u_\tau^2 / \nu} \sqrt{{\delta_{99}^+}}=0.2$ where $\mathbf{\omega}$ is the vorticity, $u_\tau$ is the friction velocity, $\nu$ is the kinematic viscosity and $\delta^+_{99}$ is the 99\% boundary-layer thickness in wall units \citep{Lee_2017}.
A levelset function $\psi$ is defined at the inflow plane and tags, or differentiates, the fluid within the boundary layer ($\psi = 1$) and the free stream ($\psi = 0$).  The transport equation of $\psi$ is
\begin{eqnarray}
\label{eqn_level}
	\frac{\partial \psi}{\partial t} + \frac{\partial u_j \psi}{\partial x_j} = 0, \label{eq:levelset}
\end{eqnarray}
and therefore $\psi$ is a diffusion-free scalar that serves as a virtual sharp interface between the two fluids; 
full details of the implementation and exhaustive validation are provided elsewhere~\citep{Jung_2015, You_2019}.
The value $\psi = 0.5$ is adopted as the threshold for conditional sampling of the boundary-layer $\psi > 0.5$ and free-stream $\psi < 0$ fluids.
Note that the interface height $\eta_I = \eta(\psi=0.5)$ provides an objective measure of the boundary-layer thickness that is less sensitive to the details of the profile than the conventional 99\% thickness. 

\begin{figure*}
	\begin{center}
		\includegraphics[width=0.55\columnwidth]{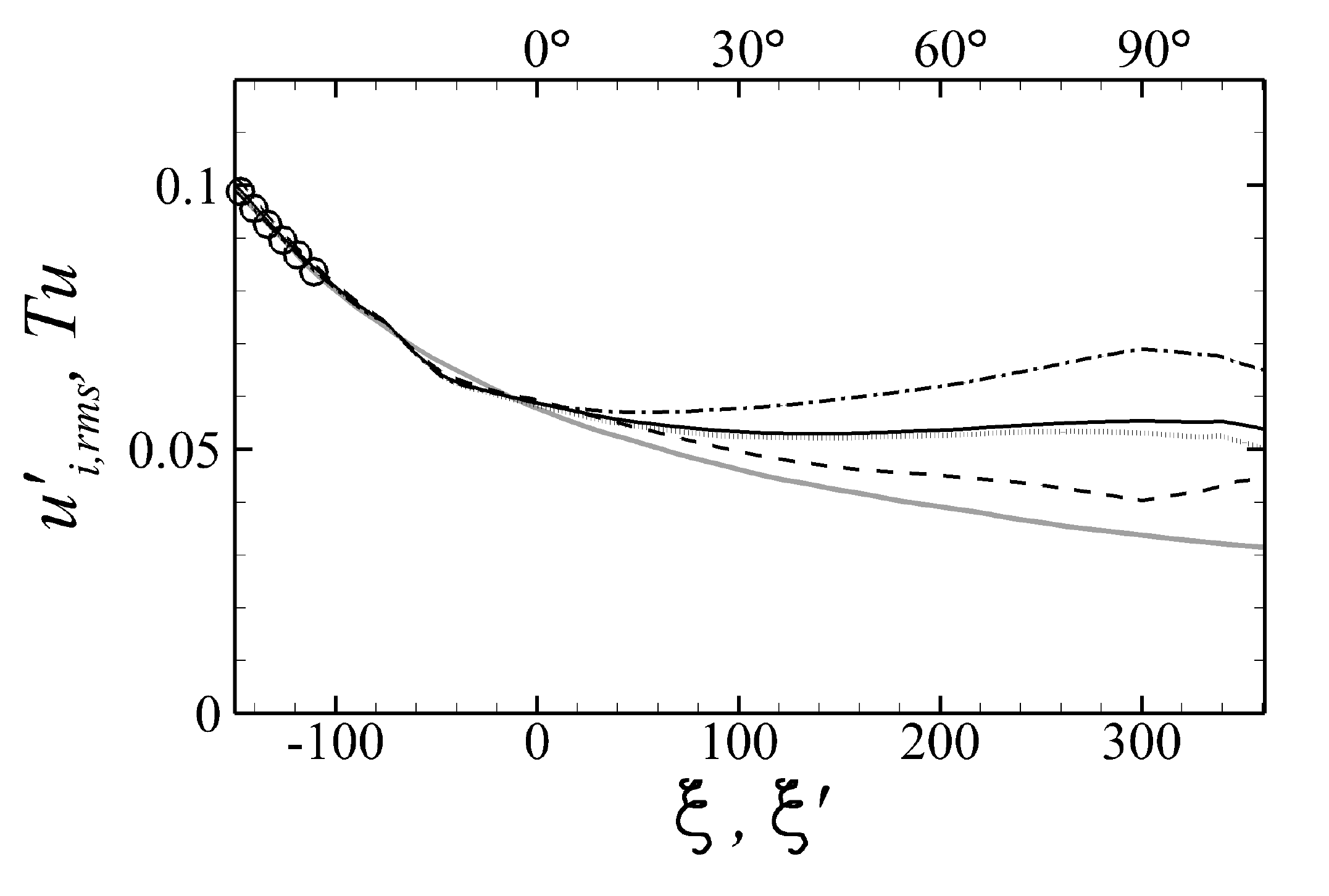}
	\end{center}
	\caption{Downstream dependence of (black solid line) free-stream turbulence intensity $Tu$, (\dashed) $ u_{\xi,rms}' $, (\chaindot) $ u_{\eta,rms}' $, (\dotted) $ w_{rms}'$, and ($\bigcirc$) the temporal evolution of $Tu$ in decaying HIT which exploits the coordinate transformation $\xi^\prime =  U_\infty t$. 
	Gray line indicates the evolution of $Tu$ on the flat plate \citep{You_2019}.
	} 
	\label{fig_tu}
\end{figure*}

Beyond an initial transient and once the flow over the curved surface has reached a statistically stationary state, statistics were collected for $T_{\scriptscriptstyle{\mathrm{stat}}}=1{,}207.5$ (REF) and $892.5$ (FRC) convective time units.  
A bar will indicate an average in homogeneous coordinates, and the prime will refer to perturbation quantities according to Reynolds decomposition, for example $u_\xi=\overline{u_\xi} + u_\xi'$.

Unlike flat-plate boundary layers, the tangential free-stream velocity on the curved section is
not uniform.  Instead, it increases linearly outside the boundary-layer edge.  That potential velocity profile is denoted $U_p$, and is linearly extended into the boundary layer to determine the wall value $U_{pw}$ \citep{Barlow_1988a}. The free-stream mean velocity profile influences the development of the FST which we quantify in figure \ref{fig_tu}. 
Upstream, within the flat section ($\xi\lesssim -100$), the decay in space in the main simulation agrees with the temporal decay of $Tu$ in the pseudo-spectral auxiliary DNS of HIT to within the Taylor's hypothesis $\xi^* =  U_\infty t$.  
In that region, the present results also agree with the previous study of free-stream turbulence over a flat-plate boundary layer \citep{You_2019}, but the two curves show the first signs of dissimilarity ahead of the flat-to-curved transition due to the pressure gradient in that region.
Farther downstream, on the curved section, the difference is more pronounced because both the mean shear in the free stream and the strain due to curvature lead to production of Reynolds shear stress, which in turn leads to production of tangential and wall-normal stresses. The outcomes are anisotropy and slower decay and of the free-stream turbulence relative to flow over a flat plate.

\begin{figure}
\centering
\includegraphics[width=0.99\textwidth]{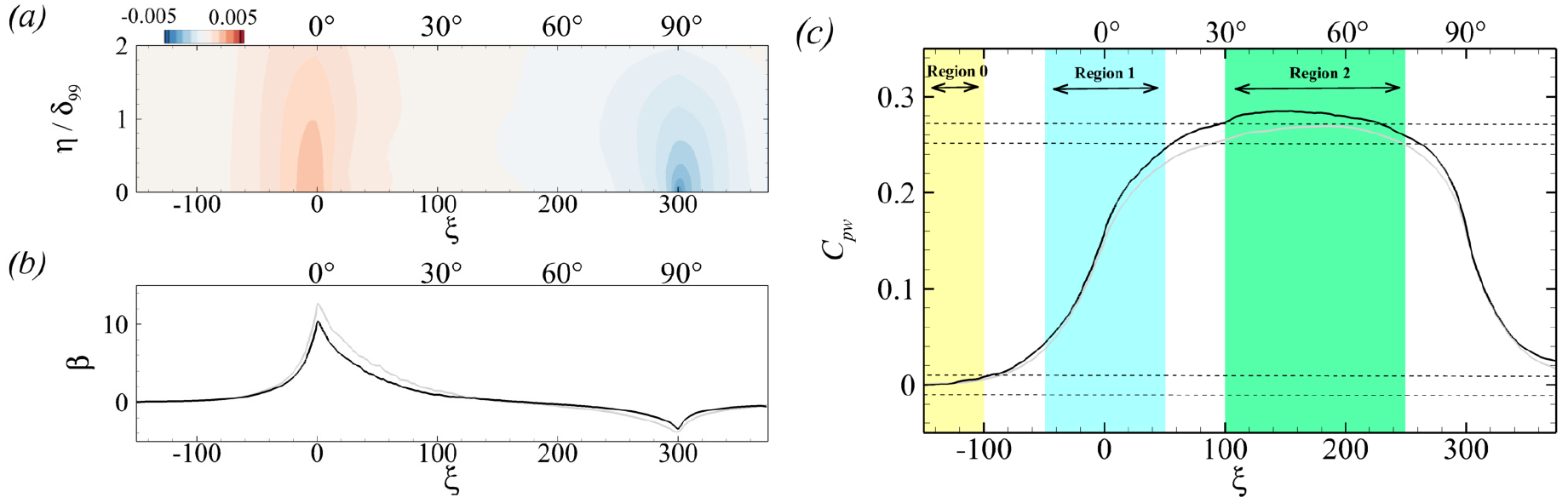}
	\caption{$(a)$ Streamwise pressure gradient~$\partial \overline{p}/\partial \xi$ in REF. $(b)$ Clauser pressure-gradient parameter $\beta$:~(gray) REF and (black) FRC. $(c)$~Wall-pressure coefficient, $C_{pw}$: (gray) REF and (black) FRC. Bottom dashed lines indicate ZPG region according to the criterion by \citet{Harun_2013}, and top dashed lines mark $C_{pw,max} - C_{pw} \leq 0.02$ for REF.
	}
\label{fig:pressureGradientCpw}
\end{figure}


Besides the FST input, the wall geometry induces a pressure gradient responsible for turning the oncoming flow, which has profound implications on boundary layer dynamics. Figure \ref{fig:pressureGradientCpw} shows the streamwise pressure gradient $\partial \overline{p}/\partial \xi$, the Clauser parameter $\beta = \frac{\delta^\star}{\overline{\tau_w}}\frac{\partial \overline{p}_e}{\partial \xi}$, and the wall-pressure coefficient $C_{pw}\equiv ({\overline{p}_{\xi,\eta=0} -\overline{p}_{\xi=0,\eta=0} })/{\tfrac{1}{2}\rho U_\infty^2}$.  In the definition of the Clauser parameter, $\delta^\star$ is the displacement thickness, $\overline{\tau_w}$ is the mean wall shear stress and $\overline{p}_e$ is the mean pressure at $\eta=\delta^\star$.  Three regions can be clearly delineated:
approximately zero-pressure-gradient (ZPG) boundary layer on the flat plate (Region 0), adverse-pressure-gradient (APG) flow near the onset of the curvature (Region 1), and
nearly ZPG flow on the concave curve (Region 2).  Table \ref{tab_cp} provides the information of stations which will be discussed in the following section.
Region 0 has been the focus of numerous previous studies of the boundary-layer response to external turbulence
\citep{Hancock_1989, Dogan_2016, You_2019}, and hence the focus herein is directed to Region 1 and Region 2.

\begin {table}
\begin{center}
\begin{tabular}{ c  c  c   c  c  c } \hline
 Zone     & Description                 & $\xi$  & $\varphi$ & $Re_{\tau,REF}$ & $Re_{\tau,FRC}$ \\\hline
 Region 0 &  ZPG on flat plate                & $-100$ &    --             & 455 & 480 \\\hline
 \multirow{2}{*}{Region 1} & \multirow{2}{*}{APG near onset of curvature}        & $0$    &  $0^\circ$       & 363 & 463 \\
          &                                   & $+50$  &  $15^\circ$      & 461 & 623 \\\hline
 \multirow{2}{*}{Region 2} & \multirow{2}{*}{ ZPG on concave curvature}         & $+100$ &  $30^\circ$      & 557 & 769 \\
          &                                   & $+200$ &  $60^\circ$      & 762 & 1027 \\
 \hline
\end{tabular}
\end{center}
\vspace*{-6pt}
\caption{Summary of main analysis locations selected based on $C_{pw}$ in figure~\ref{fig:pressureGradientCpw}. }
\label{tab_cp}
\end{table}


\section{Influence of FST on the boundary layer: a statistical perspective} 

\begin{figure}
	\begin{center}
		\includegraphics[page=1,width=0.95\textwidth]{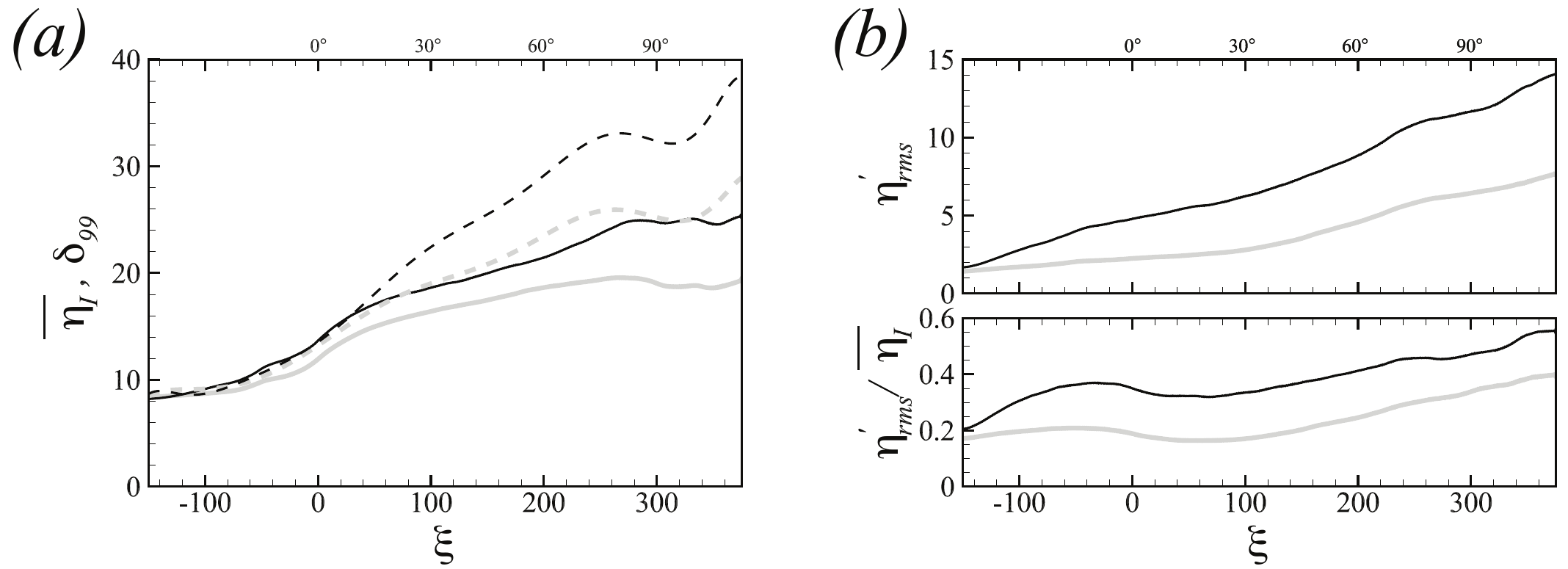}
	\end{center}
	\caption{$(a)$ Downstream development of the ($\dashed$)~boundary-layer thickness $\delta_{99}$ and (\lline)~mean interface height $\overline{\eta_I}$ based on levelset function $\psi=0.5$. 
	$(b)$ Downstream development of the root-mean-square fluctuations in the interface height $\eta'_{rms}$ and its normalized value $\eta'_{rms} / \overline{\eta_I}$. 
	(gray) REF; (black) FRC.} 
	\label{fig:thickness:etaIrms}
\end{figure}

\begin{figure}
	\begin{center}
		\includegraphics[page=1,width=0.95\textwidth]{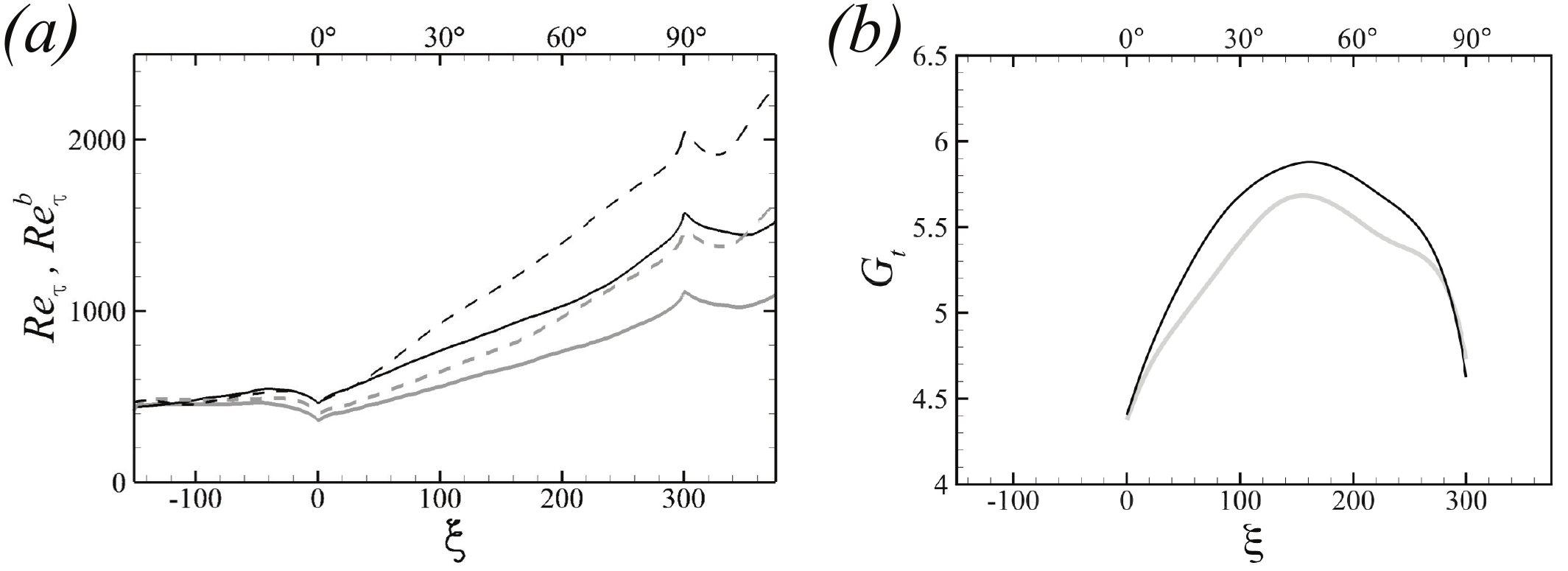}
	\end{center}
	\caption{Downstream dependence of $(a)$~the friction Reynolds numbers (\lline) $Re_\tau \equiv {u_\tau \overline{\eta_I}}/{\nu}$ and (\dashed) $Re_\tau^b \equiv {u_\tau \delta_{99}}/{\nu}$ and $(b)$ G\"ortler number $G_t = 43 \sqrt{\theta/R}$ \citep{Tani_1962}:   
	(gray) REF; (black) FRC.}
	\label{fig:ReGt}
\end{figure}

The present simulations involve the combined effects of pressure gradient, curvature and free-stream turbulence on the boundary layer.  The outcome is expectedly complex.  In this section, we examine the overall changes in the statistical state of the flow.  

Free-stream turbulence is known to enhance mixing near the edge of the boundary layer and as a result increases its thickness. 
Figure \ref{fig:thickness:etaIrms}$a$ reports two quantities that examine this effect: 
(i) $\overline{\eta_I} = \overline{\eta(\psi = 0.5)}$ which is the mean height of the virtual interface that distinguishes the boundary layer and the free stream using the levelset function and (ii) the 99\% thickness $\delta_{99}$ defined as the wall-normal location where $\overline{u_\xi} = 0.99 U_p$.
Both thickness metrics increase appreciably near the onset of the curvature due to the adverse pressure gradient. 
However, over the curve in Region 2, the mean height of the material line $\overline{\eta_I}$ has a depressed growth relative to $\delta_{99}$ which is known to be sensitive to details of the mean-velocity profile.
Since $\overline{\eta_I}$ has a physical interpretation, it will be adopted when possible in the rest of this work.
The ratio of the boundary-layer thickness to the radius of the curved wall is $\overline{\eta_I}/R<0.13$ (and $\delta_{99}/R<0.17$),	which places the present flows in the regime of moderate-curvature effect according to the criterion by \citet{Patel_1997}.

The root mean square fluctuations in the interface height $\eta'_{rms}$ and its noramlized value $\eta'_{rms}/\overline{\eta}_I$ are reported in figure~\ref{fig:thickness:etaIrms}$b$.  
For canonical, unforced flat-plate boundary layers  $\eta'_{rms}/\overline{\eta}_I$ is nearly constant.  The present results demonstrate that curvature alone (gray curve, $\xi\gtrsim 50$) promotes the undulation of the interface such that $\eta'_{rms}/\overline{\eta}_I$ increases along the curved. We therefore expect deep excursions of the free-stream fluid into the boundary layer and vice-versa along the curved section, and free-stream turbulence (black curves) enhances this effect.  This view will be reinforced in section~\ref{sec:structures} where we directly compute intermittency.

The downstream dependence of the friction Reynolds numbers, $Re_\tau \equiv {u_\tau \overline{\eta_I}}/{\nu}$ and $Re_\tau^b \equiv {u_\tau \delta_{99}}/{\nu}$ is plotted in figure \ref{fig:ReGt}$a$.
Even though $\overline{\eta_I}$ and $\delta_{99}$ increase rapidly near the onset of the curvature, the adverse pressure gradient decreases $u_\tau$ and the Reynolds numbers.  
The higher recorded values of $Re_\tau$ in the FRC case are due in part to the larger boundary-layer thickness, and also an increase in the wall stress in presence of FST.
Figure~\ref{fig:ReGt}$b$ shows the behavior of the G\"ortler number for the turbulent boundary layers 
\begin{equation}\label{eq:Gt}
    G_t = \frac{U_\infty \theta}{\nu_T} \sqrt{\frac{\theta}{R}} =  43 \sqrt{\frac{\theta}{R}},
\end{equation}
which uses an eddy viscosity $\nu_T=0.0234 U_\infty \theta$~\citep{Tani_1962}. The definition of the momentum thickness, $\theta$, for curved wall flow and associated discussion are provided in Appendix~\ref{app:momentumthickness}.
Near the onset of curvature, the G\"ortler number is $G_t \gtrsim 4$, which is large enough to promote instability~\citep{smith1955growth,Tani_1962}, and it is larger in the FRC case. Based on this metric, free-stream turbulence enhances the curvature effects.

\begin{figure}
\centering
\includegraphics[width=0.99\textwidth]{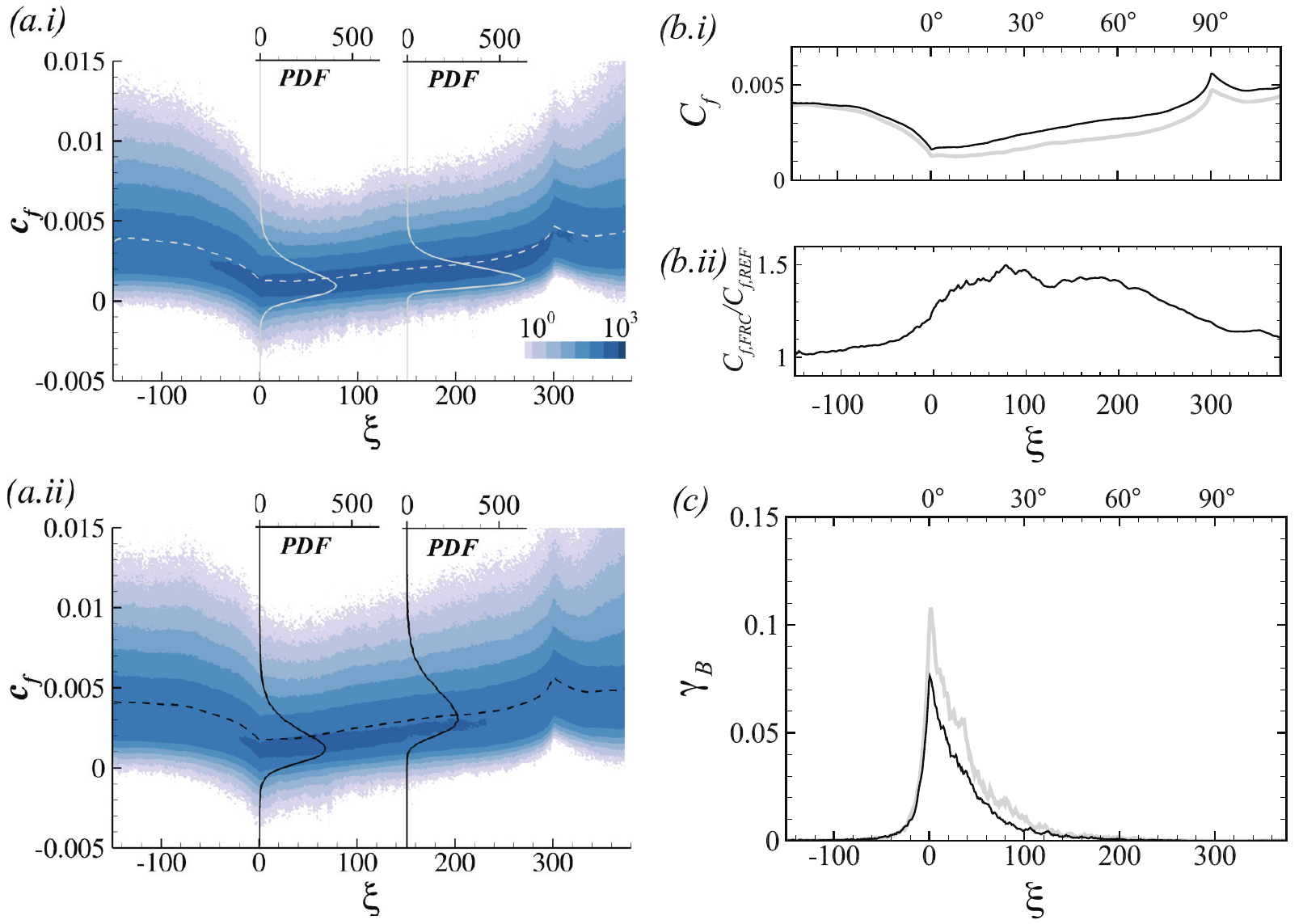}
\caption{$(a)$ Streamwise development of the probability density function (PDF) of instantaneous skin-friction coefficient, $c_f$, in ($i$) REF and ($ii$) FRC. Dashed lines are the mean values and the solid lines are the PDF at $\xi=0$ and $\xi=150$. $(b)$ Downstream dependence of $(i)$ the mean skin-friction coefficient, $C_f$ and $(ii)$ $C_{f,FRC}/C_{f,REF}$.
$(c)$ Intermittency of backward flow, $\gamma_B$. 
 (Gray) REF; (black) FRC.}
\label{fig:cfPDFwithXi:option2}
\end{figure}

Much of the interest in concave-wall boundary layers beneath vortical forcing has been dedicated to the impact on skin friction~\citep{Kestoras_1995, Kestoras_1998}.
The probability density function (PDF) of the skin-friction coefficient, $c_f \equiv \tau_w /\tfrac{1}{2} \rho U_\infty^2$, is reported in figure~\ref{fig:cfPDFwithXi:option2}, both $(a.i)$ REF and $(a.ii)$ FRC. The later case has a broader PDF which is skewed toward intense $c_f>0$ and, as a result, has a larger mean value $C_f$.
The largest relative increase in $C_f$ when the flow is forced is on the order of $49\%$, which is appreciably higher than the $15\%$ observed for the same flow conditions over a flat plate \citep{You_2019}.  
Note that the increase is not limited to the low $C_f$ region near the onset of curvature; instead it is sustained above $40\%$ over the majority of Region 2 on the curved wall.
While the effect of forcing appears relatively modest at the onset of curvature, an important change in the state of the flow takes place in that region due to the flow deceleration.
As the curvature is approached, the skin friction drops precipitously due to the adverse pressure gradient in both REF and FRC cases.
In the latter, FST enhances momentum mixing, which leads to a more moderate reduction of $C_f$.
The positive values of mean $C_f$ should not, however, mask the intermittent separation clearly captured by $c_f<0$ events in figure~\ref{fig:cfPDFwithXi:option2}$a$. 
The intermittency, or probability, of negative instantaneous wall-shear stress is denoted $\gamma_B$ and is plotted in figure \ref{fig:cfPDFwithXi:option2}$c$; its reduced value in FRC is noteworthy due to the qualitative change in boundary layers at separation.

\begin{figure}
	\begin{center}
		\includegraphics[width=0.45\columnwidth]{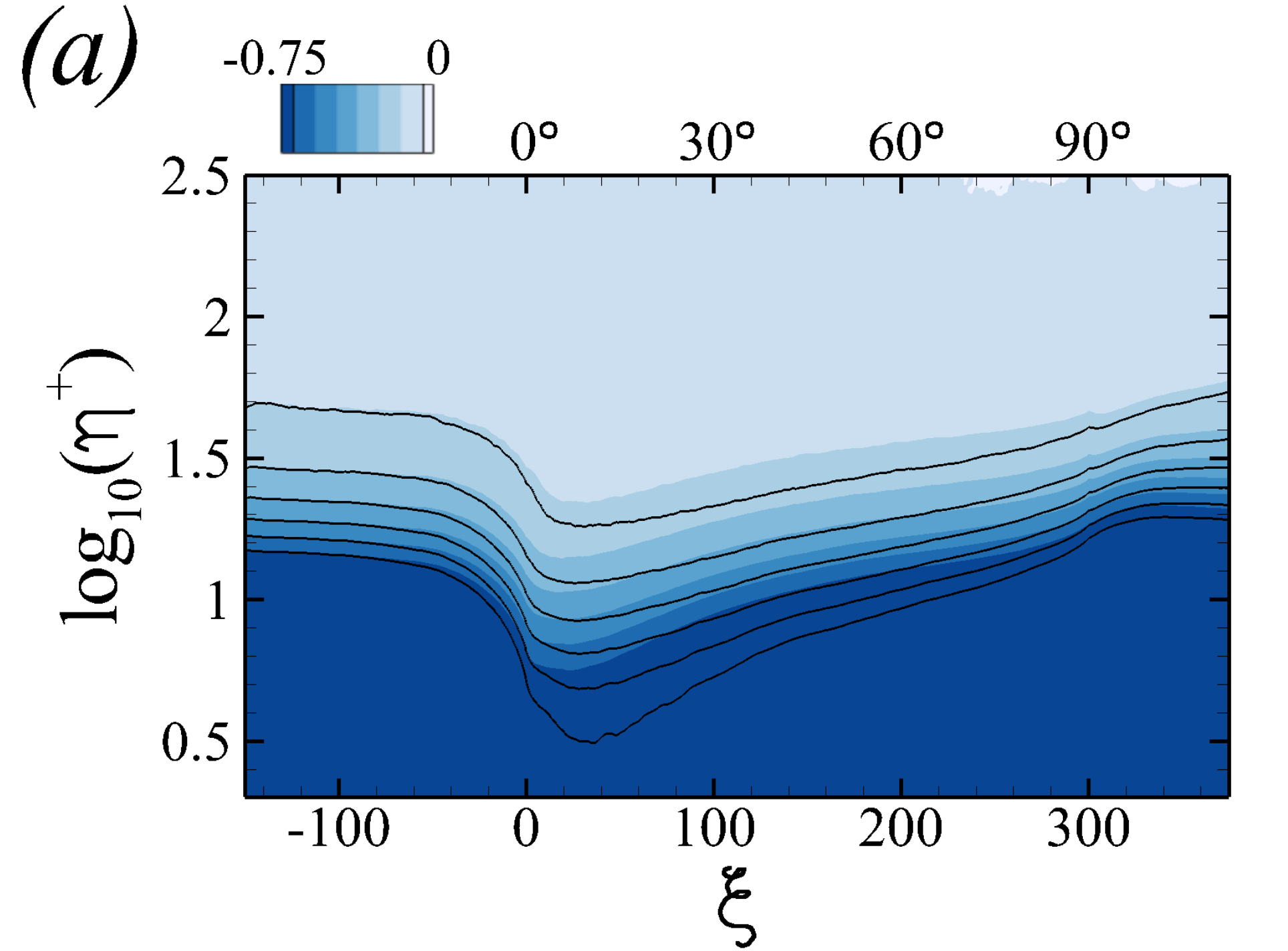}
		\includegraphics[width=0.53\columnwidth]{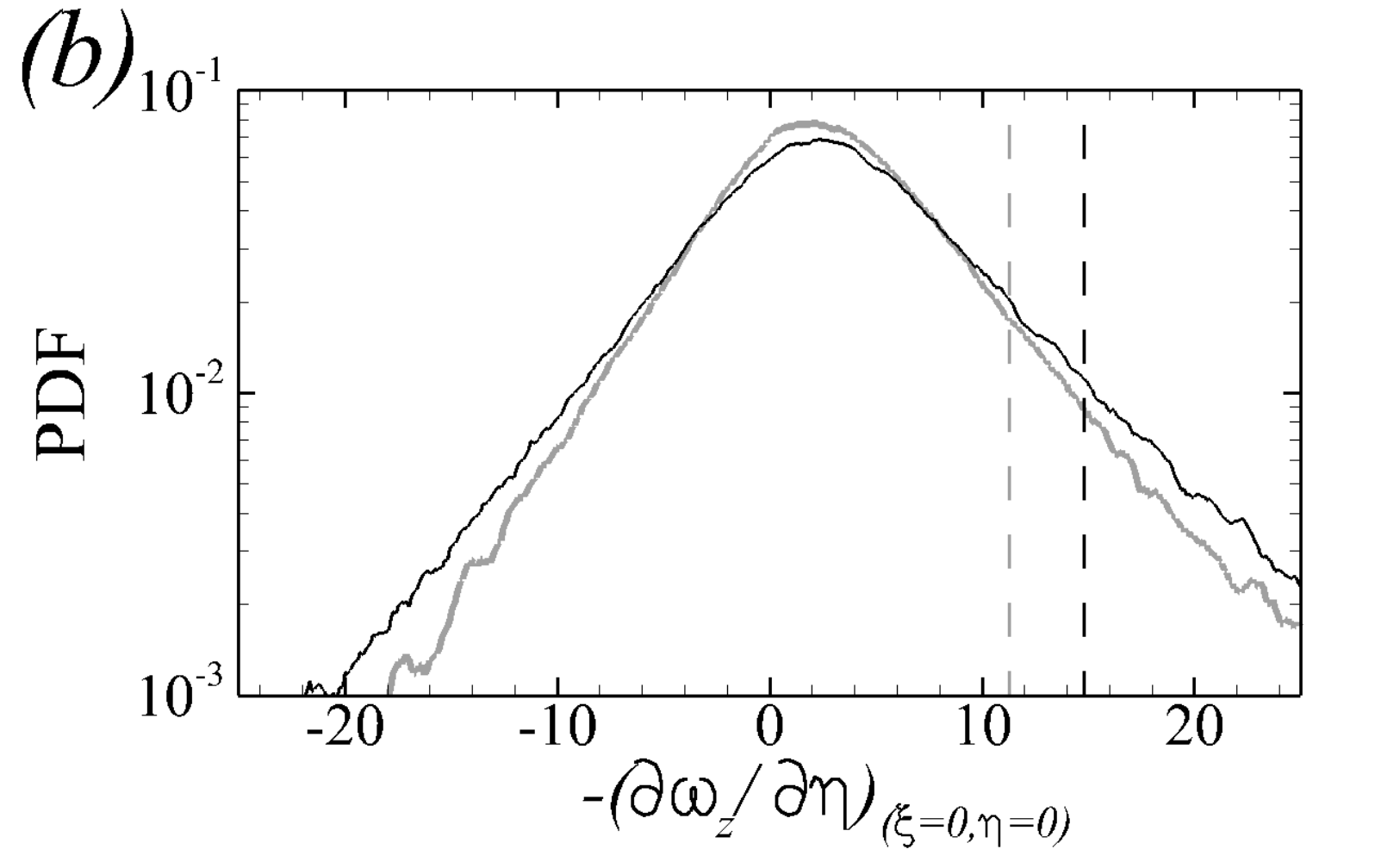}
	\end{center}
	\vspace{-0.1in}
	\caption{$(a)$ Mean spanwise vorticity $-0.75 \leq \overline{\omega_z} \leq 0$ of (lines) REF and (color) FRC cases.
		$(b)$ PDF of $-\partial \omega_z/\partial \eta$ at the wall at the onset of the curvature, $(\xi, \eta) = (0,0)$.
			Dashed lines mark thresholds for higher amplitude events having the same probability as intermittent separation. (Gray) REF and (black) FRC. }
	\label{fig_sp_vrt}
\end{figure}

\begin{figure}
	\begin{center}
		\includegraphics[width=0.49\columnwidth]{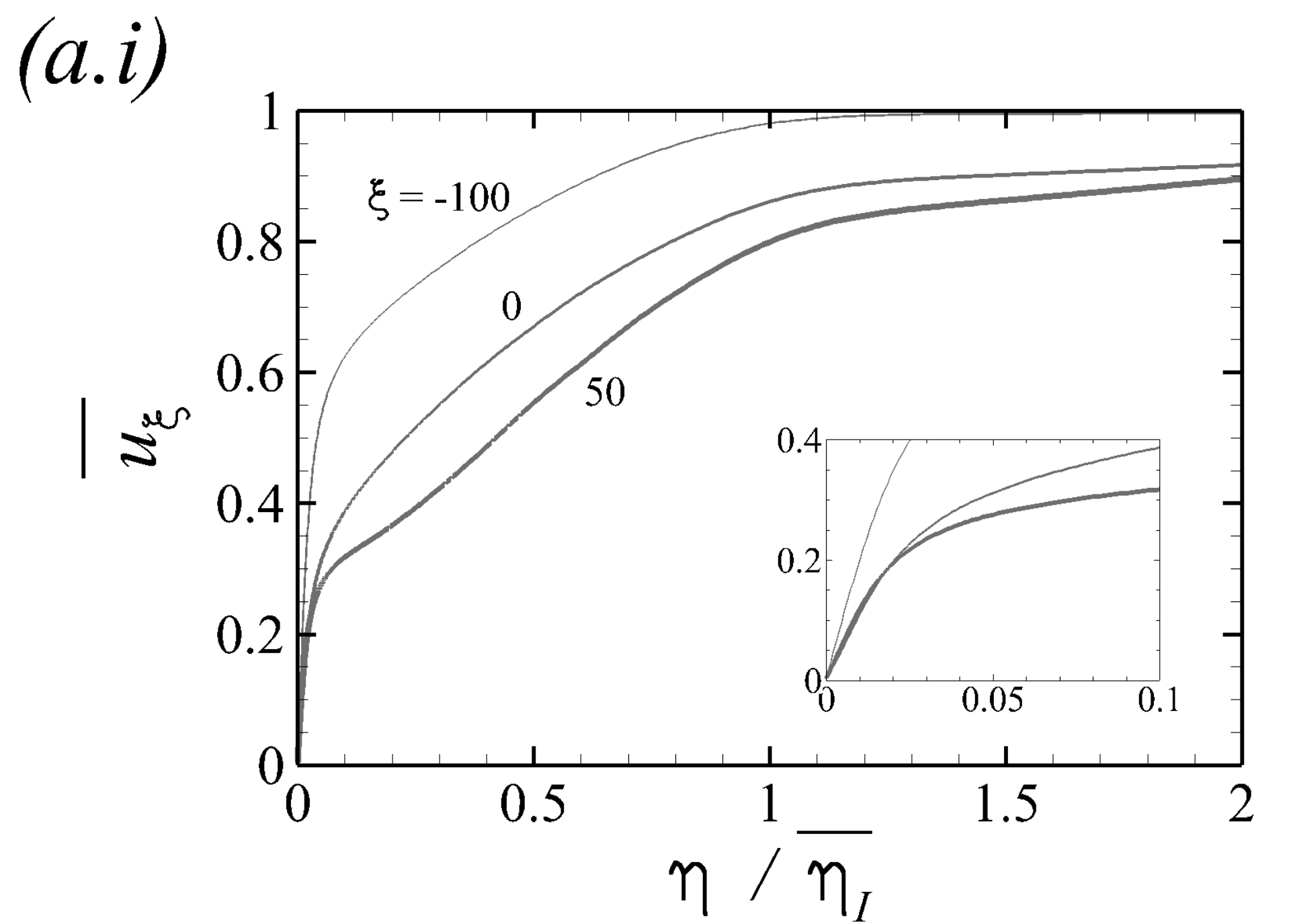}
		\includegraphics[width=0.49\columnwidth]{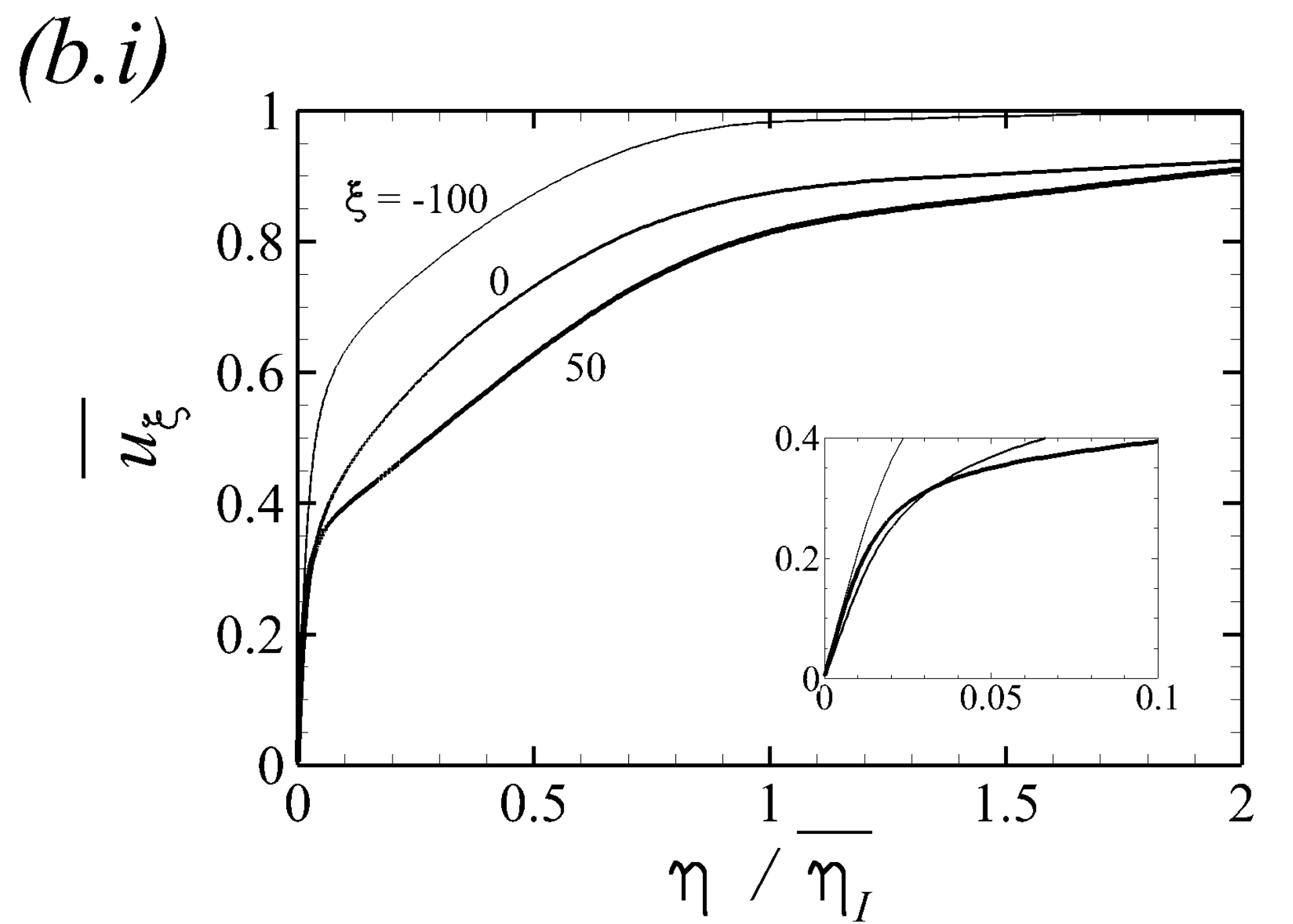}
		\includegraphics[width=0.49\columnwidth]{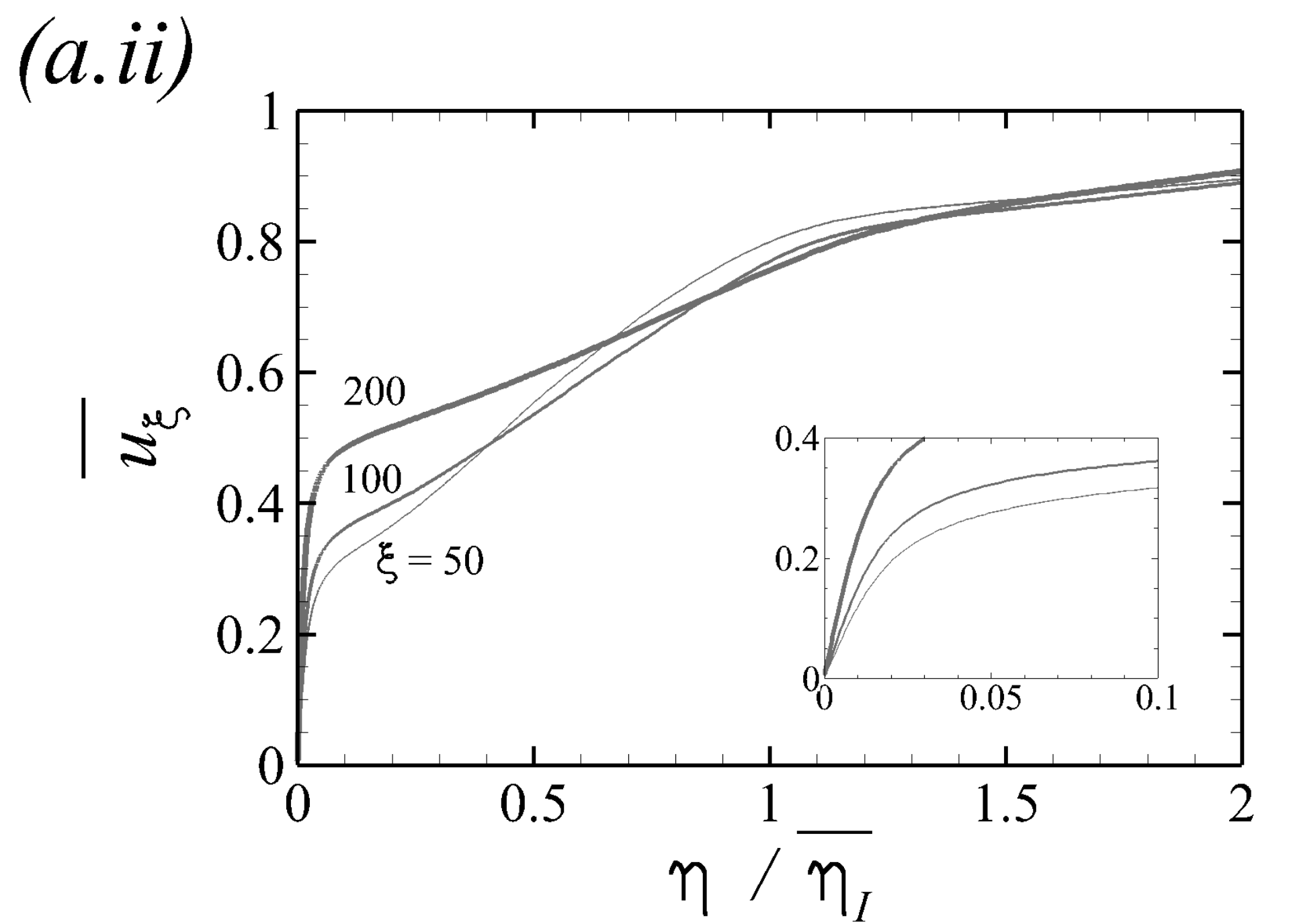}
		\includegraphics[width=0.49\columnwidth]{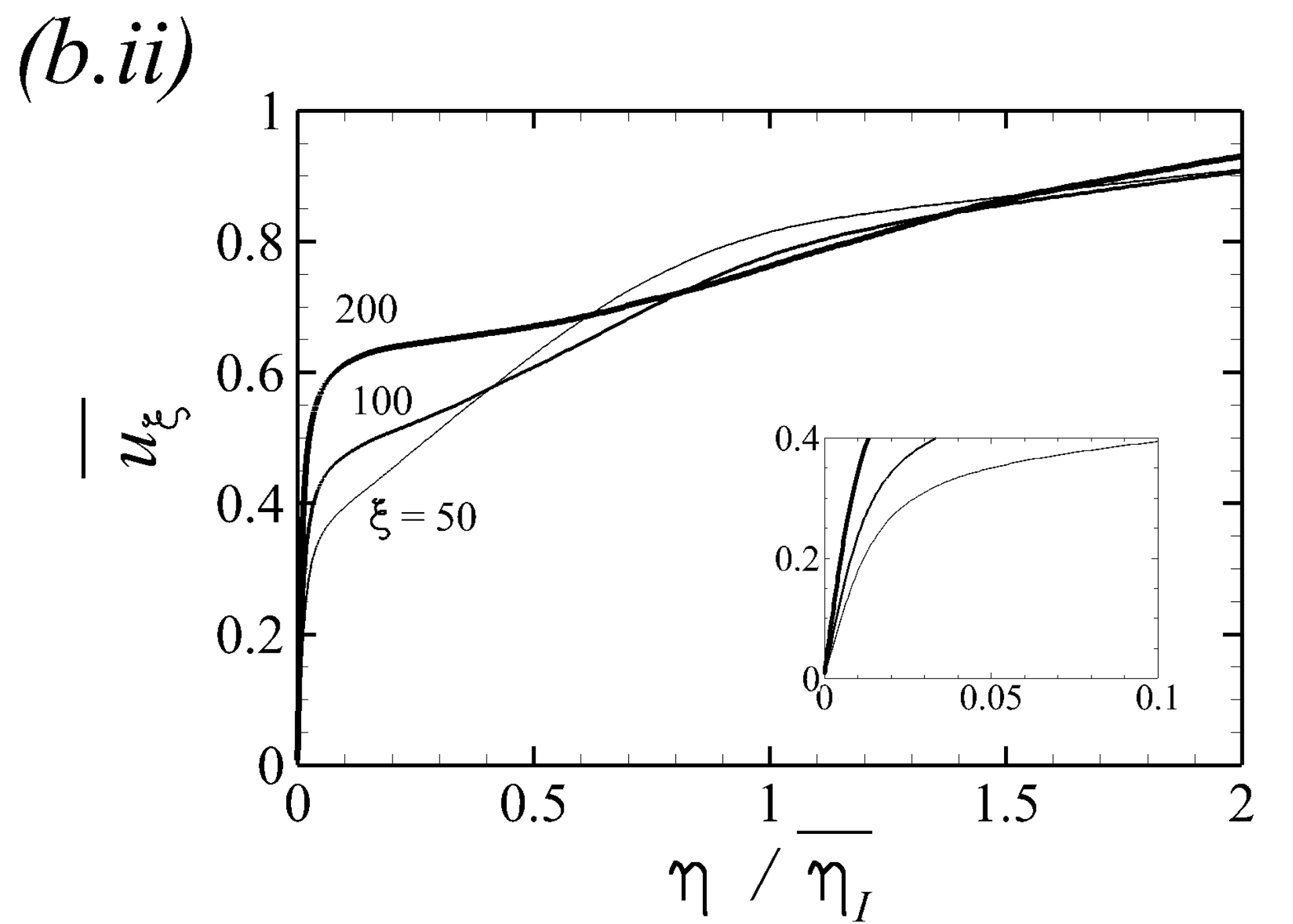}
	\end{center}
	\caption{Downstream development of the mean streamwise velocity in $(a)$~REF and $(b)$~FRC. 
	$(i)$ Progression from the flat-plate ZPG Region 0 to APG Region 1, and $(ii)$ from APG Region 1 to ZPG Region 2 along the curve.}
	\label{fig_u}
\end{figure}

An elegant interpretation of separation is in terms of the spanwise vorticity and its wall flux;
the latter is due to the streamwise pressure gradient \citep{Lighthill_1963}.  
Figure \ref{fig_sp_vrt}$a$ shows the spanwise vorticty distribution near the wall.  
At the onset of curvature, the depletion of negative vorticity is less pronounced in the FRC case, which can be interpreted in terms of a reduction in its mean outflux at the wall.  
A more detailed view is provided in figure \ref{fig_sp_vrt}$b$ which reports the PDF of instantaneous outflux of negative vorticity at the onset of curvature.  
The integral of the PDF yields a smaller value for the FRC case, consistent with figure \ref{fig_sp_vrt}$a$.      
The PDF of the forced case also has larger positive and negative tails, i.e.\,stronger instantaneous outflux of negative vorticity (positive values) and also influx (negative values).  
The former alone would be at odd with reduced frequency of separation. 
However, the strong influx of negative vorticity (negative values) renders the state of the boundary layer less prone to separation.  
Intuition may suggest that this effect is associated with enhanced momentum mixing on the flat upstream section, due to the additional vortical motions from the free stream that may breach the boundary layer.  
However, recent results for flat-plate flows showed that free-stream turbulence does not reach the buffer layer within such short streamwise distance. 
Instead, the external forcing has an indirect effect of modulation of the near-wall region that enhances the near-wall turbulent shear stresses \citep{You_2019}.  
The resulting energetic near-wall flow is then less prone to separation at the onset of curvature.  
Two vertical lines are marked on figure \ref{fig_sp_vrt}$b$; events with higher amplitudes than these thresholds have the same probability as separation ($9.73\%$ for REF and $7.39\%$ for FRC).  
Based on this simple conceptual model, the results indicate that the threshold required for the forced boundary layer to undergo local intermittent separation is approximately 30$\%$ higher than REF.  
The mean-flow profile $\overline{u_\xi}$ is therefore anticipated to be fuller in the forced flow, and is shown in figure \ref{fig_u}.
Indeed, at $\xi = 0$ through $\xi=50$, the near-wall region of the FRC boundary layer carries more momentum than the reference case.

The change in the mean profile as the flow traverses from the APG Region 1 to the ZPG Region 2 on the curve is noteworthy (figure \ref{fig_u}): 
The figure shows that the mean shear in $(b.ii)$ is slightly reduced in the outer portion of the BL and the mean profile becomes fuller, which is indicative of enhanced mixing due to the free-stream turbulence.

\begin{figure}
	\begin{center}
		\includegraphics[width=0.95\columnwidth]{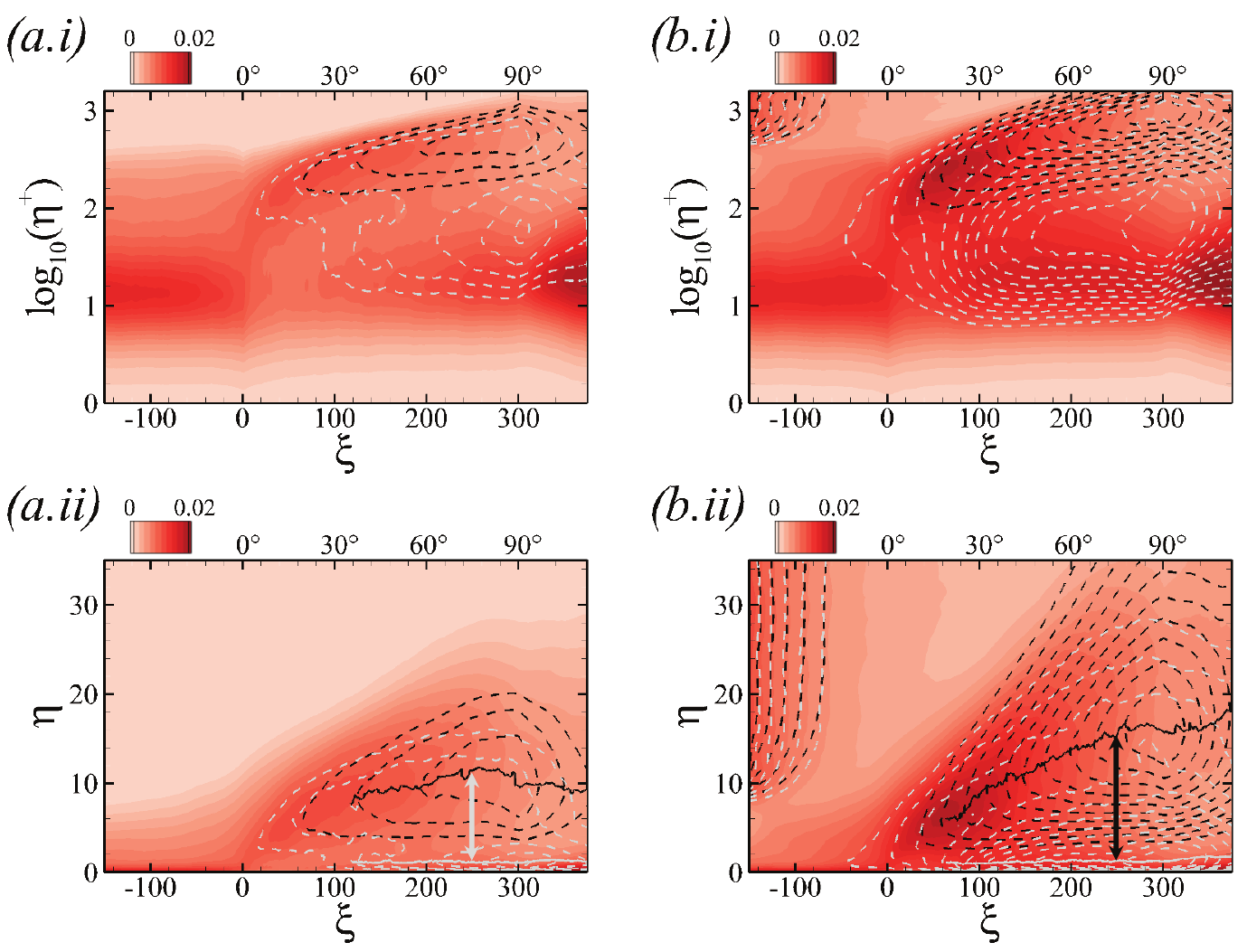}
	\end{center}
	\caption{
	Reynolds normal stresses for $(a)$ REF and $(b)$ FRC. The wall-normal coordinate is normalized using $(i)$~viscous and $(ii)$~outer scales.  
	Colours correspond to the streamwise stress, and lines show the (black) wall-normal $\overline{u_\eta' u_\eta'}$ and (gray) spanwise $\overline{w' w'}$ stresses,  from $5 \times 10^{-3}$ with increment $1 \times 10^{-3}$.
} 
	\label{fig_stress_cont}
\end{figure}

Evidence of persistent, or statistically relevant, G\"ortler structures is sought by plotting the turbulence stresses in figure~\ref{fig_stress_cont}$(a,b)$, for both the quiescent and turbulent free stream cases.
The colour contours show the tangential component, and the lines show the (black) wall-normal and (gray) spanwise ones.  
The first observation is a sudden change in the contours of $\overline{u_\xi' u_\xi'}$ across the onset of curvature: 
The upstream wall-normal profile of $\overline{u_\xi' u_\xi'}$ has only one maximum in the buffer layer, but beyond $\xi = 0$ two maxima can be detected: 
The inner peak retains its original height within the buffer layer $\eta^+ \approx 11$,
decays quickly due to adverse pressure gradient, and shows faster recovery in the forced flow; 
In contrast the outer peak is in the logarithmic layer and shifts away from the wall with the downstream growth of the boundary layer. 
The emergence of the outer peak at that onset of curvature is consistent with APG \citep[see e.g.][]{Hickel_2008}, which is sufficiently large to induce intermittent separation.  
In addition, relative to the reference case, free-stream turbulence enhances the intensity of these structures\textemdash an effect that is anticipated based on previous studies of forced flat-plate boundary layers.

\begin{figure}
	\begin{center}
		\includegraphics[width=0.60\columnwidth]{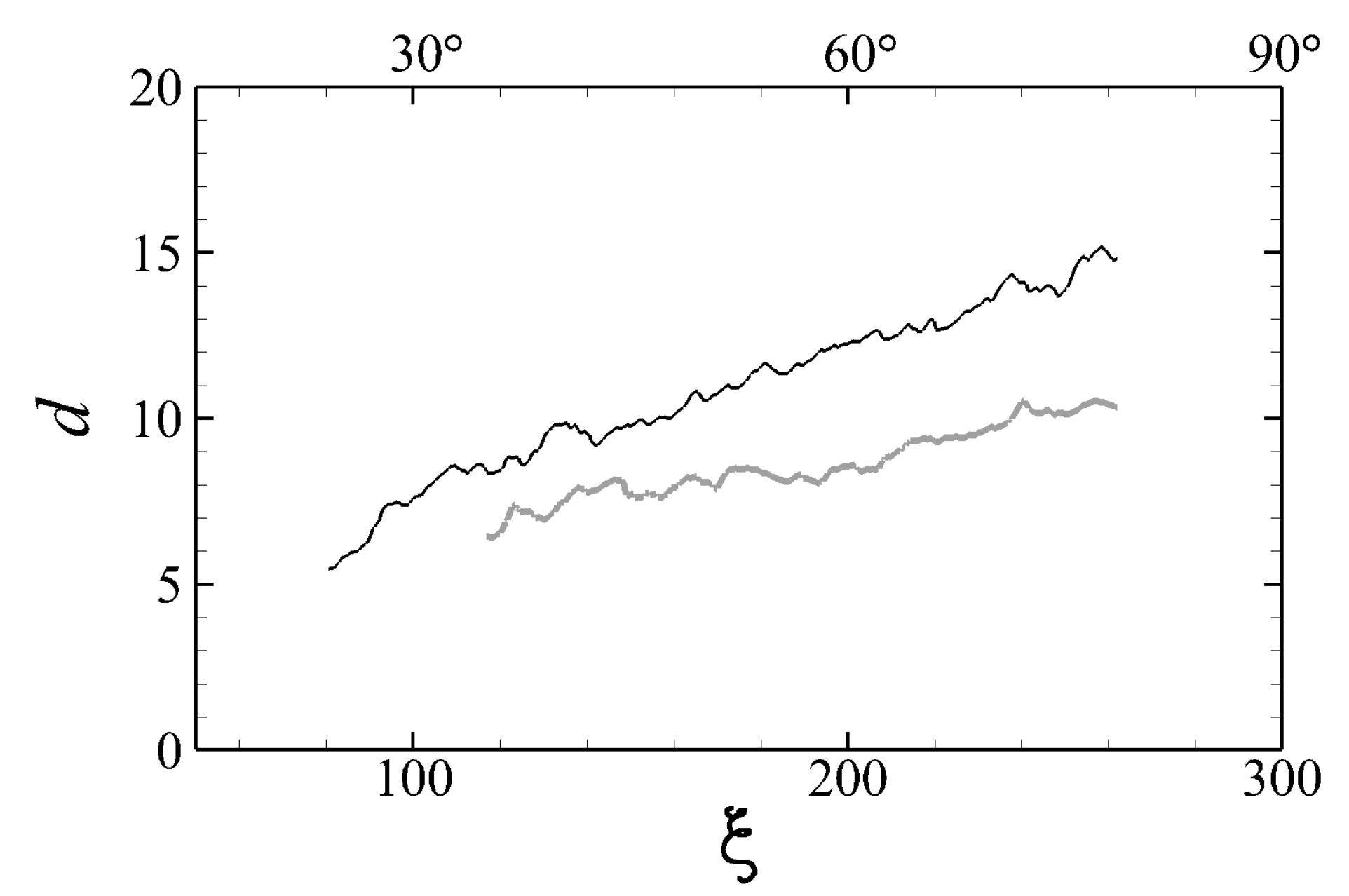}
	\end{center}
	\caption{Separation distance $d$ between the peaks of $\overline{u_\eta' u_\eta'}$ and $\overline{w' w'}$. (Gray) REF and (black) FRC.}
	\label{fig_ip_op}
\end{figure}

Observations in connection with the tangential stress are not, however, the most important to note from this figure if interest is in the G\"ortler structures. 
Instead, attention is drawn to the wall-normal $\overline{u_\eta' u_\eta'}$ and spanwise $\overline{w' w'}$ stresses in Region 2.  
Both stresses amplify on the curved wall, which is consistent with earlier studies \citep{Barlow_1988a, Lund_1996, Arolla_2015}, and the present results show that the effect of free-stream forcing is rather pronounced. 
In the lower panels ($a.ii$ and $b.ii$), the wall-normal coordinate is normalized by the inlet boundary-layer momentum thickness, which highlights that the separation between the peak $\overline{u_\eta' u_\eta'}$ and $\overline{w' w'}$ increases downstream.  
We interpret the increase in $\overline{u_\eta' u_\eta'}$ and $\overline{w'w'}$ as the first, perhaps indirect evidence of G\"ortler structures (further evidence provided in \S\ref{sec:structures}).
In reality, instantaneous structures may form at various heights in the boundary layer, meander, decay or be overtaken by other structures.  
For these reasons, in the statistical interpretation, we regard the peaks of $\overline{u_\eta' u_\eta'}$ and $\overline{w' w'}$ as only indicative of the locations of the G\"ortler structures, and hence the separation distance $d=\textrm{arg max}\left[\overline{u_\eta'u_\eta'}(\eta)\right]-\textrm{arg max}\left[\overline{w'w'}(\eta)\right]$ between their peaks as indicative of the size of the vortices (see figures~\ref{fig_stress_cont}$a.ii$ and $b.ii$).  
Figure~\ref{fig_ip_op} shows the increase of that size $d$ with downstream distance, more so in the forced flow in light of the stronger amplification of the outer large-scale structures and the faster growth of the boundary layer.

\begin{figure}
	\begin{center}
		\includegraphics[width=0.45\columnwidth]{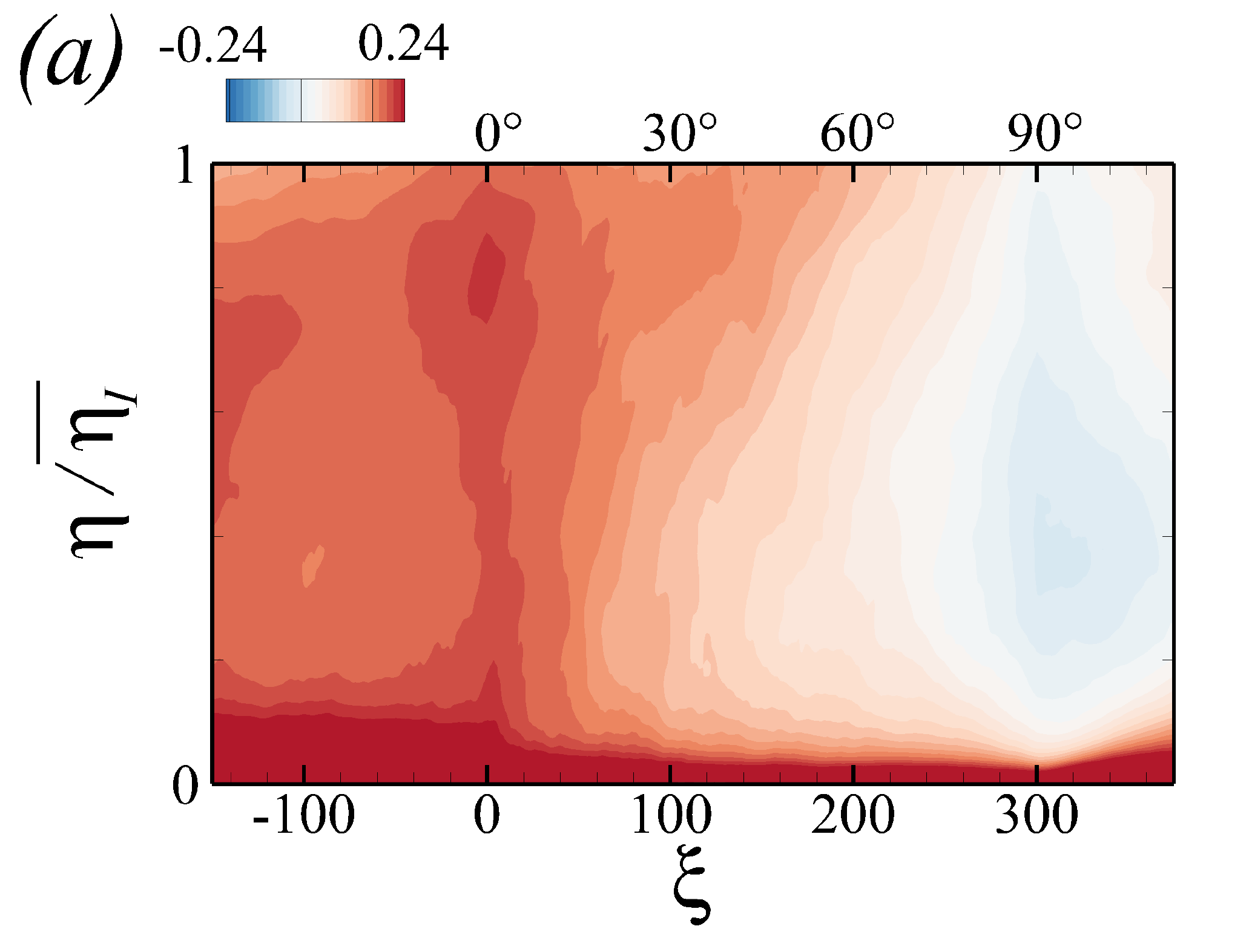}\hspace*{12pt}
		\includegraphics[width=0.45\columnwidth]{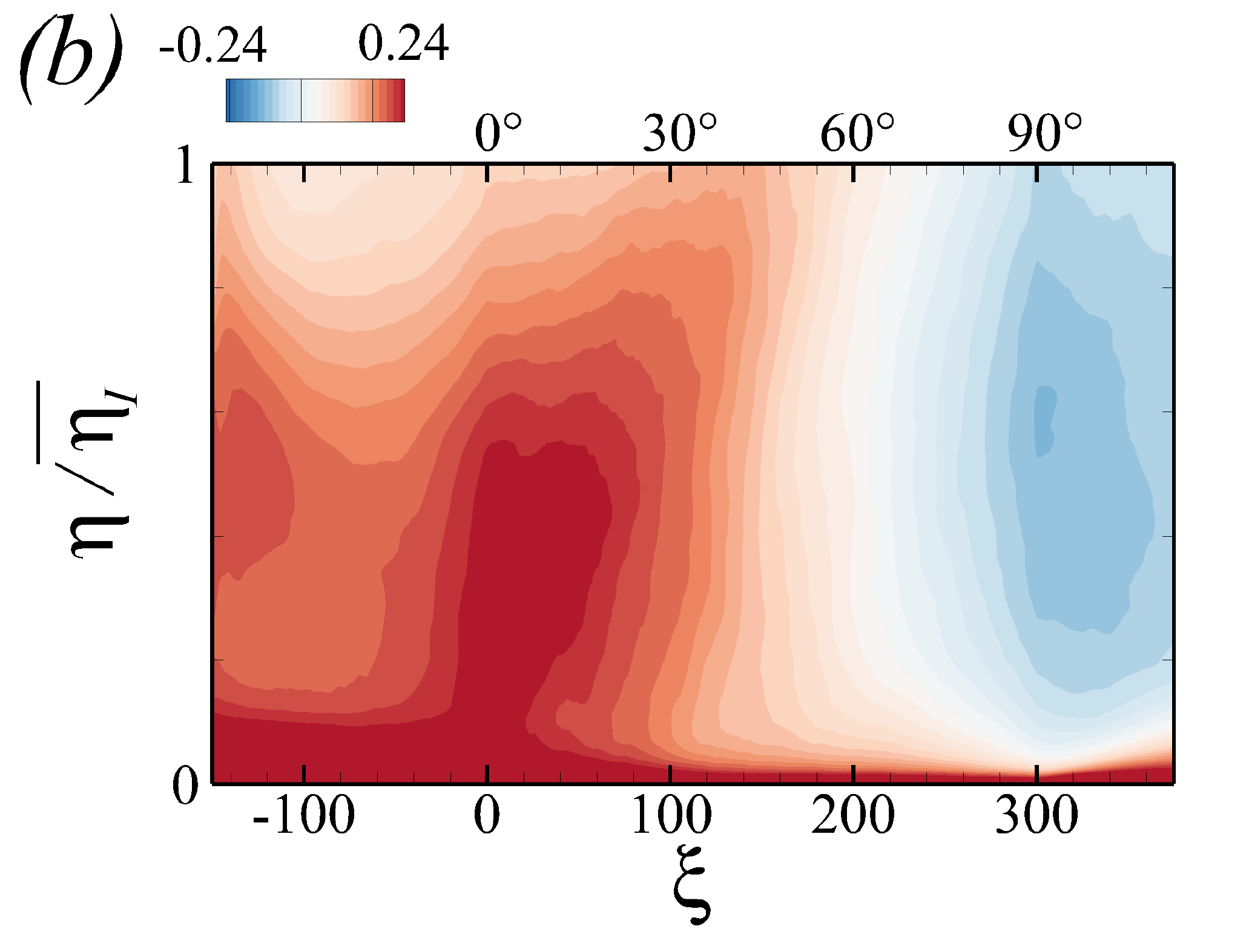}
	\end{center}
	\caption{Deviation of Reynolds normal stresses from isotropy, $I = \overline{u_\xi' u_\xi'}/2k - \frac{1}{3}$.
	$(a)$ REF and $(b)$ FRC.
} 
	\label{fig_iso}
\end{figure}

The departure from isotropy $I = \overline{u_\xi' u_\xi'}/{2k} - \frac{1}{3}$ succinctly captures the changes of the perturbation field within the boundary layer, across the onset of curvature and on the curved wall.
Figure \ref{fig_iso} shows positive values of $I$ in the initial flat section, which are consistent with the streamwise Reynolds stress being dominant in flat-plate boundary layers. Note that near the BL edge, $(b)$ shows that mixing with isotropic free-stream turbulence reduces $I$.
At the onset of curvature, the appreciable increase in $I$ reflects the increase in $\overline{u'_\xi u'_\xi}$ in the logarithmic layer (see figures~\ref{fig_stress_cont}$a.i$ and $b.i$).  Along the curved section, $I$ decreases with the amplification of $\overline{u_\eta' u_\eta'}$ and $\overline{w' w'}$ associated with the formation of outer G\"ortler structures. The decay in $I$ is more precipitous in the forced flow in the region $50 \lesssim \xi \lesssim 300$, which is symptomatic of the larger amplification of these structures.  

\begin{figure}
\begin{center}
	\includegraphics[width=0.49\columnwidth]{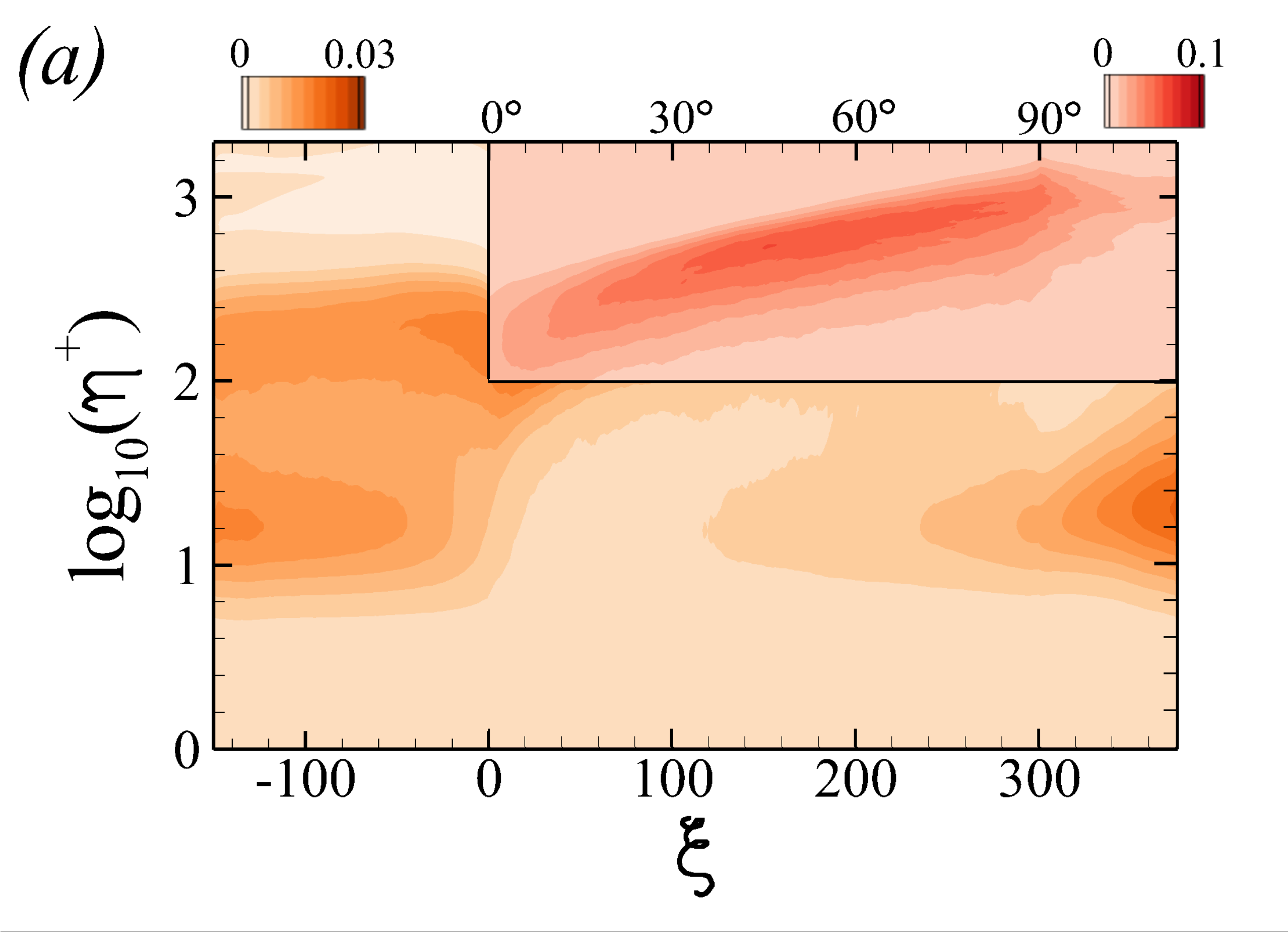}
	\includegraphics[width=0.49\columnwidth]{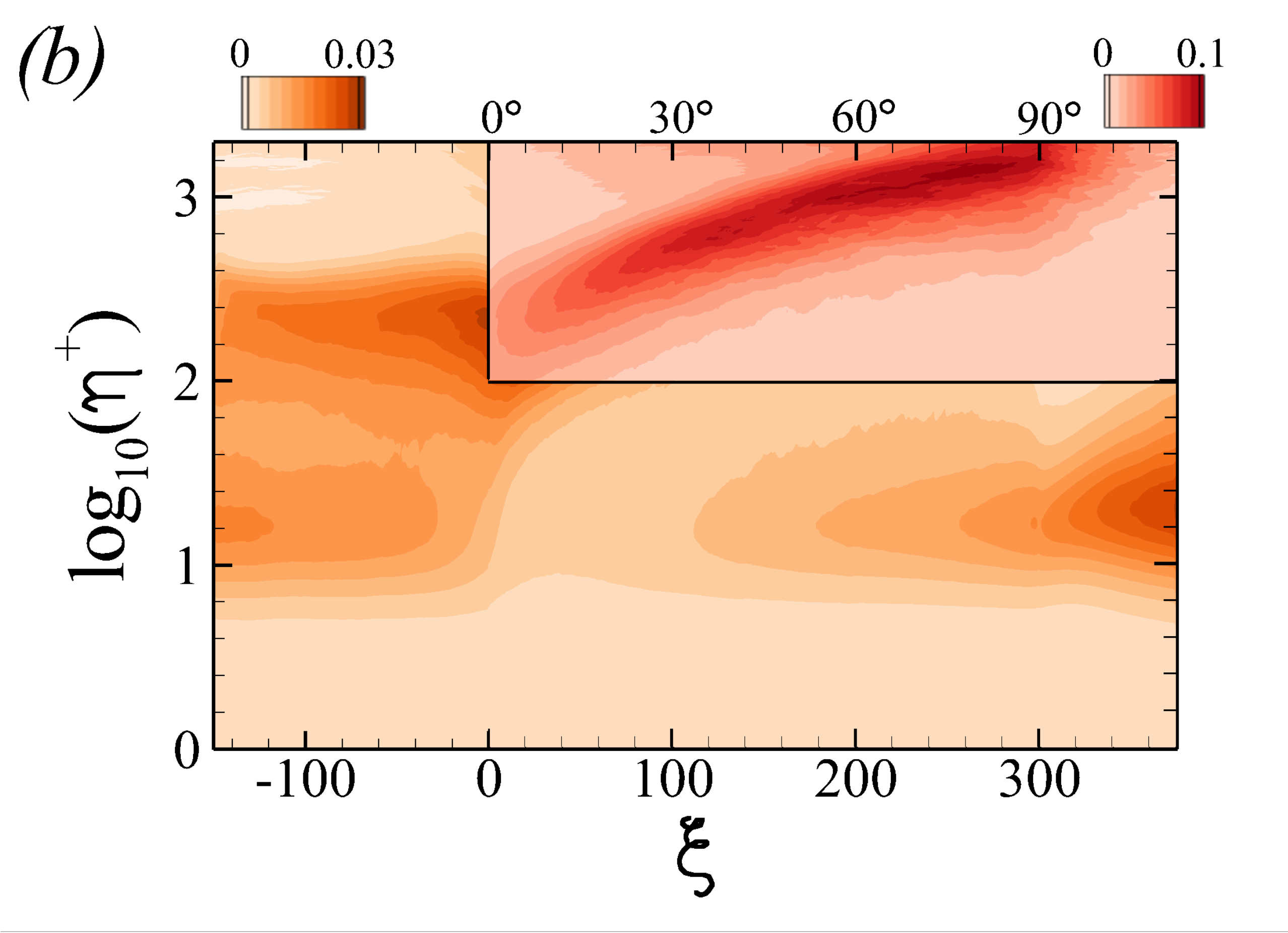}
\end{center}
\vspace{-0.1in}
\caption{Contours of pre-multiplied turbulence kinetic-energy production, $\eta^+\mathcal{P}$. $(a)$ REF and $(b)$ FRC.
}
	\label{fig_prod}
\end{figure}

The changes in the normal stresses are paralleled by changes in the production of turbulence kinetic energy, $\mathcal{P}=-\mathbf{R}:{\nabla \overline{\mathbf{u}}}$ where $\mathbf{R}$ is the Reynolds stress tensor. 
In figure \ref{fig_prod}, since a logarithmic scale is adopted in the wall-normal direction, the contours show the pre-multiplied quantity $\eta^+\mathcal{P}$ in order to reflect the contribution to the wall-normal integral.  
Also note that two contour levels are adopted in each figure, with larger range for the outer region of the boundary layer above the curved surface.
On the flat upstream section, the inner and outer peaks make comparable contributions to the integrated production, and both are enhanced by free-stream turbulence\textemdash an effect that has been detailed in previous studies \citep{You_2019}.
The adverse pressure gradient at the flat-to-curved transition suppresses the near-wall production, and recovery is slow on the curved region albeit faster in presence of free-stream turbulence.
In the outer region of the curved-wall boundary layer ($\xi\gtrsim0$), the production peak is significantly enhanced. 
While its initial amplification near the onset of curvature coincides with the lifting of near-wall streaks due to APG, its continued amplification downstream coincides with the amplification of the outer stresses and potentially the formation and amplification of G\"ortler structures.  
In presence of free-stream forcing, the magnitude of that outer peak is nearly twice its value in the reference configuration.

\begin{figure}
	\begin{center}
		\includegraphics[width=0.49\columnwidth]{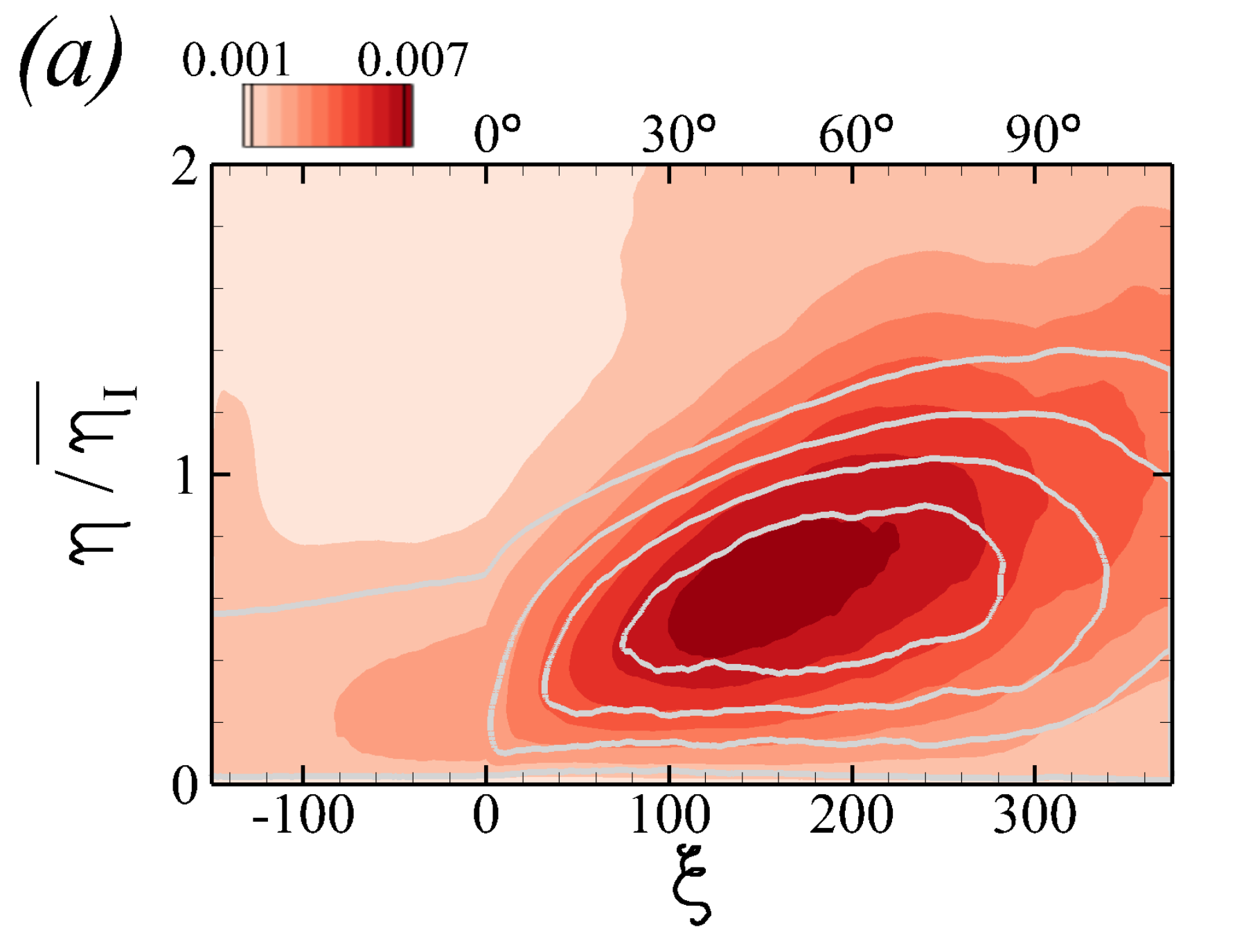}
		\includegraphics[width=0.49\columnwidth]{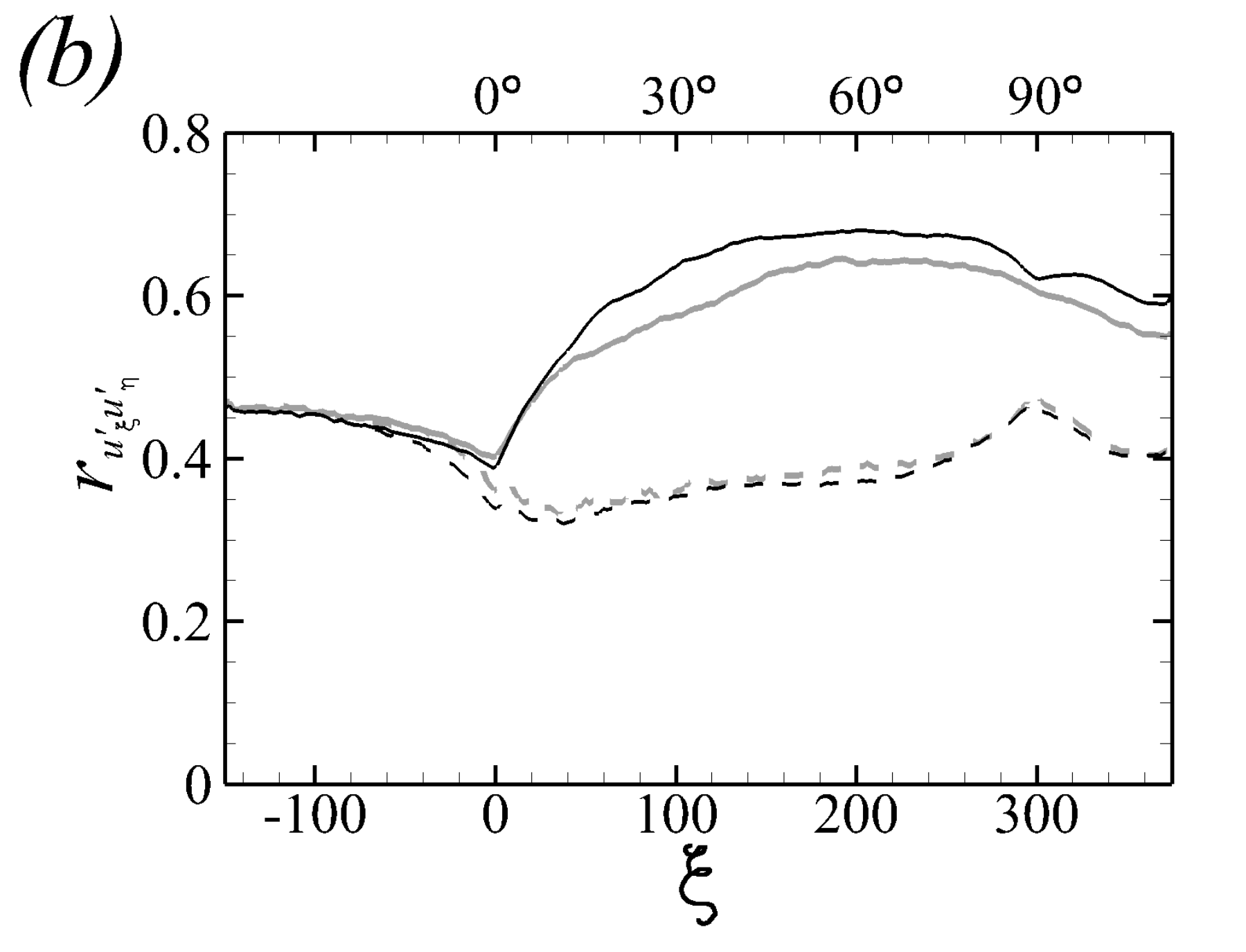}
	\end{center}
	\caption{$(a)$ Reynolds shear stress $-\overline{u_\xi' u_\eta'}$ for (line) REF and (color) FRC cases. 
			Contours lines correspond to levels from $1 \times 10^{-3}$ with increment $1 \times 10^{-3}$. 
			$(b)$ Downstream development of shear-stress correlation coefficient $r_{u_\xi' u_\eta'}$,  extracted at (\dashed) $\eta^+ = 10$ and (\lline) wall-normal location of the maximum. (Gray) REF and (black) FRC.
}
	\label{fig_uv_coeff}
\end{figure}

The TKE production in the outer region is significantly affected by Reynolds shear stresses.
We first recall results from forced flat-plate boundary layers for comparison:   Even when the free-stream turbulence is isotropic, and hence free of average shear stress, it enhances $-\overline{u_\xi' u_\eta'}$ within the boundary layer although it reduces the stress correlation coefficient $r_{u_\xi' u_\eta'} \equiv {-\overline{u_\xi' u_\eta'}}/{u_{\xi,rms}' u_{\eta,rms}'}$ \citep{Hancock_1989, Thole_1996}.  
The shear stress and its correlation coefficient for the present curved-wall boundary layers are reported in figure \ref{fig_uv_coeff}.  
The former quantity is plotted throughout the boundary layer, and the latter is extracted at select locations. 
Over the curved section, $-\overline{u_\xi' u_\eta'}$ increases appreciably and reaches larger values for the FRC case. 
Figure \ref{fig_uv_coeff}$b$ shows the correlation coefficient at $\eta^+ = 10$ and at the wall-normal height where it is maximum.  
The figure shows that the peak occurs at $\eta^+ = 10$ on the flat section of the wall, but the correlation at that location decays due to the pressure gradient \citep{Gungor_2016}.  
Note that this effect was not reported in the previous experimental curved-wall studies by \citet{Barlow_1988a} who removed the effect of pressure gradient by contouring the top convex wall.
On the curved wall, however, the peak shifts higher in the boundary layer and is much larger in magnitude due to the coherence of the turbulence structures in that region.  
In addition, the coefficient of the forced case is markedly greater in the outer region than that of the reference flow.
We anticipate that the FST strengthens the outer roll motion on the curved region, thereby enhancing the correlation coefficient.

This section has demonstrated that the free-stream turbulent forcing mitigates the intermittent separation that can take place at the onset of curvature, enhances mixing of mean momentum along the curved wall and leads to appreciable and sustained increase in skin friction.  The forcing also alters the state of the turbulence within the boundary layer along the curved wall.  The distribution of the turbulent stresses suggests the formation of naturally triggered G\"{o}rtler structures, which are more energetic and larger in size when the boundary layer is buffeted by the external turbulence. 
The following section upholds these interpretations by directly probing the flow structures on the curved wall without and with FST.

\section{Modification of boundary-layer structures}
\label{sec:structures}

\begin{figure}
	\begin{center}
		\includegraphics[width=0.49\columnwidth]{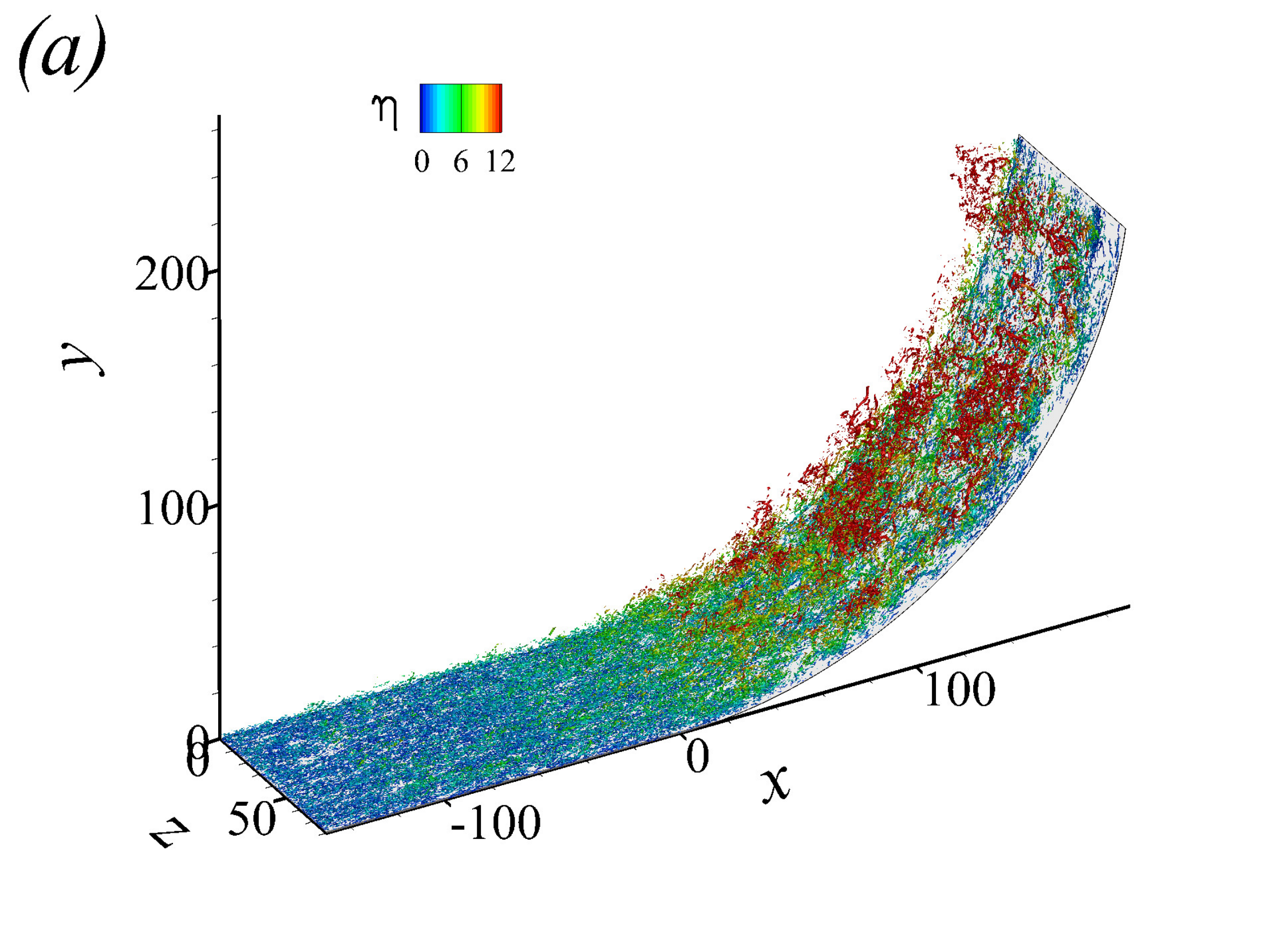}
		\includegraphics[width=0.49\columnwidth]{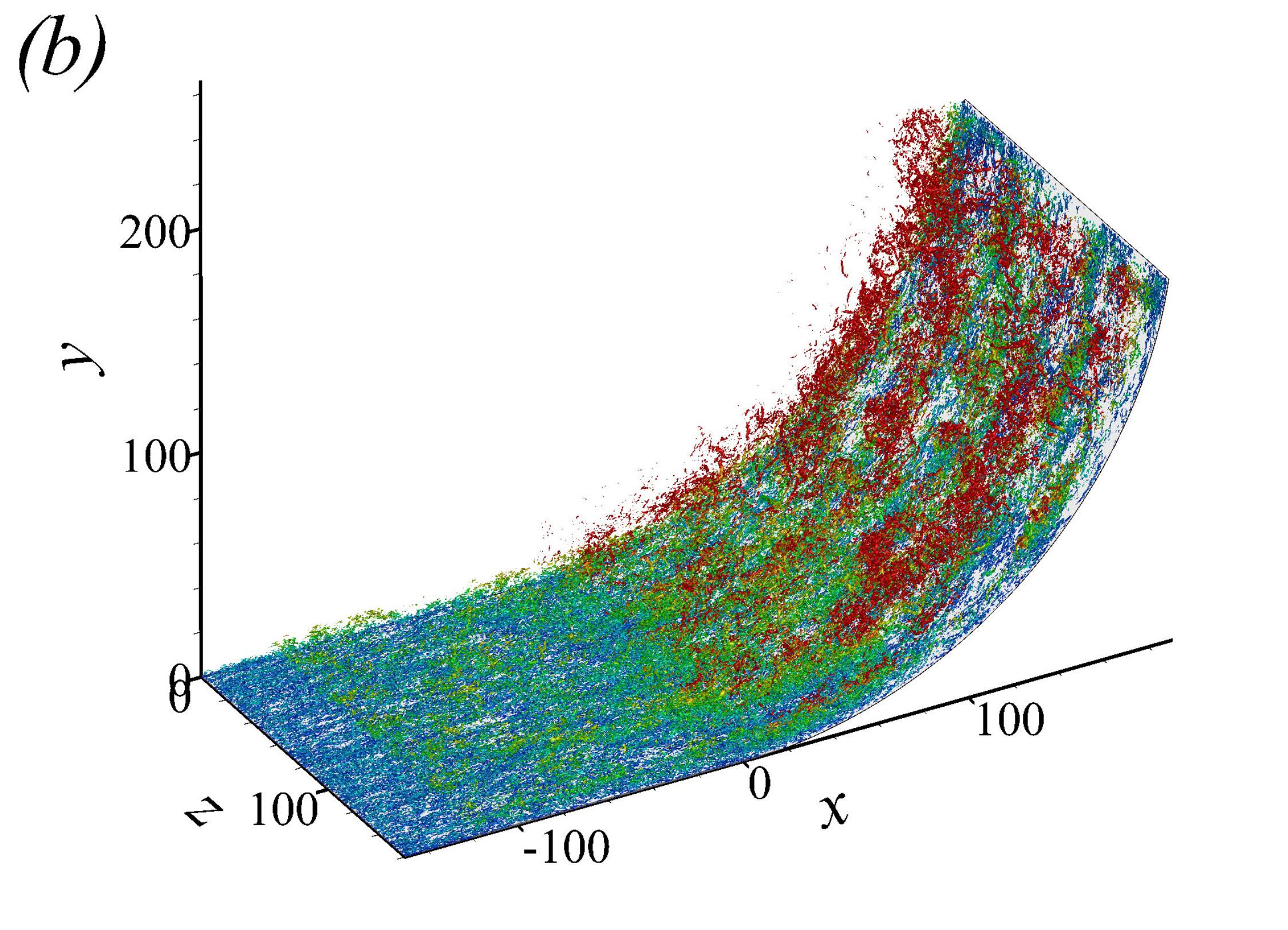}
	\end{center}
	\caption{Iso-surface of $Q$-criterion with threshold $Q= 0.03$, coloured by $0\leq \eta \leq 12$. $(a)$ REF and $(b)$ FRC.
} 
	\label{fig_qcri}
\end{figure}

A commonly adopted approach to identifying coherent vortical motions is to evaluate an invariant of the velocity gradient tensor, for example $Q \equiv \Omega_{ij} \Omega_{ij} - S_{ij} S_{ij}$ where $S_{ij}$ is the symmetric, rate-of-deformation tensor and $\Omega_{ij}$ is the anti-symmetric, spin tensor.
Figure \ref{fig_qcri} shows iso-surfaces of the $Q$-criterion coloured by their wall-normal distance.  
In both REF and FRC, the structures grow and spread in $\eta$ as they travel into the concave curvature; similar observations were made for FST-free flows~\citep{Arolla_2015}. 
This trend is consistent with the development of the outer peak in the profiles of Reynolds stresses (figures \ref{fig_stress_cont}). 
The increase in vortical activity has previously been attributed to G\"ortler vortices due to the centrifugal effect, although coherent large-scale vortical structures are not apparent in the figure. 
It is also important to note that the response in presence of FST is hardly distinguishable from the reference flow, based on the instantaneous $Q$ iso-surfaces. 
While streamwise organization is visually discernible in the figure, G\"ortler vortices which are associated with coherent longitudinal roll motions can not be easily identified, and attempts to adopt filtering techniques were not successful.

\begin{figure}
\begin{center}
		\includegraphics[width=1\columnwidth]{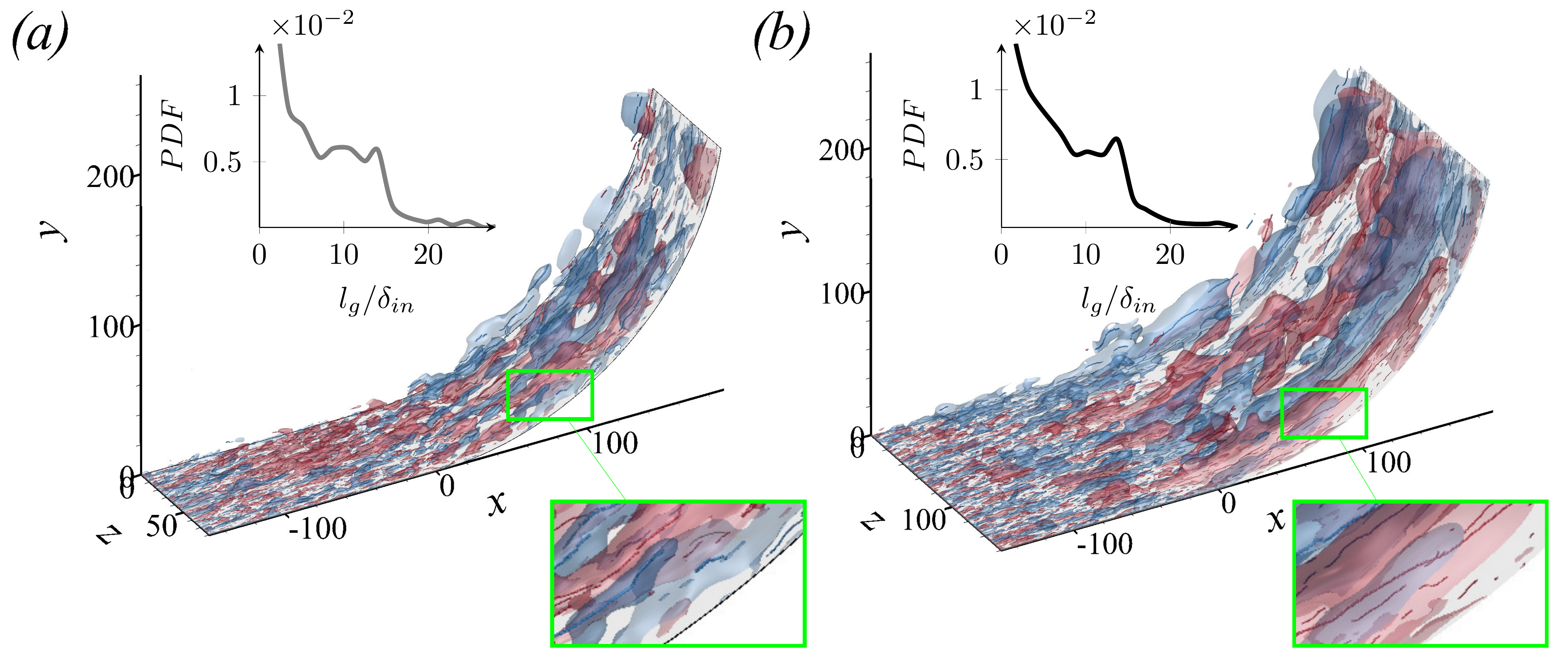}
	\end{center}
	\caption{Iso-surfaces of filtered streamwise velocity, (red) $\hat{u}_\xi' = +0.06$ and (blue) $\hat{u}_\xi' = -0.06$. 
	Lines within the transparent surfaces mark the cores of the structures.
	$(a)$ REF and $(b)$ FRC. Insets show the PDF of the lengths of the cores in the outer region $30<\eta^+<\eta_I^+$ and which cross $\xi=100$ ($\varphi=30^\circ$).} 
	\label{fig_ftr}
\end{figure}

Direct identification of naturally emerging G\"ortler motions in turbulent flows over curved walls is difficult.  This challenge led \citet{Barlow_1988b} to artificially impose them using vortex generators.  For streamwise homogeneous flows, \cite{moser1987effects} performed streamwise averaging of the velocity field to visualize vortical structures in a curved channel flow. 
Due to streamwise inhomogeneity in the present configuration, however, we consider the Gaussian-filtered velocity fields as done in previous studies~\citep{Hutchins_2007,lee2014spatial,Hwang_2016}.  The displacement of momentum effected by the G\"ortler motions generates coherent tangential velocity perturbations that are readily observable in the Gaussian-filtered $\hat{u}'_\xi$. 
Following the approach described by \citet{Lee_2017}, we evaluate iso-surfaces of $\hat{u}'_\xi$ and identify the cores of those structures (see figure \ref{fig_ftr}).  In order to interpret the present results, it is helpful to recall those for a flat plate at the same Reynolds numbers \citep{You_2019}:  In that case, a canonical boundary layer without free-stream forcing does not develop outer large-scale motions; under free-stream turbulence, large-scale structures form and amplify, but a much longer streamwise extent is required than the flat region in the present configuration.  
Contrasting REF and FRC for $\xi<0$, the velocity structures appear similar in that region, and hence any differences downstream on the curved wall are due to the interaction in that regime. 
There, the outer structures are clearly visible, and the iso-surfaces are larger in presence of free-stream forcing which also implies that the tangential velocities within their cores are higher in amplitude.

Figure \ref{fig_ftr} also shows that we can identify the cores of the large-scale structures, which can be classified into positive  $\mathcal{C}^P$~($\hat{u}'_\xi>0$) and negative $\mathcal{C}^N$~($\hat{u}'_\xi<0$) ones.
The PDF of the lengths $l_g$ of these cores is reported in the inset, for structures that are detected in the outer region $30<\eta^+<\eta_I^+$ at $\xi=100$ ($\varphi=30^\circ$). The PDF and shows an increased probability at lengths $O(5-6\,\delta_{99})$.
In addition to their streamwise advection, these cores exhibit a weak spanwise drift velocity with equal probability of positive or negative value. The average $w$ along the core of a structure is typically one order of magnitude less than the local $w_\textrm{rms}$.

\begin{figure}
	\begin{center}
		\includegraphics[width=1.0\columnwidth]{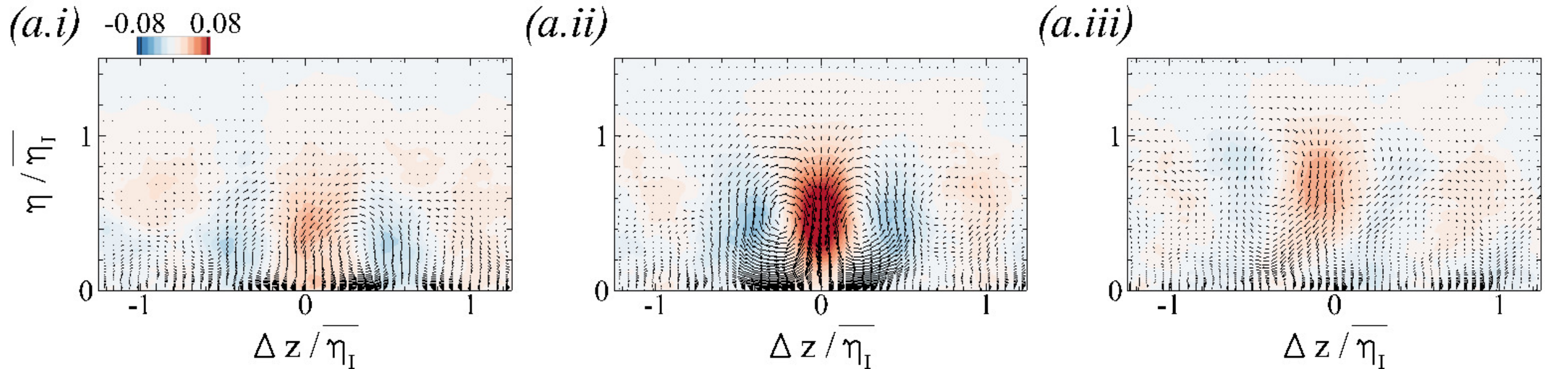}
		\includegraphics[width=1.0\columnwidth]{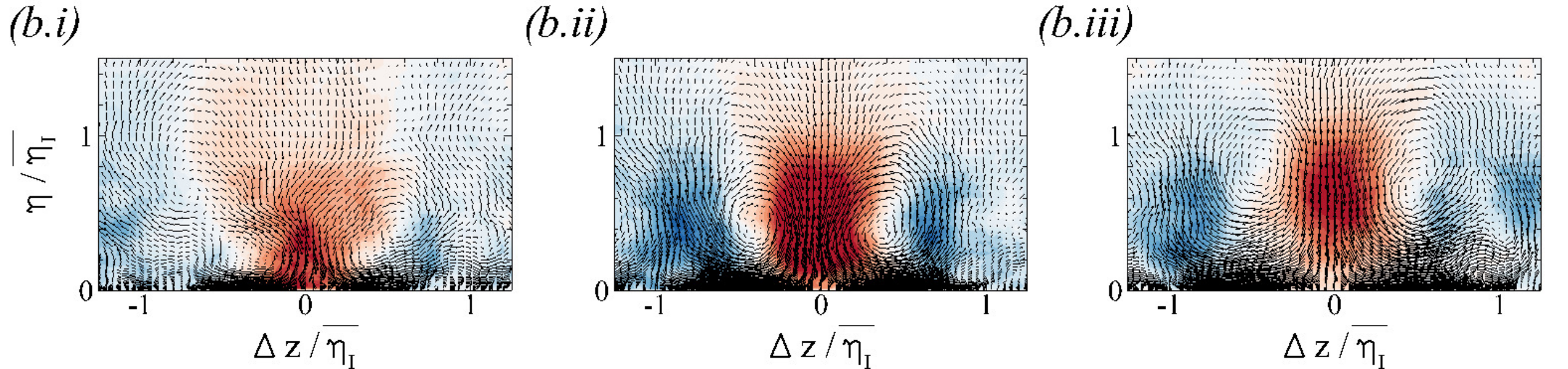}
	\end{center}
	\caption{Conditionally averaged positive streamwise velocity on the curve with reference position $\xi_o = 100$ ($30^\circ$ station), and coloured by ${u_\xi'}^P_\mathcal{L}$.  Black vectors are $({u_\eta'}^P_\mathcal{L}, {w'}^P_\mathcal{L})$.
	$(a)$~REF, $(b)$~FRC, $(i)$~$\Delta \xi= -25$, $(ii)$~$\Delta \xi = 0$,
	and $(iii)$~$\Delta \xi = +25$.
} 
	\label{fig_cond_str1}
\end{figure}

\begin{figure}
	\begin{center}
		\includegraphics[width=1.0\columnwidth]{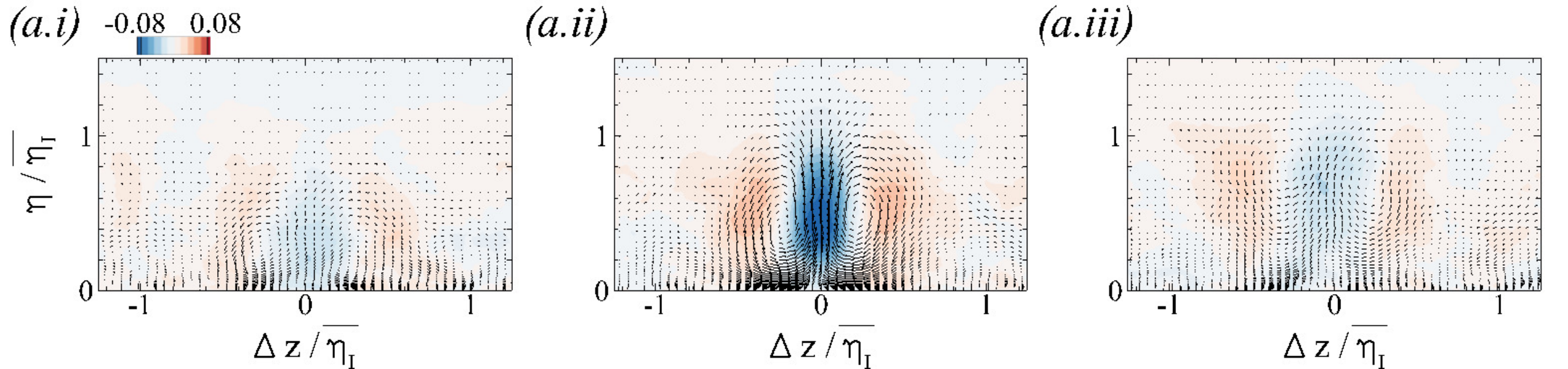}
		\includegraphics[width=1.0\columnwidth]{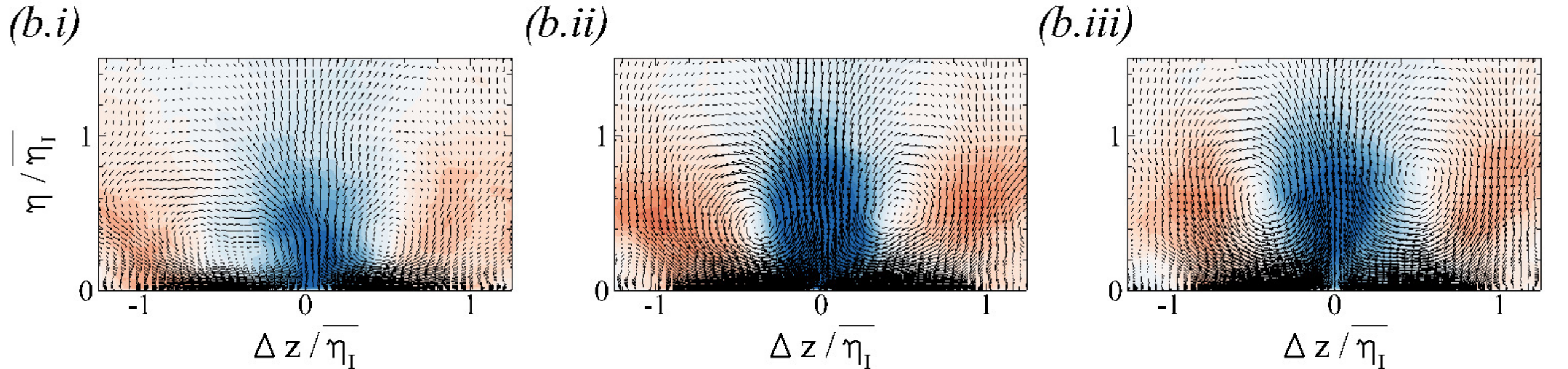}
	\end{center}
	\caption{Conditionally averaged negative streamwise velocity on the curve with reference position $\xi_o = 100$ ($30^\circ$ station), and coloured by ${u_\xi'}^N_\mathcal{L}$.  Black vectors are $({u_\eta'}^N_\mathcal{L}, {w'}^N_\mathcal{L})$.
	$(a)$~REF, $(b)$~FRC, $(i)$~$\Delta \xi= -25$, $(ii)$~$\Delta \xi = 0$,
	and $(iii)$~$\Delta \xi = +25$. 
} 
	\label{fig_cond_str2}
\end{figure}

Definitive evidence of the outer roll, or G\"ortler, motions and their influence on the near-wall flow are sought by computing conditionally averaged velocity perturbation fields.  
The condition adopted for averaging is the existence of a core of an outer tangential velocity structure, and is further differentiated into positive and negative ones similar to the instantaneous visualization (c.f.\,figures \ref{fig_ftr}).
The conditionally averaged perturbation fields are therefore,
\begin{align}
	\label{eqn_03_1}
	 \textbf{u}^P_\mathcal{L} (\Delta \xi,\eta,\Delta z) &= \overline{ \textbf{u} (\xi_o+\Delta \xi,\eta,z+\Delta z) ~~|~~ {\mathcal{C}^P}_{|30<\eta^+<\eta_I^+}  },  \\
	\label{eqn_03_2}
	 \textbf{u}^N_\mathcal{L} (\Delta \xi,\eta,\Delta z) &= \overline{ \textbf{u} (\xi_o+\Delta \xi,\eta,z+\Delta z) ~~|~~ {\mathcal{C}^N}_{|30<\eta^+<\eta_I^+}  }.  
\end{align}
Figures \ref{fig_cond_str1} and \ref{fig_cond_str2} show the contours of $\textbf{u}^P_\mathcal{L}$ and $\textbf{u}^N_\mathcal{L}$, with the reference streamwise position $\xi_o = 100$ ($30^\circ$ station).
Vectors represent the in-plane velocity components.
While the outer roll motions are visible at $\Delta \xi =0$ in the canonical boundary layer, they are much more pronounced (larger in size and amplitude) beneath free-stream turbulence.   
The more important observation, however, is the persistence of the roll motion and associated tangential velocity response upstream and downstream of the reference position.  
In the quiescent flow, the perturbation field ${u_\xi'}^{\{P,N\}}_\mathcal{L}(\Delta \xi = \pm 25)$ is decorrelated appreciably from the reference point (e.g.~figures \ref{fig_cond_str1}-\ref{fig_cond_str2}, $a.i$ and $a.iii$).
In contrast, in the FRC case, the structures identified at the reference location persist beyond $\Delta \xi = \pm 25$ (e.g.~figures \ref{fig_cond_str1}-\ref{fig_cond_str2}, $b.i$ and $b.iii$).
The improved coherence underscores that free-stream turbulence does not decorrelate the turbulence within the logarithmic layer on the curved section, but rather promotes it.

The outer roll motion also increases in size with downstream distance, from $\Delta \xi= - 25$ to $+25$. 
This view was also supported by further evidence (not shown) where we evaluated the conditional velocity fields using equations (\ref{eqn_03_1}) and (\ref{eqn_03_2}) at reference locations $\xi_o=75$ and $\xi_o=125$.
The large roll motions which are growing with downstream distance in the forced flow are consistent with earlier statistical evidence based on the Reynolds stress, namely that the gap between the peak $\overline{u_\eta' u_\eta'}$ and $\overline{w'w'}$ increases along the curved wall (c.f.\,figure \ref{fig_ip_op}).

\begin{figure}
	\begin{center}
		
		\includegraphics[width=1.0\columnwidth]{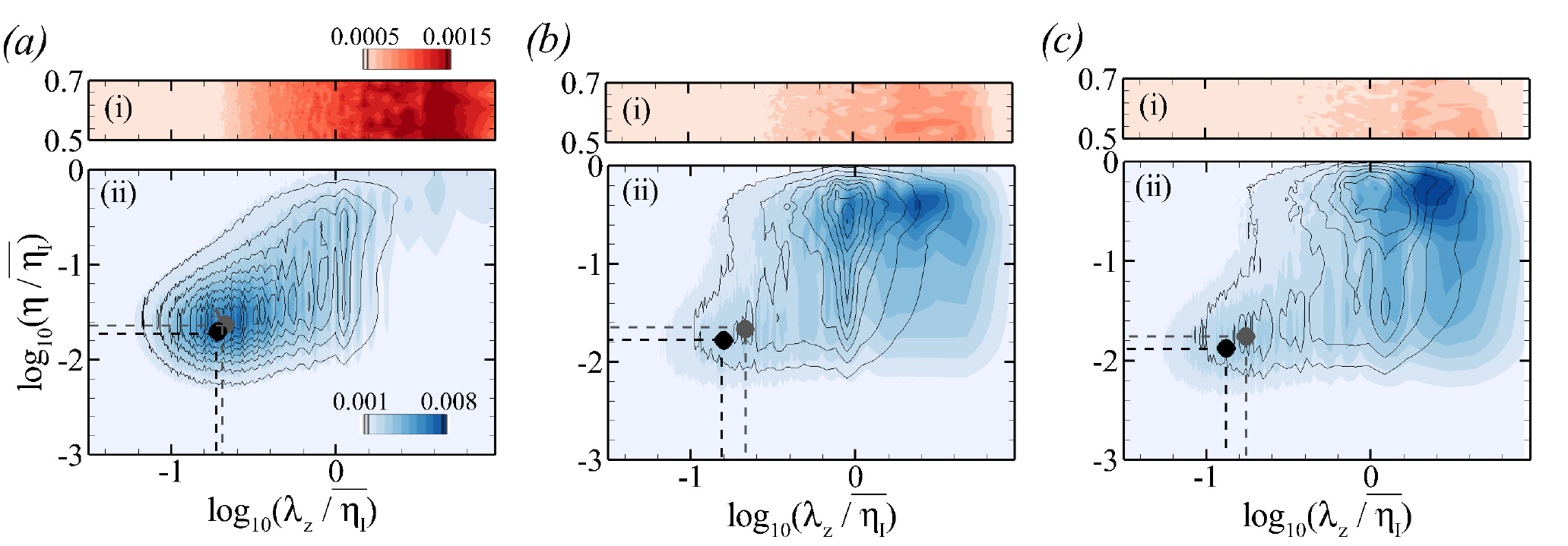}
	\end{center}
	\caption{Pre-multiplied spanwise energy spectra at $(a)$ $\xi=-100$,
	$(b)$ $\xi=50$ and $(c)$ $\xi=100$, with levels ($i$)~$0.0005\leq \kappa_z \Phi_{u_\xi' u_\xi'}\leq 0.0015$ in the free stream and ($ii$)~$0.001\leq \kappa_z \Phi_{u_\xi' u_\xi'}\leq 0.008$ in the boundary layer. (Black lines) REF and (color) FRC.
	Filled circles mark $\eta^+=10$ and $\lambda_z^+=100$ in (gray) REF and (black) FRC.
	} 
	\label{fig_spec}
\end{figure}

The amplification of G\"ortler motions depends on the stability characteristics of the boundary layer and the upstream forcing~\citep{hall1982taylor}. In the REF case, the forcing is provided solely by the turbulence in the upstream boundary layer, and an associated G\"ortler response develops on the curve. In the FRC case, there is the additional free-stream forcing which interacts with the developing boundary layer.  As early as the flat section, the results by \citet{You_2019} showed that the low-frequency perturbations from the free stream are effective at penetrating the logarithmic layer. The adverse pressure gradient which reduces shear sheltering can promote such penetration \citep{zaki_durbin_2006}. As a result, the dominant G\"ortler response on the curved section may differ from the REF case.
We evaluate the spectra to examine the energy at different scales in the perturbation fields. 
Figure \ref{fig_spec} shows the spanwise energy spectra of the streamwise velocity,
\begin{eqnarray}
\label{eqn_spec}
	\Phi_{u_\xi' u_\xi'} (\lambda_z) =  \int_{-\infty}^{\infty} \overline {u_\xi'(z) u_\xi'(z+\zeta)} e^{-i \kappa_z \zeta} \ d \zeta,
\end{eqnarray}
pre-multiplied by spanwise wavenumber $\kappa_z = 2 \pi / \lambda_z$ where $\lambda_z$ is the wavelength. Three streamwise positions are considered, and at each the spectrum is reported $(i)$ in the free stream  and $(ii)$ within the boundary layer where line and colour contours correspond to the REF and FRC case, respectively.
On the flat section (figure \ref{fig_spec}$a$), only an inner peak is visible, be that in the REF or FRC boundary layer.  A faint increase of energy is observed near the edge of the FRC boundary layer at large spanwise scales that are commensurate with the most energetic wavelengths in the free stream in $(a.i)$.
On the curved section, the energy in the small wavelengths decreases (figures \ref{fig_spec}$b$-$c$), which is consistent with reduced TKE dissipation (not shown).
In the reference flow, a clear outer peak emerges at $\lambda_z \approx \overline{\eta_I}$.
In contrast, in FRC at $15^\circ$ (color contour of figure~\ref{fig_spec}$b$), there are two outer peaks and both are associated with G\"ortler motions: 
The first, at the smaller spanwise wavelength (left), coincides with the contours of the canonical boundary layer and is the naturally forming G\"ortler vortices.  
The second emerges at the wavelength of the free-stream turbulent forcing (see panel $(a.i)$); this peak is the boundary layer G\"ortler response to the external forcing.  
The two peaks ultimately merge downstream $\xi = 100$, with the larger G\"ortler vortices becoming the most dominant. 
Similar to earlier studies of turbulence on curved walls~\citep{Tani_1962}, comparison can be made to results from linear stability. In our configuration, at $\xi=50$, the local G\"ortler number $G_t=5.4$ (c.f. figure~\ref{fig:ReGt}$b$) and the most energetic spanwise wavenumber is $k_z \theta=0.43$ which is also the most unstable mode according to linear theory \citep{smith1955growth,Tani_1962}.
An important observation is the extent to which that wavenumber persists, or remains energetic, deep into the boundary layer, which is suggestive of a stronger modulation of the near-wall region than in flat plates, perhaps even a direct influence.

\begin{figure}
	\begin{center}
		\includegraphics[width=0.6\columnwidth]{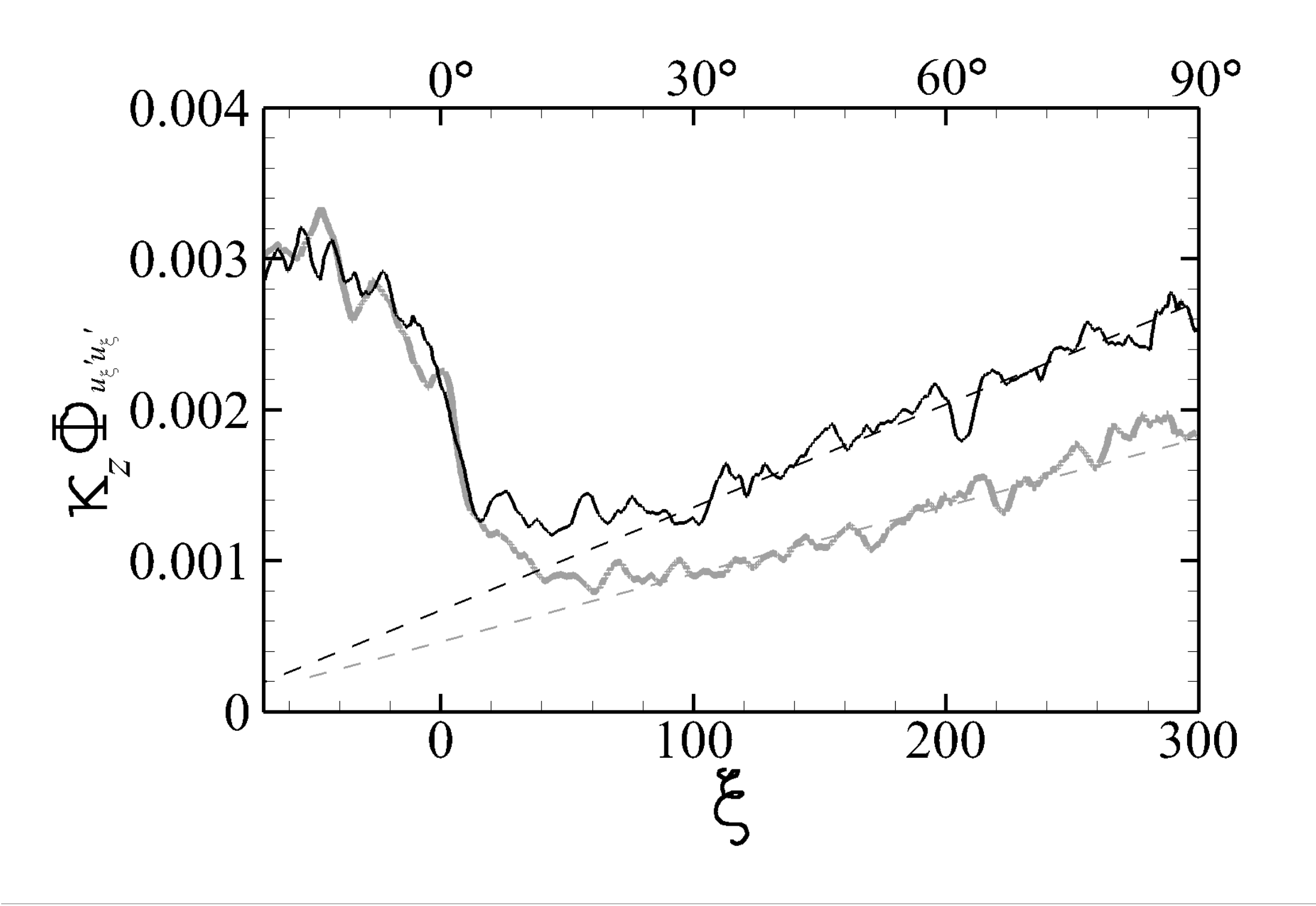}
	\end{center}
	\caption{Downstream development of $\kappa_z \Phi_{u_\xi' u_\xi'}$ at $\eta^+ = 10$ and $\lambda_z^+ = 100$:
		(gray) REF and (black) FRC.
		Dashed lines indicate the recovery rate in Region 2.
	}
	\label{fig_spec2}
\end{figure}

The small-scale energy in the near-wall region is also modified by the free-stream forcing.
Figure \ref{fig_spec} shows that, along the curved wall, energy shifts towards smaller spanwise wavenumbers in FRC which is consistent with enhanced dissipation.  
The inner peak in the spectra is generally located at $\eta^+ \approx 10$ and $\lambda_z^+ \approx 100$, and thus the downstream dependence of $\kappa_z \Phi_{u_\xi' u_\xi'}$ at the position is provided in figure \ref{fig_spec2}.
Downstream of the fast decay associated with the adverse pressure gradient at the onset of curvature, $\kappa_z \Phi_{u_\xi' u_\xi'}$ recovers at a faster rate for FRC.

\begin{figure}
	\begin{center}
		\includegraphics[page=1,width=0.875\textwidth]{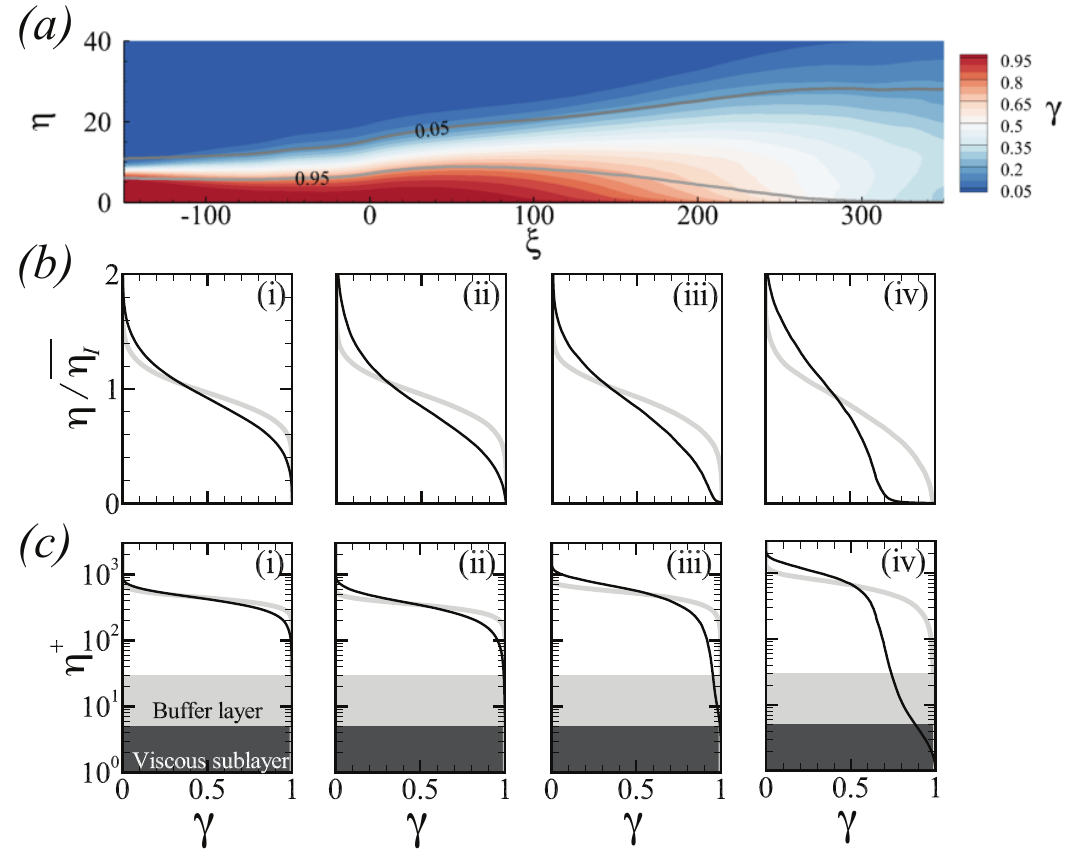}
	\end{center}
	\caption{$(a)$ Intermittency $\gamma$ for (lines, $\gamma=\{0.05, 0.95\}$) REF  and (colour, $0.05 \leq \gamma \leq 0.95$) FRC. Intermittency profiles in $(b)$~outer and $(c)$~viscous units, at (i-iv) $\xi = \{-100,0, 100, 200\}$; (Gray) REF and (black) FRC.}
	\label{fig:gamma:new}
\end{figure}

The coherent outer G\"ortler motions have important implications, including their ability to enhance mixing and, in the forced configuration, potentially transport the free-stream turbulence deep into the boundary layer towards the wall. An objective measure of penetration of the free-stream fluid into the boundary layer is the intermittency $\gamma$.  
Starting from the levelset function that provides an objective instantaneous virtual interface between the two regions, an indicator function is defined $\Gamma = \{0,1\}$ for $\psi \gtrless 0.5$, and then averaged to obtain $\gamma = \overline{\Gamma}$.  
We recall that in a flat-plate boundary layer, \cite{You_2019} adopted the same measure and demonstrated that the free-stream turbulence does not breach the buffer layer within the streamwise domain length of interest here.
The results for the present curved wall configuration are reported in figure \ref{fig:gamma:new}. 
Panel ($a$) captures the difference between the flat and curved sections: even for the unforced flow the intermittency iso-lines spread more quickly on the curve which is indicative of enhanced mixing of the free-stream and boundary-layer fluids (see also the wall-normal profiles in outer scaling in panel ($b$)). 
Forcing by free-stream turbulence amplifies these trends significantly, and in panels ($c.iii$-$iv$) we observe remarkable levels of penetration of free-stream turbulence inside the buffer layer at $\xi\gtrsim 100$~($\varphi>30^\circ$, Region 2). 
This enhanced ``mixing" is unique to the present configuration and is precipitated by the large-scale outer roll motions, transporting the free-stream fluid deep towards the wall and ejecting near-wall fluid outwards. 
The interaction of the outer and inner regions is therefore direct. 
Ultimately, however, this large-scale effect is also accompanied by further dispersion of the ingested turbulence due to the various flow scales within the boundary layer, and by molecular diffusion acting on the smallest scales.   
Thus, we next consider the recovery of the near-wall small-scale structures on the curved wall, and in particular their modulation by the outer large scales that can spur that recovery.

\begin{figure}
\centering
\includegraphics[width=0.45\textwidth]{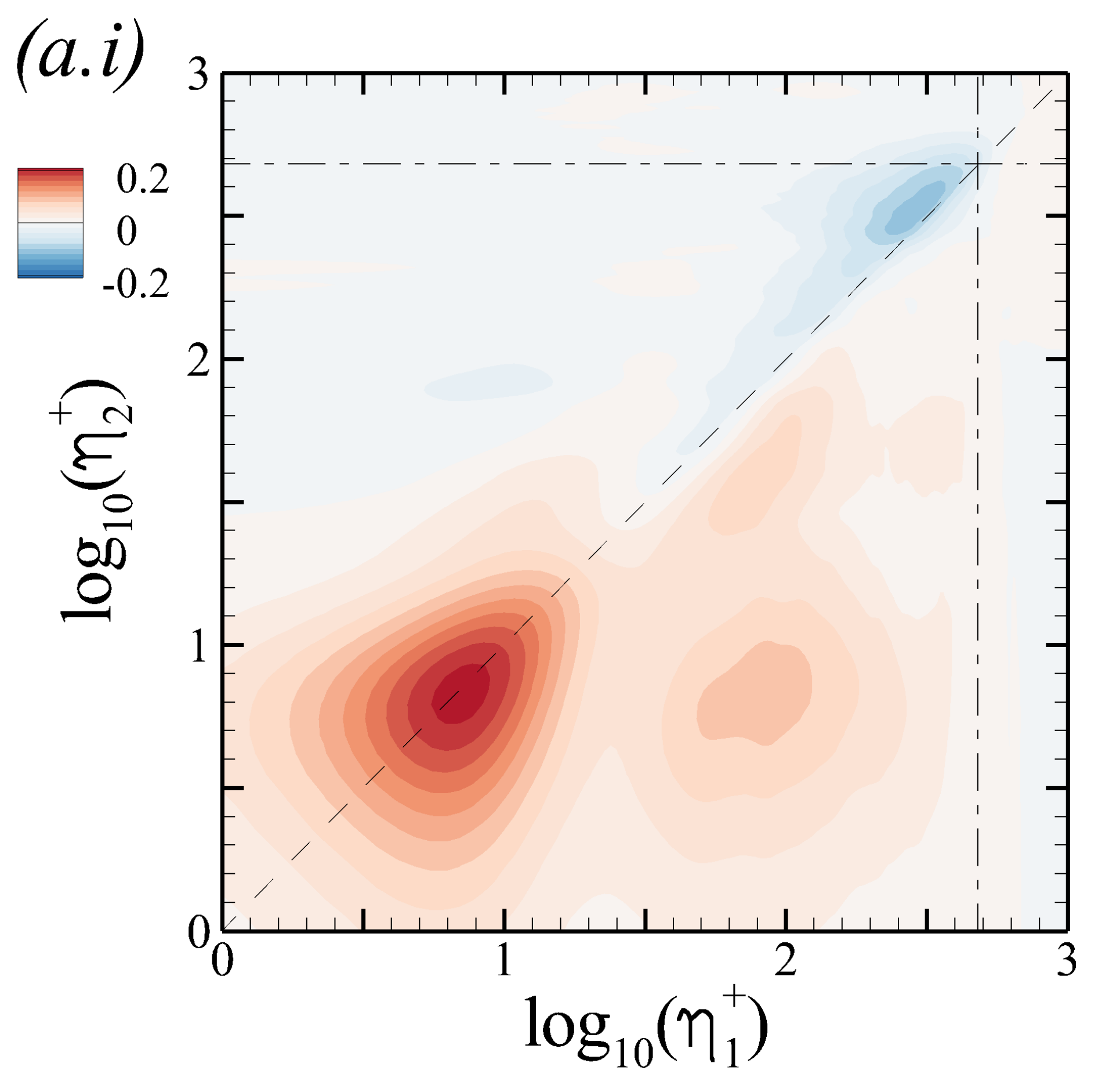}
\includegraphics[width=0.45\textwidth]{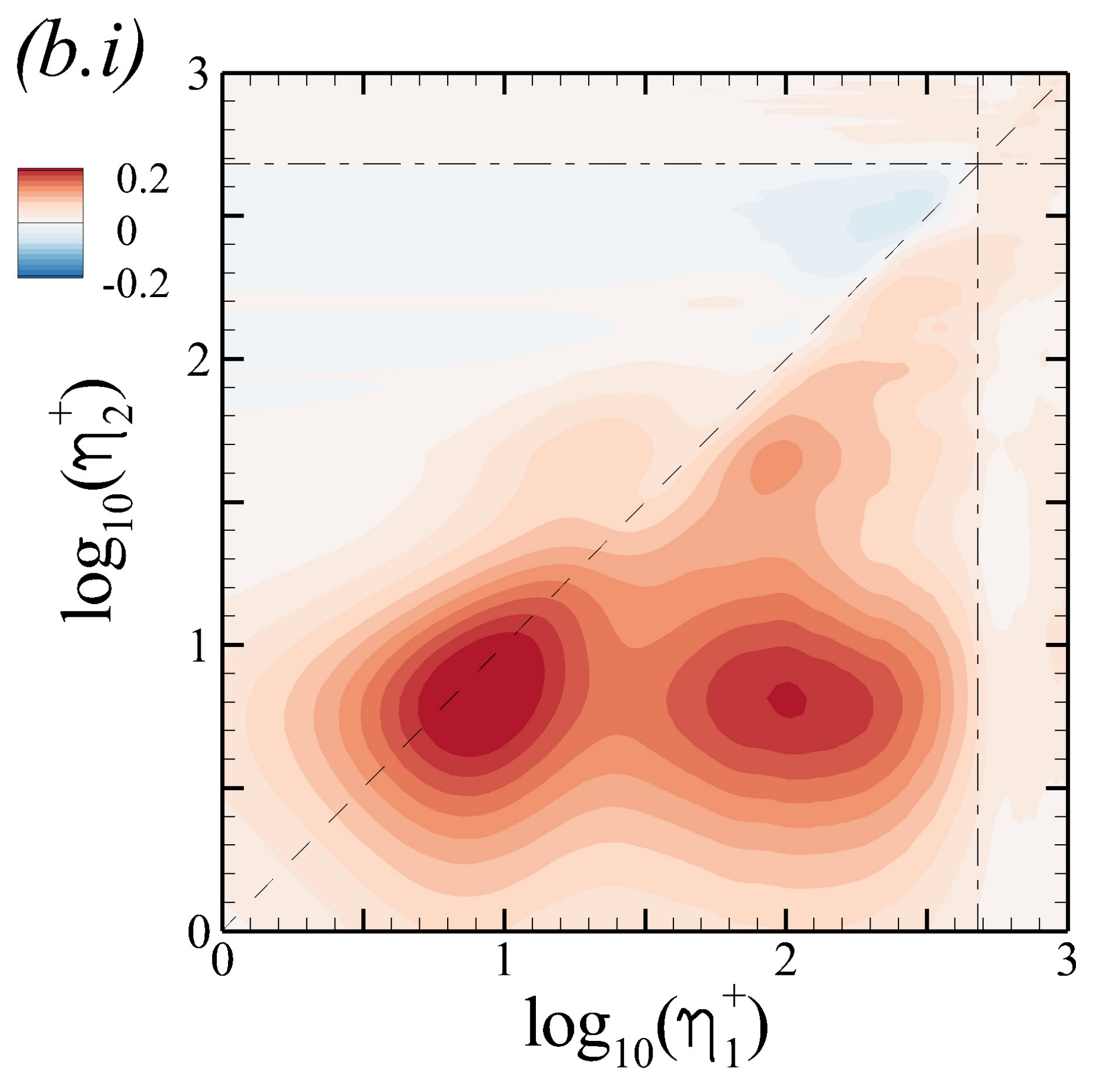}
\includegraphics[width=0.45\textwidth]{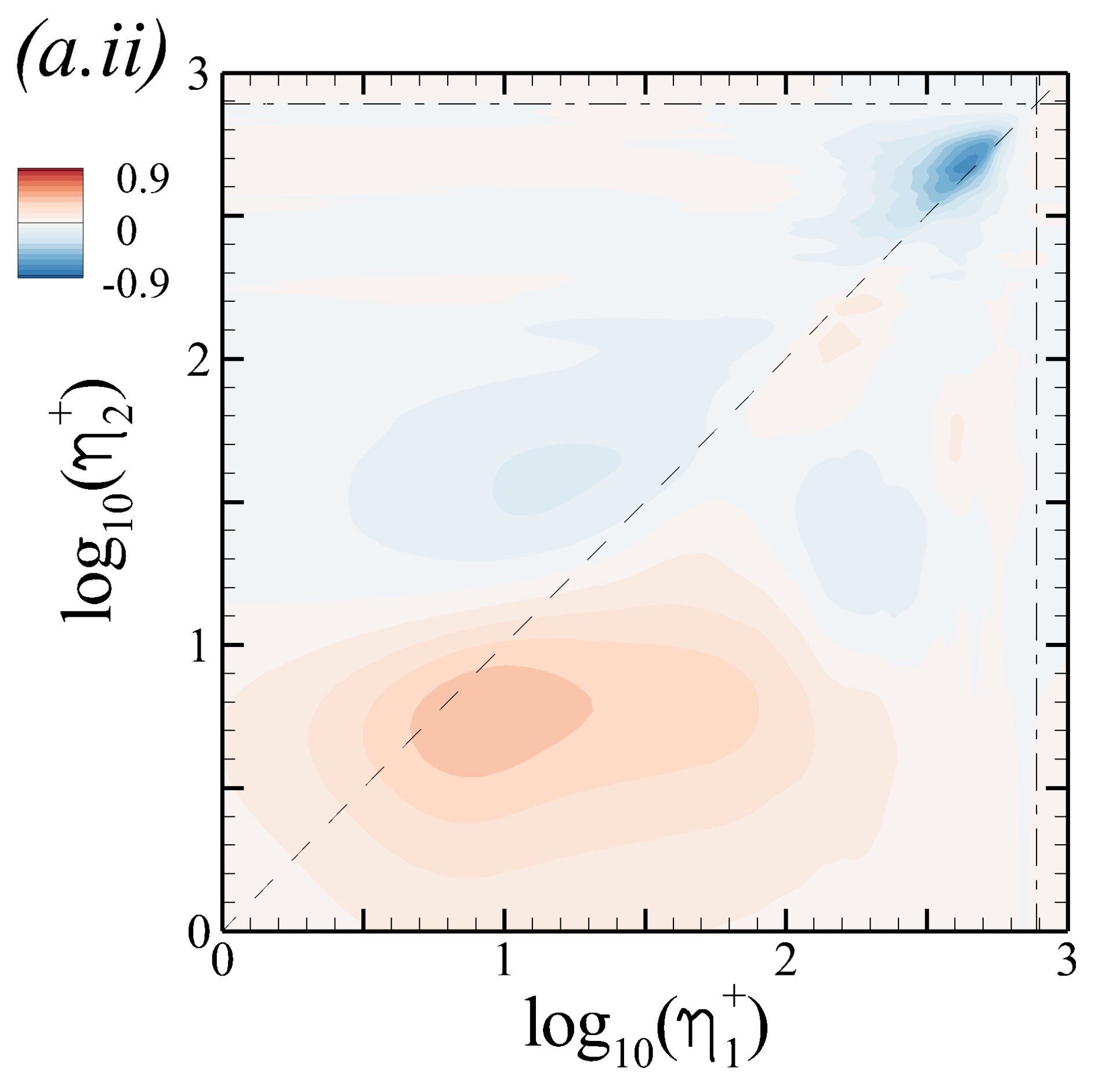}
\includegraphics[width=0.45\textwidth]{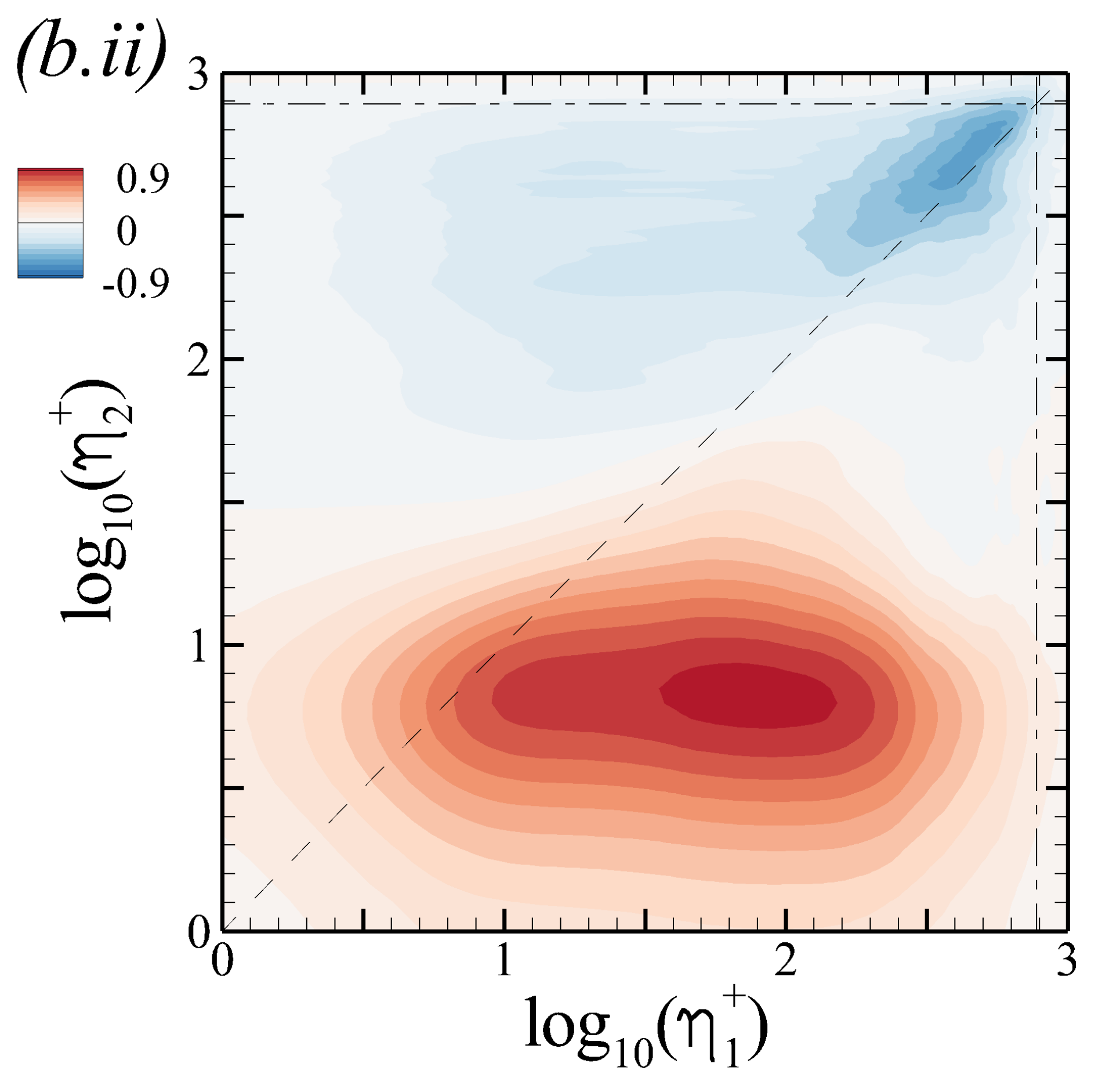}
\caption{Amplitude-modulation coefficient $C_{u_\xi,u_\xi}/u_\tau^2$ for $(a)$ REF and $(b)$ FRC cases.
	$(i)$~Flat section ($\xi= -50$) and $(ii)$ curved section ($\xi= 100$; $30^\circ$ station). 
	Vertical and horizontal dash-dotted lines (\chaindot) mark the edge of the boundary layer, $\eta = \overline{\eta_I}$. 
	}
\label{fig_am}
\end{figure}

\citet{Bernardini_2011} defined the two-point amplitude modulation (AM) coefficient $C_{u_\xi,u_\xi} (\eta_1,\eta_2)$ in order to quantify the influence of the large-scale motion at $\eta_1$ on the near-wall motion at $\eta_2$, where
\begin{eqnarray}
\label{eqn_amp}
	C_{u_\xi,u_\xi} (\eta_1,\eta_2) = \overline{{u_\xi}'_L(\eta_1) {u_\xi}'_{EL} (\eta_2)},
\end{eqnarray}
${u_\xi}'_L$ is a large-scale velocity, and ${u_\xi}'_{EL}$ is a filtered envelope of a small-scale part.
The cut-off filter for the large- and small-scale signal is set to $0.5 \overline{\eta_I}$, which from the spectra shown in figure~\ref{fig_spec} should discriminate these scales.
In the flat section ($\xi = -100$), figure \ref{fig_am}$i$ shows that upstream boundary layer exhibit a weak level of modulation beneath a quiescent free stream, and this effect is enhanced under FST.  
In the latter case, the outer large-scale motions at $\eta_1^+ \approx 100$ show relatively higher level of modulation of the small scales at $\eta_2^+ \approx 7$. These results are typical of flat plates, and the values of $C_{u_\xi,u_\xi} (\eta_1,\eta_2)$ would increase downstream if the flat section were extended. 
A qualitative change takes place over the curved section:  
Figure \ref{fig_am}$(ii)$ highlights a much stronger modulation (note the change in the contour levels), especially in the forced flow where the coefficient at  $(\eta_1^+, \eta_2^+) \approx (100, 7)$ is now the most dominant and is nearly twice as high as without FST.  
The results for FRC are the outcome of the energetic footprint of the large scales that persists near the wall (figure \ref{fig_spec}) and leads to a faster recovery of near-wall small scales (see figure  \ref{fig_spec2}).

\begin{figure}
	\begin{center}
		\includegraphics[width=0.49\columnwidth]{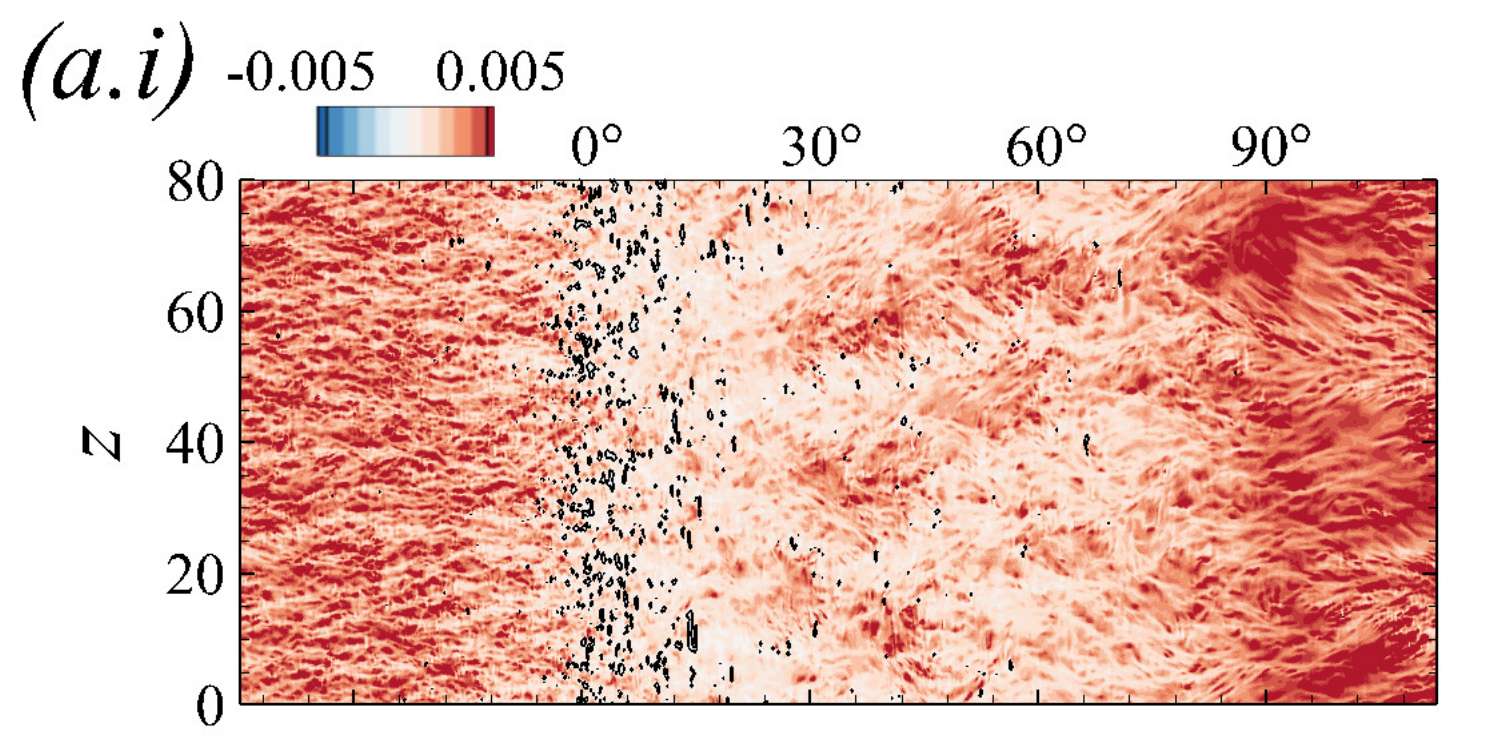}
		\includegraphics[width=0.49\columnwidth]{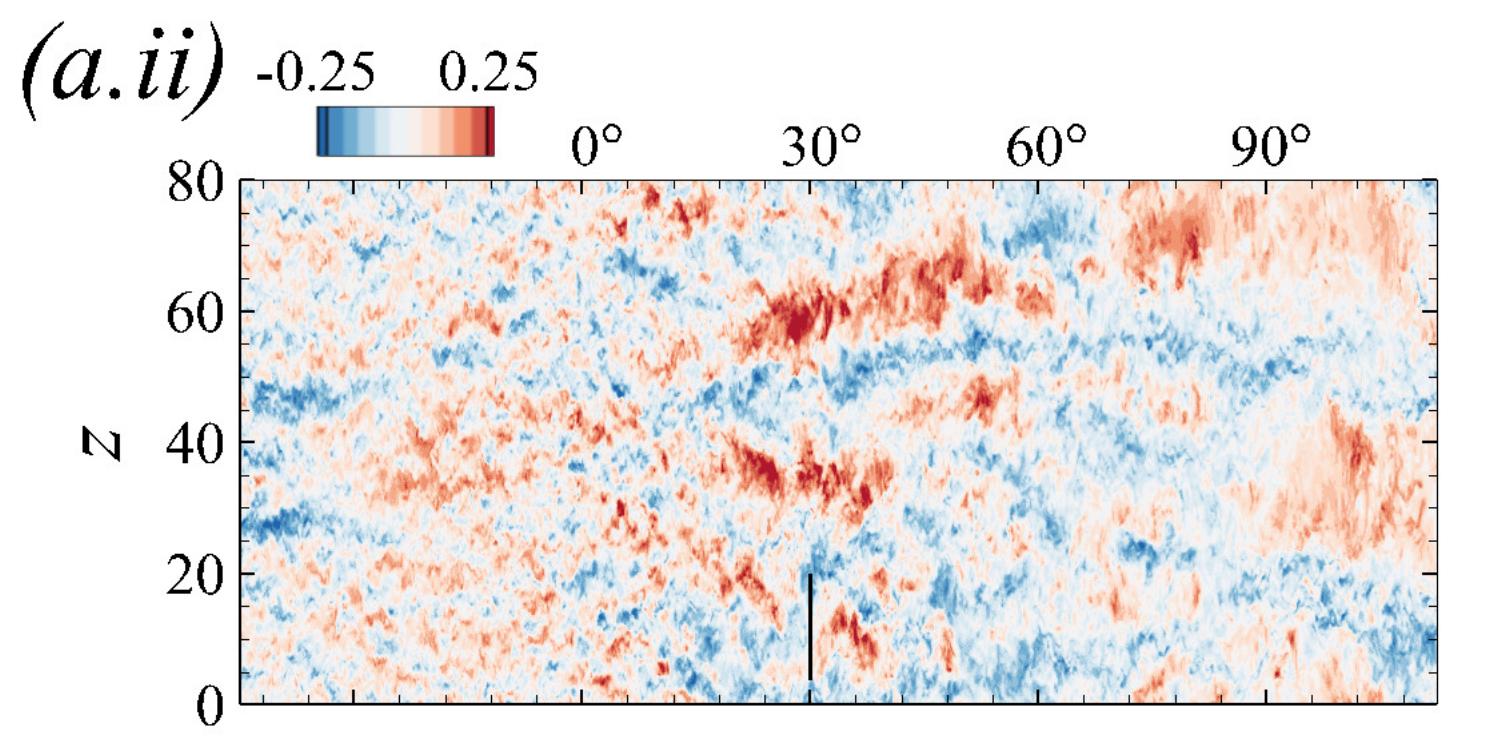}
		\includegraphics[width=0.49\columnwidth]{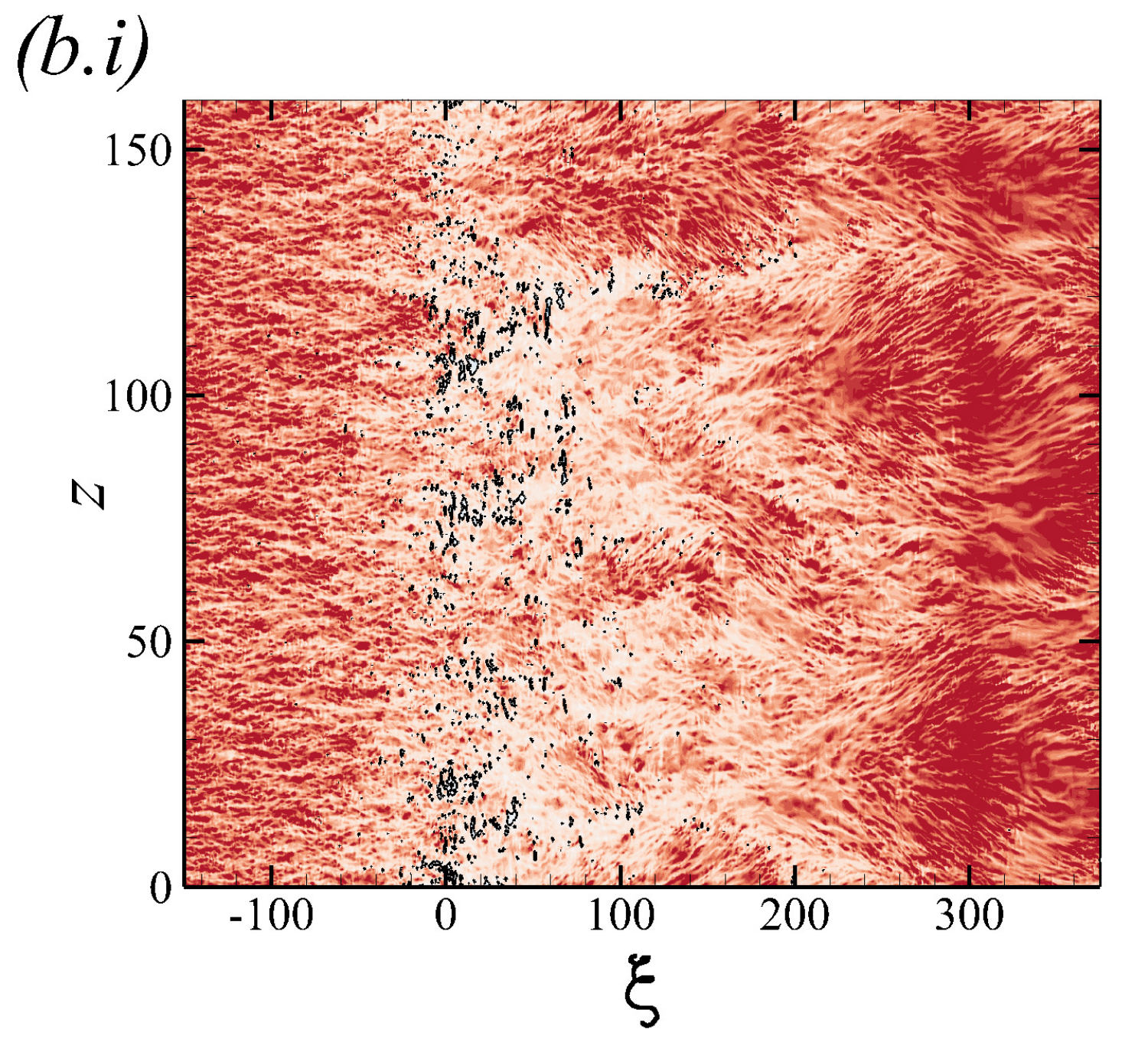}	
		\includegraphics[width=0.49\columnwidth]{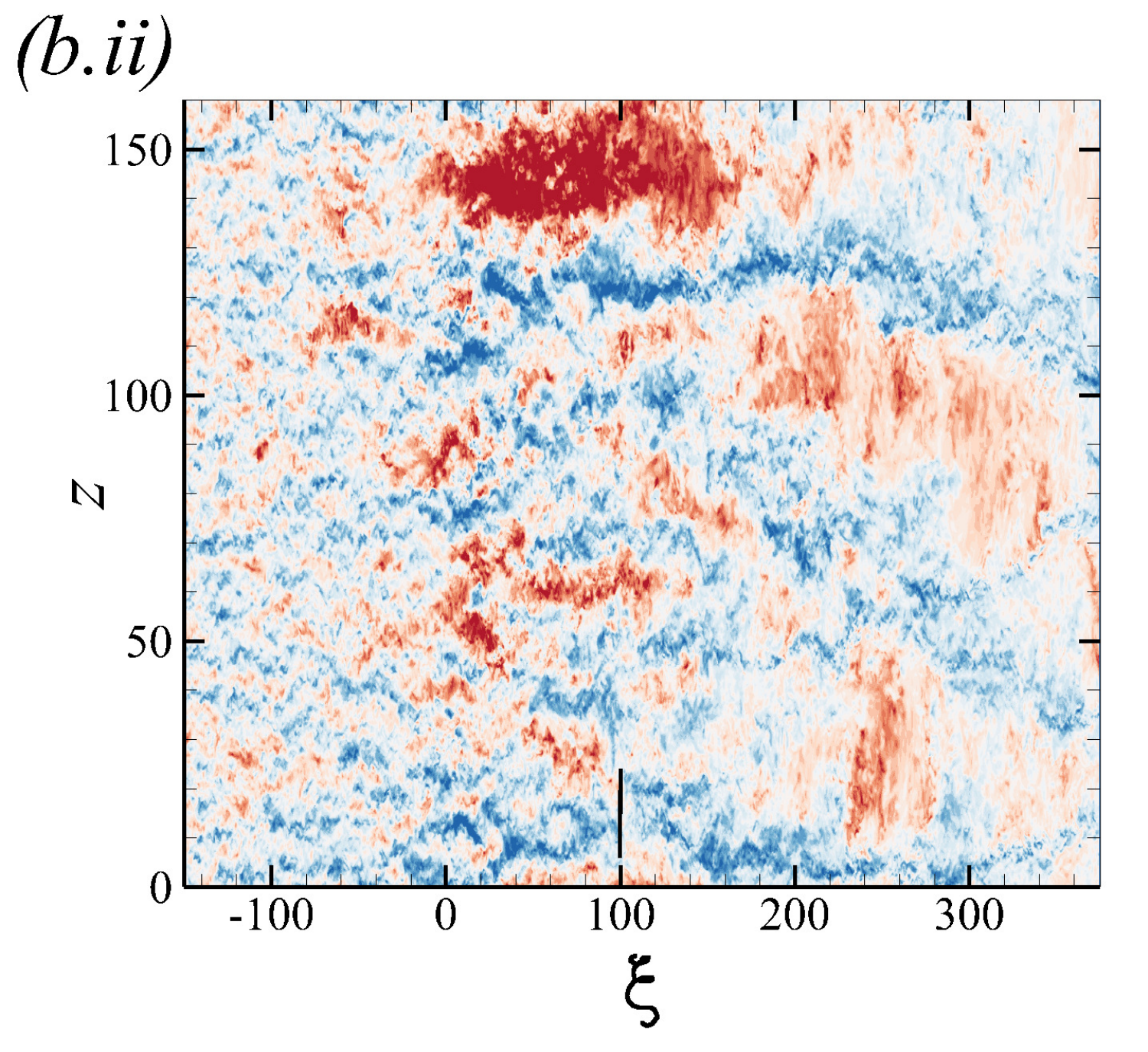}
	\end{center}
	\caption{$(i)$ Top view of instantaneous skin-friction coefficient, $-0.005\leq c_f \leq 0.005$ and black isocontour marks $c_f = 0$.
	$(ii)$ Instantaneous streamwise-velocity fluctuations $-0.25\leq u_\xi'\leq 0.25$ at $\eta \approx 3$ ($\eta^+ \approx 100$ at $\xi=0$). 
	$(a)$ REF; $(b)$ FRC.  
	The short black line in $(ii)$ is a graphical length scale equal to $\overline{\eta_I}$ at $\xi=100$.
	}
	\label{fig_cf_uf}
\end{figure}

The deep ingestion of outer fluid by the G\"ortler structures not only modulates the near-wall scales but also influences the stress at the wall which is of practical interest.   
In order to empirically demonstrate this connection, figure \ref{fig_cf_uf}$(i)$ shows an instantaneous realization of the skin-friction coefficient $c_f$, and panels $(ii)$ show the corresponding tangential velocity fluctuations $u_\xi'$ in the outer region. 
The contours of $c_f$ initially have a streaky pattern that is disrupted by the APG at the onset of curvature.
However, the most important observation is downstream, where localized large-scale regions of high $c_f$ are observed on the curved wall beneath outer large-scale and high-amplitude $u_\xi'$ perturbations.

Close inspection of figure \ref{fig_cf_uf}$(i)$ also provides visual support of faster recovery of the small-scale near-wall structures. 
For both REF and FRC, the streamwise-aligned near-wall streaks have an intense signature in the wall stress on the flat plate, which is abruptly disrupted near the onset of curvature $\xi \approx 0$.
The re-emergence of this signature across the span is clear near the end of the curved wall, $\xi \approx 300$. 
The key observation, however, is free-stream turbulence re-establishes these streaky structures much earlier upstream on the curve, albeit intermittently, as early as $\xi\approx 50$ for this particular flow realization.

\begin{figure}
	\begin{center}
		\includegraphics[width=0.49\columnwidth]{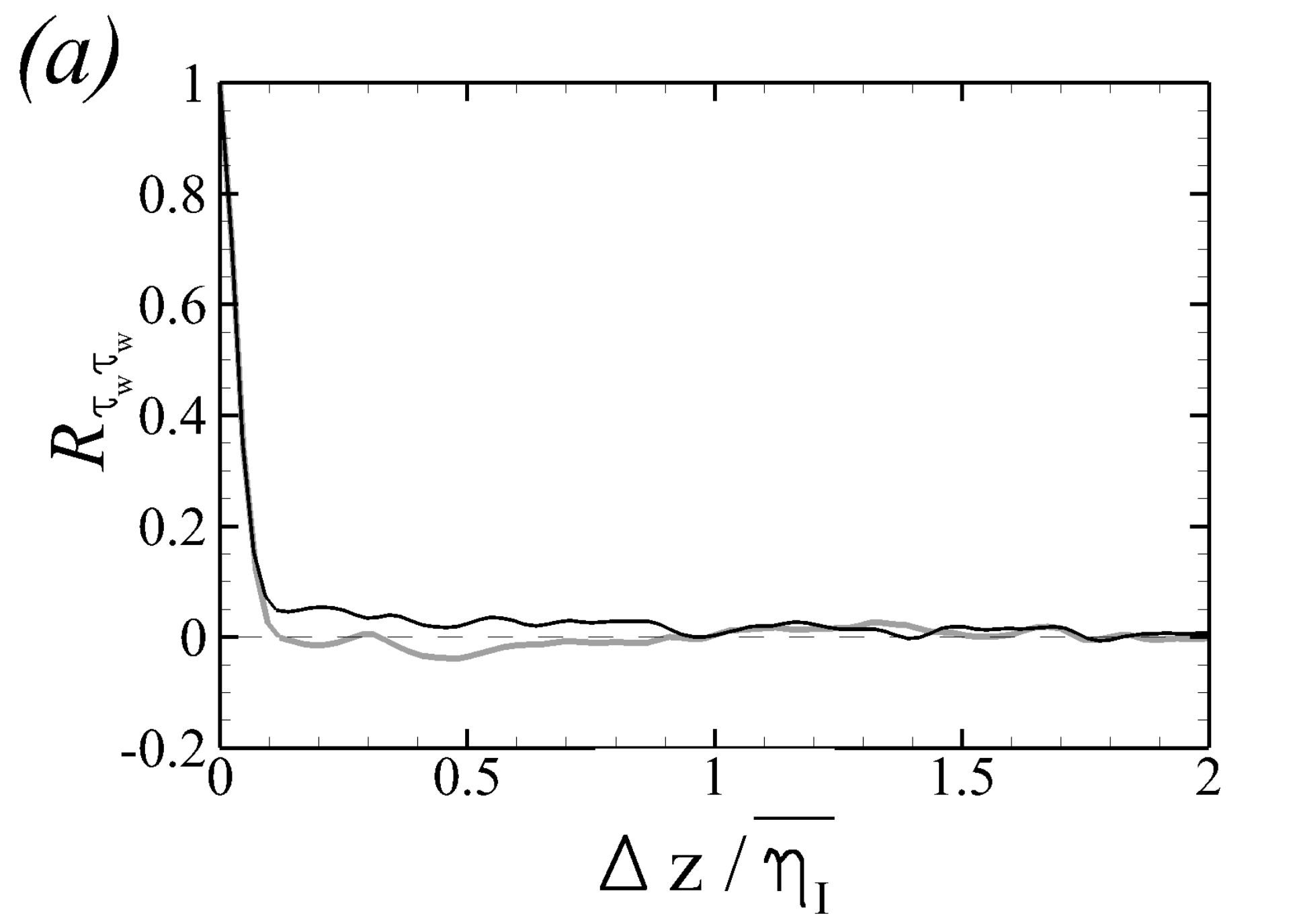}
		\includegraphics[width=0.49\columnwidth]{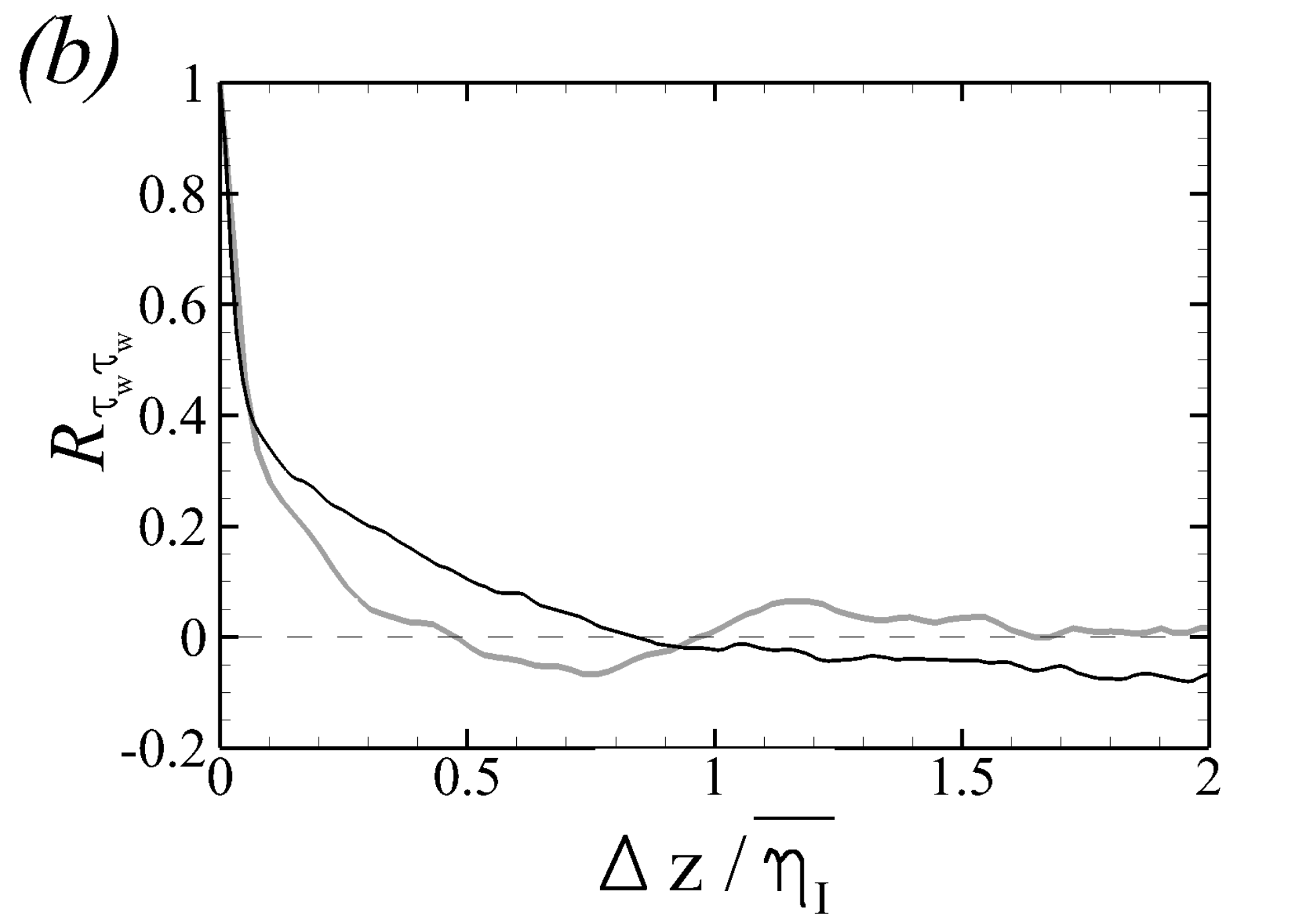}
	\end{center}
	\caption{Spanwise two-point correlation of the wall-shear stress $R_{\tau_w \tau_w}$ at $(a)$ $\xi= -100$ and $(b)$ $\xi = 100$ ($30^\circ$ station). (Gray) REF and (black) FRC. 
} 
	\label{fig_Rt}
\end{figure}

The above empirical observations are quantified, firstly using the normalized two-point correlation of the wall stress in the span, 
\begin{equation}
	R_{\tau_w\tau_w}(\xi,\Delta z)=\frac{\overline{\tau_w(\xi,z) \tau_w(\xi,z+\Delta z)}}{\overline{\tau_w(\xi,z)\tau_w(\xi,z)}}.
\end{equation}
Figure \ref{fig_Rt}$a$ shows that, on the flat section ($\xi=-100$), the wall stress $\tau_w$ decorrelates within a fraction of the boundary layer thickness. 
In contrast, on the curve, the stress is much more correlated in the span, and the correlation length is much wider in presence of free-stream forcing potentially due to the larger size of the outer G\"ortler motions.

\begin{figure}
	\begin{center}
		\includegraphics[width=0.49\columnwidth]{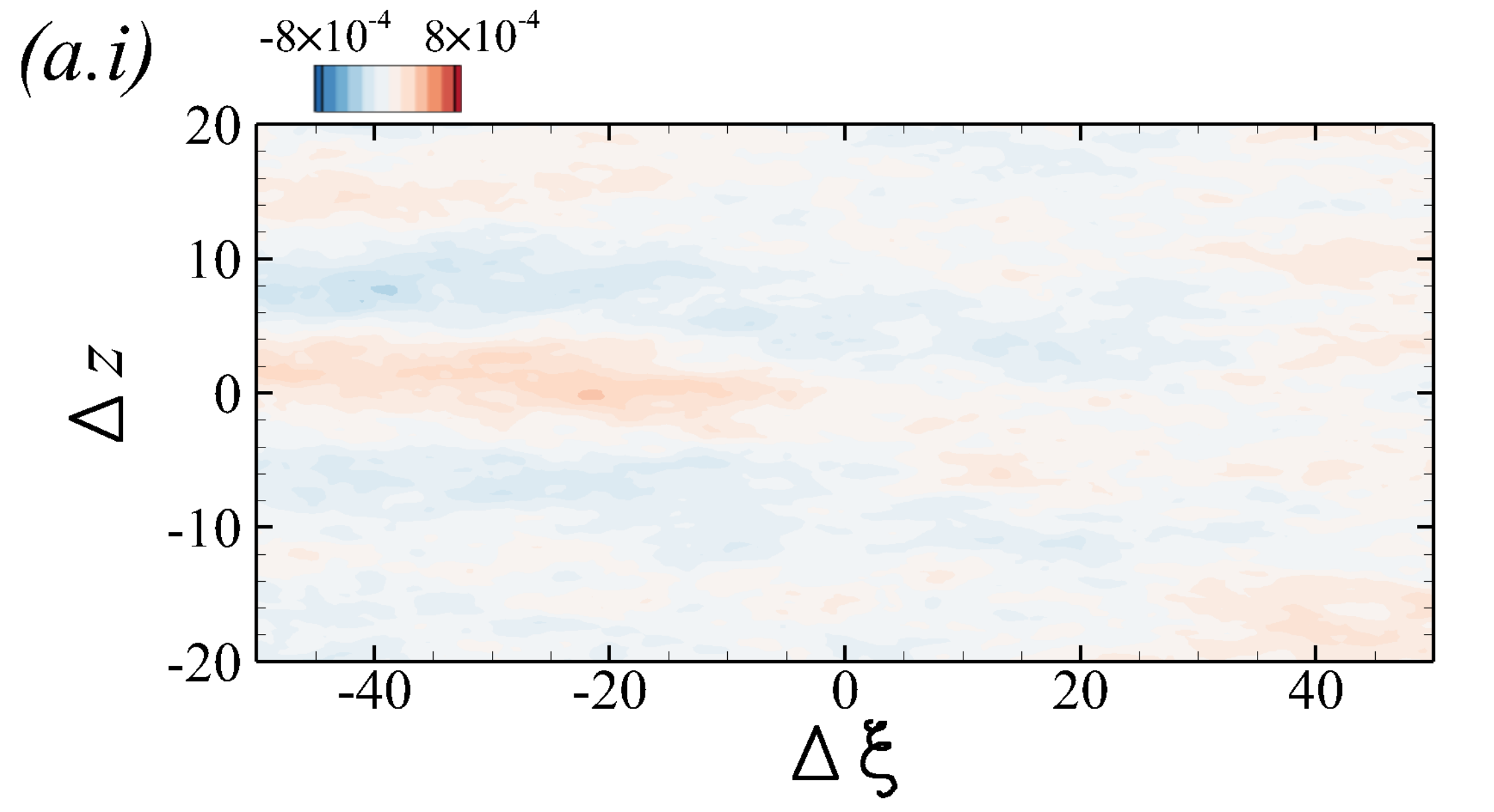}
		\includegraphics[width=0.49\columnwidth]{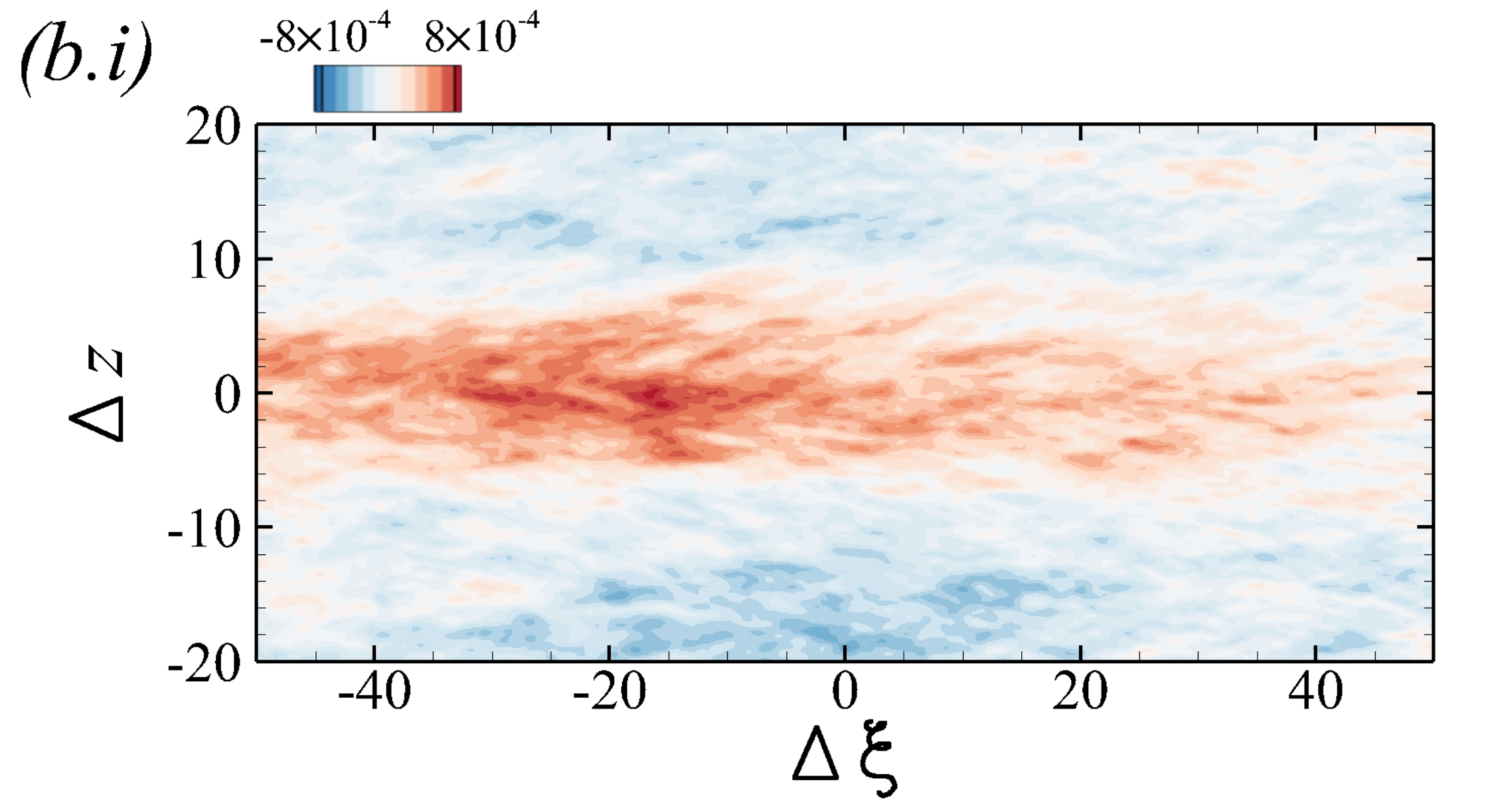}
		\includegraphics[width=0.49\columnwidth]{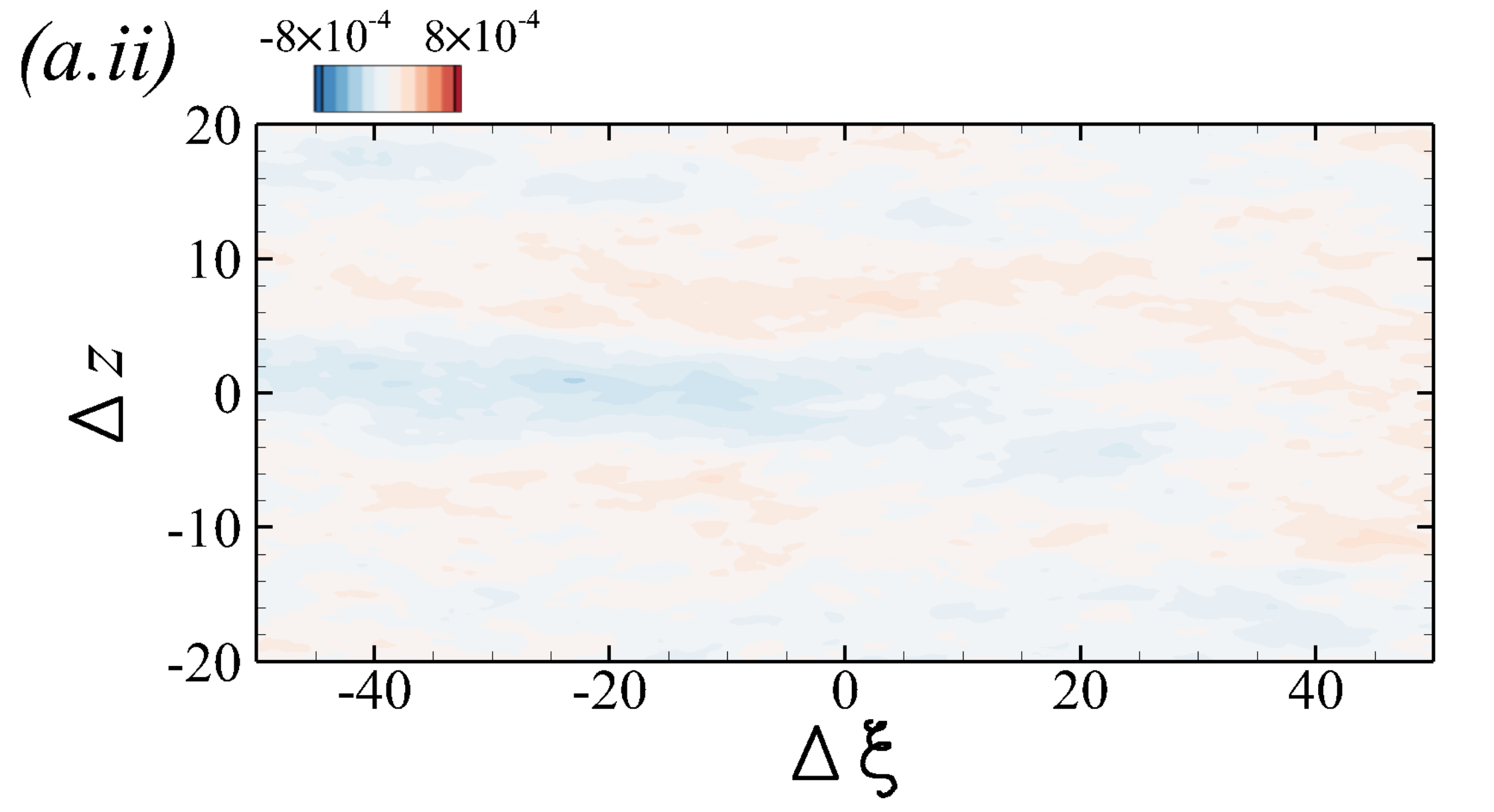}
		\includegraphics[width=0.49\columnwidth]{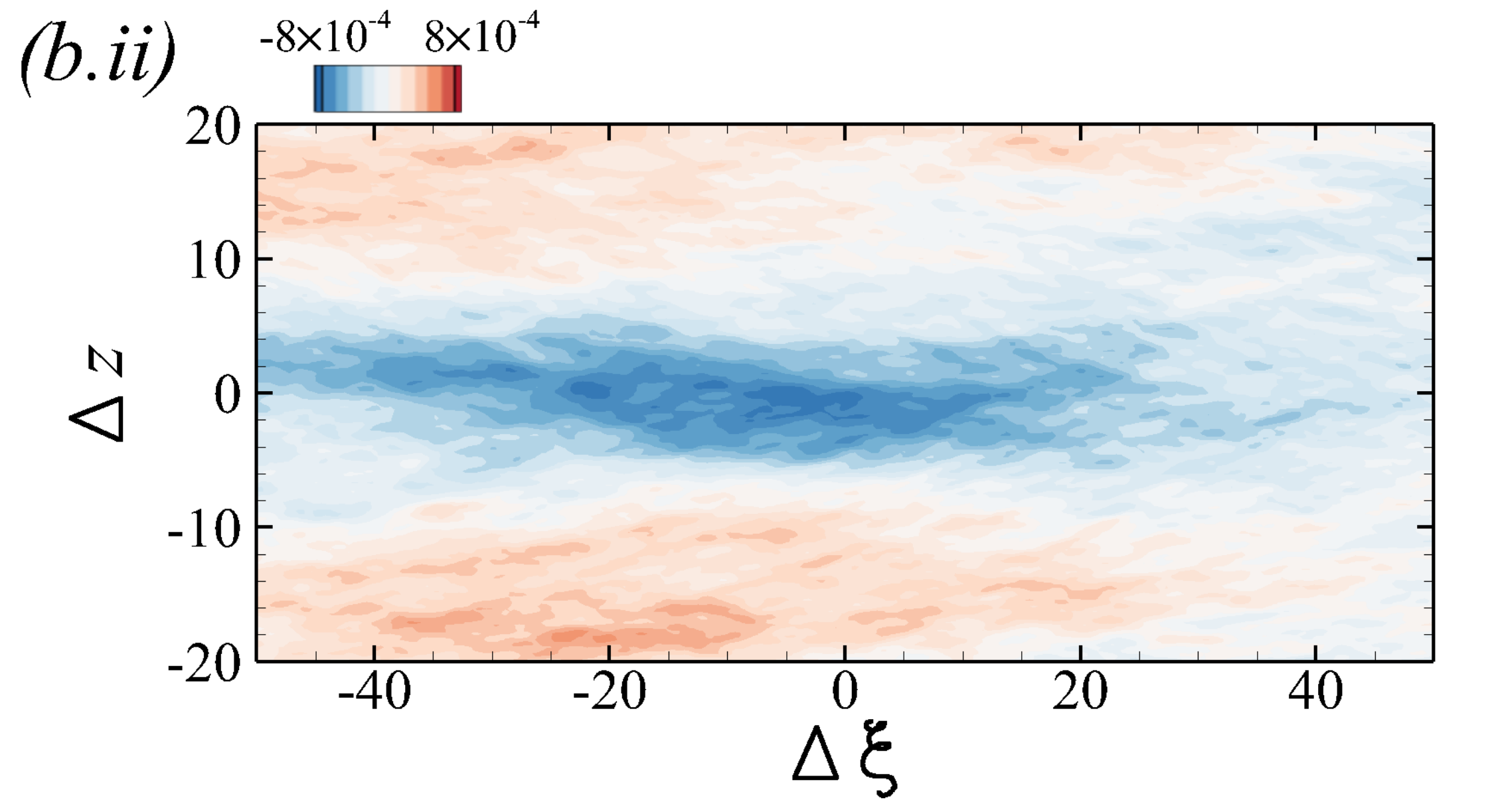}
	\end{center}
	\caption{Conditionally averaged skin-friction coefficient from $(a)$~REF and $(b)$~FRC at $\xi=100$ ($30^\circ$ station). $(i)$~$|c_f^P| \leq 8\times 10^{-4}$ and $(ii)$~$|c_f^N|\leq 8\times 10^{-4}$.} 
	\label{fig_cf_up_cores}
\end{figure}

Here we examine whether the wall stress beneath the outer roll motions is indeed statistically large in scale and magnitude, not only in a single realization as shown in figure \ref{fig_cf_uf} but in a statistically significant fashion.  
This point is upheld by directly evaluation the average perturbation wall stress, 
\begin{align}
	 {c^P_f} (\Delta \xi,\Delta z) &=\overline{c_f' (\xi+\Delta \xi,z+\Delta z) ~~|~~ {\mathcal{C}^P}_{|30<\eta^+<\eta_I^+} } ~~\text{and} \\ 
	 {c^N_f} (\Delta \xi,\Delta z) &=\overline{c_f' (\xi+\Delta \xi,z+\Delta z) ~~|~~ {\mathcal{C}^N}_{|30<\eta^+<\eta_I^+} }, 
\end{align}
conditional on the presence of an outer large-scale tangential velocity structure as a surrogate for the G\"ortler structures. 
The results are reported in figure~\ref{fig_cf_up_cores}. 
In both the REF and FRC flows, the peak perturbation stresses are recorded at  $\Delta \xi < 0$, which is consistent with an inclination of the outer structures relative to the wall.
However, in the forced case, the magnitude of perturbations in the wall stress is larger, and the footprint of the outer large scales is greater in extent.  
In fact, the affected region extends over a longer streamwise distance $\Delta \xi$ (both positive and negative $\Delta \xi$) due to the coherence of the outer large scales in this case (c.f.\,figures \ref{fig_cond_str1} and \ref{fig_cond_str2}). 
In addition, the opposite wall stress perturbations are observed in the span which is consistent with the influence of roll motions displacement momentum away and towards the wall at spanwise locations that are separated by their width.

The results presented here provide a structural interpretation of the statistical changes of the boundary layer on the curved surface, without and with free-stream turbulence. 
While G\"ortler structures may be elusive in the former case, they are more prominent in the forced flow.
Their higher amplitude, larger size and strong coherence are evident in the spectra and in conditional averages.  
The more energetic G\"ortler structures also promote the recovery of the boundary layer downstream of the APG at the onset of curvature, increase  mixing in the wall-normal direction, and appreciably enhance the wall stress in their footprint.

\section{Conclusion}
Direct numerical simulations of turbulent boundary layers on a concave curve without and with free-stream turbulence were contrasted. 
In the latter case, the inlet turbulence was homogeneous and isotropic with intensity $Tu = 10\%$.  
The boundary layer and the free-stream turbulence were differentiated using a levelset approach which provides an objective, virtual interface. 

Near the onset of the curvature, the drag significantly drops due to the adverse pressure gradient which induces intermittent flow separation. The forcing by free-stream turbulence reduces the probability of separation and promotes the downstream recovery of the flow along the ZPG section of the curve.
The external forcing also increases the skin friction by up to 49$\%$ relative to the reference flow (figure \ref{fig:cfPDFwithXi:option2}).

The curved wall has a clear impact on the wall-normal and spanwise normal stresses, both indicative of the development of G\"ortler vortices in the outer region (figure \ref{fig_stress_cont}).
The wall-normal separation between the peaks of these components of stress increases with downstream distance, signaling the growth of the vortices as the boundary layer expands (figure \ref{fig_ip_op}).
When the FST is present, the outer roll motions are more energetic and larger is size. 
In addition, and in contrast to flat-plate boundary layers, introducing FST increases the peak shear-stress correlation coefficient in the curved wall flow (figure \ref{fig_uv_coeff}$b$).

Above the curved section, the turbulence structures are significantly modified.
Curvature spurs the formation of outer roll motions which increase in size on the concave wall as seen in the conditionally averaged flow fields (figures \ref{fig_cond_str1} and \ref{fig_cond_str2}).
In the forced flow, these longitudinal structures, which we interpret as G\"{o}rlter vortices, are enlarged in size, strengthened in intensity, display a clear streamwise coherence, and enhanced wall-normal mixing. 
As a result, the probability of observing free-stream fluid remains finite within the buffer layer (figure \ref{fig:gamma:new}), in contrast to earlier results for flat-plate boundary layers where external turbulence could not reach the buffer layer \citep{You_2019}.  
The impact on the near-wall region is direct (figure \ref{fig_am}), in particular in the wall stress at their footprint (figure \ref{fig_cf_up_cores}).

\section*{Acknowledgements}
This work is sponsored by the National Science Foundation (grant 1605404).
Computational resources were provided by the Maryland Advanced Research Computing Center (MARCC).


\appendix
\section{The momentum thickness}\label{app:momentumthickness}
\begin{figure}
	\begin{center}
		\includegraphics[width=0.48\columnwidth]{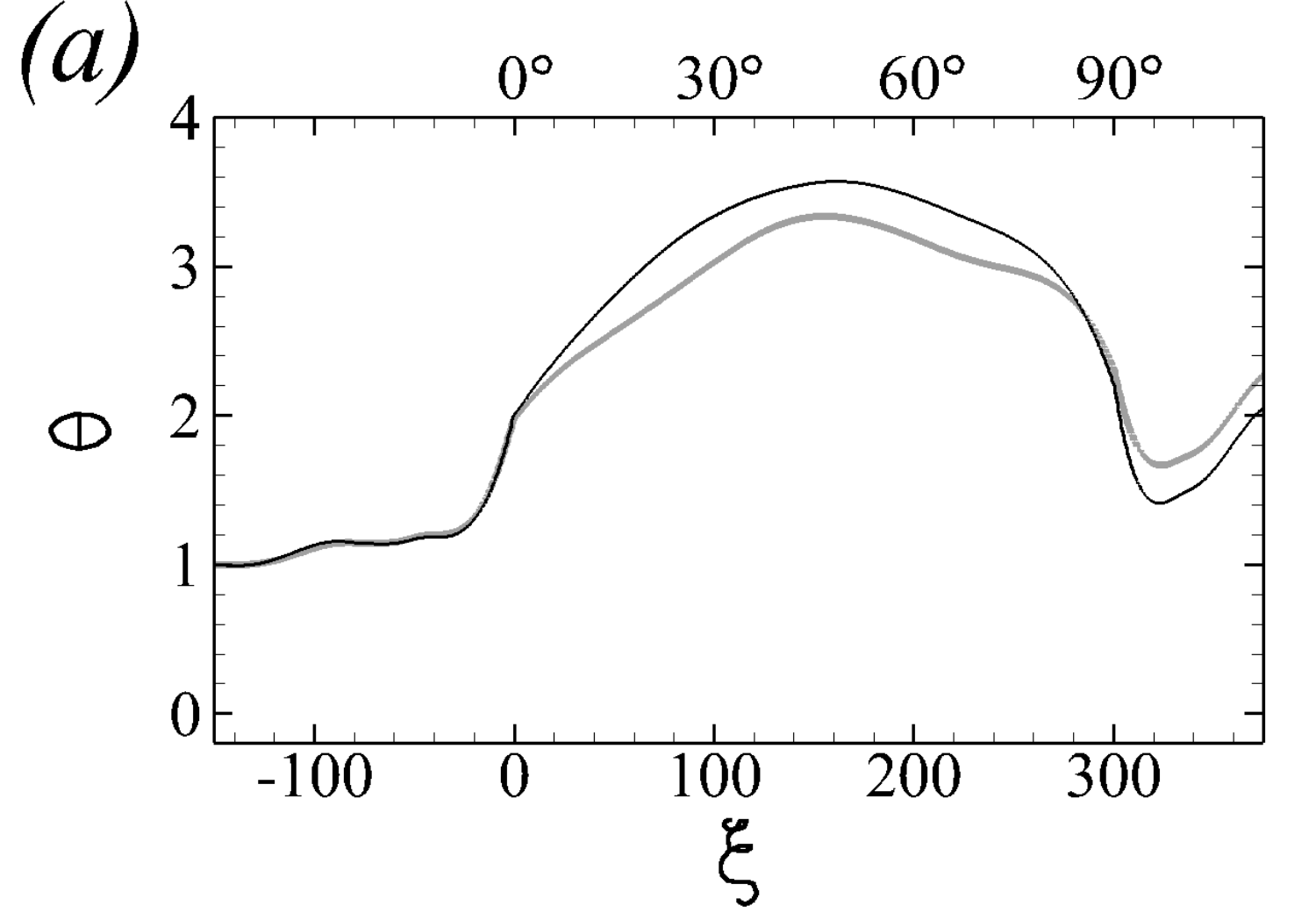}
		\includegraphics[width=0.48\columnwidth]{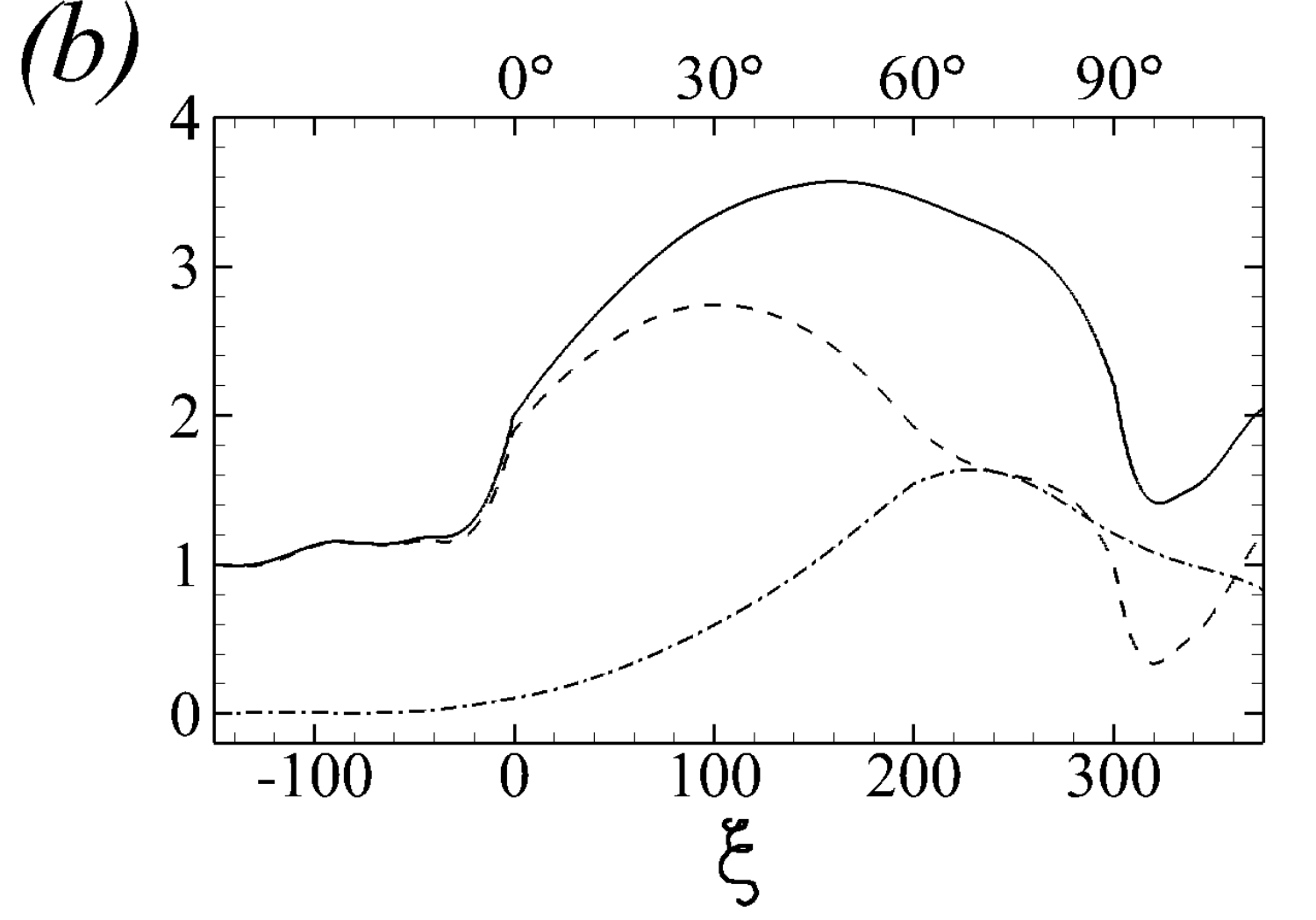}
	\end{center}
		\caption{$(a)$ Development of momentum thickness $\theta$ in (gray) REF and (black) FRC.
	$(b)$ For FRC, (\lline)~$\theta$ is further decomposed into the two contributions (\dashed~and~\chaindot) on the right hand side of equation (\ref{eqn_theta}).}
	\label{fig_theta}
\end{figure}

The momentum thickness is an important proxy of the state of the boundary layer. 
Various definitions have been proposed for curved-wall flows, and we here consider the form adopted by \citet{Patel_1969},  
\begin{equation}\label{eq:momthick}
    \theta =   \int_{0}^{\eta_p}\frac{\overline{u_\xi}}{U_{pw,R}} \left( \frac{U_{p,R}}{U_{pw,R}} - \frac{\overline{u_\xi}}{U_{pw,R}} \right)  d\eta.
\end{equation}
The subscript $R$ indicates that we are adopting the profile of the potential velocity from the REF case, which is not necessarily preserved in the FRC case.  For this reason, we also perform the integration up to the wall-normal position $\eta_p$ where the potential-velocity profiles from both REF and FRC intersect.  
The downstream dependence of $\theta$ is reported in figure \ref{fig_theta}$a$ which shows the anticipated increase in the momentum thickness near $\xi=0$ due to flow deceleration. 
On the curved wall $\theta$ is larger in the FRC case; this effect is seemingly consistent with flat-plate boundary layers where the change in  $\theta$ can be directly related to increase in skin friction under free-stream turbulence forcing.
The reality is, however, more complex as can be shown by re-expressing the momentum thickness as, 
\begin{align}
\label{eqn_theta}
\theta 
&= \frac{(U_{pw})^2}{(U_{pw,R})^2} \underbrace{\int_0^{\eta_p} \frac{\overline{u_\xi}}{U_{pw}} \left( \frac{U_p}{U_{pw}}- \frac{\overline{u_\xi}}{U_{pw}} \right) d\eta}_{= \tilde{\theta}}
           + \int_0^{\eta_p} \frac{\overline{u_\xi}}{U_{pw,R}} \left( \frac{U_{p,R}}{U_{pw,R}}- \frac{U_p}{U_{pw,R}} \right)  d\eta. 
\end{align}
The first term involves a momentum thickness $\tilde{\theta}$ defined using the potential velocity of each flow (instead of adopting $U_{p,R}$ for both REF and FRC), and the second term is due to the difference of the potential velocities in the two simulations. 
Figure \ref{fig_theta}$b$ shows that the increase in $\theta$ in FRC is due to this second term.  
The thicker boundary layer in the forced case increases the effective curvature, specifically $\overline{\eta_I}/R$ or similarly $\delta_{99}/R$, and hence turning of the outer flow.  The result is a steeper $U_p$ profile in the free stream, a lower $U_p$ inside the boundary layer and increased deficit $U_{p,R}-U_p$.
Interpreting the momentum thickness must therefore take this effect into account.


\bibliographystyle{jfm}
\bibliography{references}

\end{document}